\documentclass[3p,times,preprint,letterpaper]{elsarticle}

\usepackage{ecrc}

\volume{00}

\firstpage{1}

\journalname{Nuclear Instruments and Methods in Physics Research A}

\runauth{C.J. Glaser et al.}

\jid{procs}

\jnltitlelogo{Nucl. Instr. and Methods A}

\usepackage{amssymb}

\usepackage[mathlines,displaymath]{lineno}

\usepackage[figuresright]{rotating}

\usepackage{comment}
\usepackage{graphicx}
\usepackage{epstopdf}
\usepackage{amsmath}
\usepackage{graphicx}
\usepackage{multirow}

\usepackage{color}
\usepackage[symbol]{footmisc}
\usepackage{hyperref}
\usepackage{float}
\usepackage[percent]{overpic}

\usepackage{hyperref}
\hypersetup{
  colorlinks,
  citecolor=black,
  filecolor=black,
  linkcolor=black,
  urlcolor=black
}

\usepackage{color,soul}
\definecolor{RED}{rgb}{1,0,0}
\setulcolor{RED}
\usepackage{xcolor}

\newcommand{\pieii}{\ensuremath{\pi_{\text{e}2}}}
\newcommand{\pimii}{\ensuremath{\pi_{\mu2}}}
\newcommand{\pieiig}{\ensuremath{\pi_{\text{e}2\gamma}}} 
\newcommand{\pieiisg}{\ensuremath{\pi_{\text{e}2(\gamma)}}}  
\newcommand{\pimiisg}{\ensuremath{\pi_{\mu2(\gamma)}}}
\newcommand{\Rpiemu}{\ensuremath{R^{\mkern2mu\pi}_{\text{e}/\mu}}}
\newcommand{\RpiSMemu}{\ensuremath{R^{\mkern2mu\pi,\text{SM}}_{\text{e}/\mu}}}
\newcommand{\RpiEXPemu}{\ensuremath{R^{\mkern2mu\pi,\text{exp}}_{\text{e}/\mu}}}
\newcommand{\rstop}{\ensuremath{r_{\text{stop}}}}
\newcommand{\fwfr}{1.0}

\makeatletter
\def\ps@headings{%
    \def\@oddhead{\parbox{\textwidth}{\itshape\footnotesize%
        \hfill\@runauth~/~\@journalname~\@vol~(\the\year)~%
          \@firstpage--\lastpage%
         \hfill{\rm \thepage}}}%
    \def\@evenhead{\parbox{\textwidth}{\itshape\footnotesize%
         {\rm \thepage}\hfil\@runauth~/~\@journalname~\@vol~(\the\year)~%
         \@firstpage--\lastpage\hfil}}%
    \let\@oddfoot\@empty%
    \let\@evenfoot\@oddfoot}
\makeatother

\begin{document}

\begin{frontmatter}
  \setlength{\baselineskip}{18pt}
  


\dochead{}
\title{Beam particle tracking with a low-mass mini time projection
  chamber in the PEN experiment\tnoteref{label1}}
\tnotetext[label1]{Dedicated to the memory of Andrey Korenchenko}

  \author{C.J.~Glaser}
  \author{D.~Po\v{c}ani\'c},
  \address{Institute for Nuclear and Particle Physics, University
    of Virginia, Charlottesville, VA 22904-4714, USA} 
  \author{A.~van~der~Schaaf}
  \address{Physik-Institut, Universit\"at Z\"urich, CH-8057 Z\"urich,
    Switzerland}   
  \author{V.A.~Baranov},
  \author{N.V.~Khomutov},
  \author{N.P.~Kravchuk},
  \author{N.A.~Kuchinsky}
  \address{Joint Institute for Nuclear Research, Dubna, Moscow Region, Russia} 

  \author{(10 May 2021)}

\begin{abstract}  \setlength{\baselineskip}{18pt}
The international PEN collaboration aims to obtain the branching ratio
for the pion electronic decay $\pi^+ \to e^+\nu_e(\gamma)$, aka \pieii,
to a relative precision of $5\times 10^{-4}$ or better.  The PEN
apparatus comprises a number of detection systems, all contributing
vital information to the PEN event reconstruction.  This paper discusses
the design, performance, and Monte Carlo simulation of the mini time
projection chamber (mTPC) used for pion, muon, and positron beam
particle tracking.  We also review the use of the extracted trajectory
coordinates in the analysis, in particular in constructing observables
critical for discriminating background processes, and in maximizing the
fiducial volume of the target in which decay event vertices can be
accepted for branching ratio extraction without introducing bias.
\end{abstract}


\end{frontmatter}

\newpage
\tableofcontents


\setlength{\baselineskip}{18pt}
\pagestyle{headings}
\newpage

\section{Introduction and motivation\label{sec:intro}}
Since the 1930s, the fields of experimental particle and nuclear physics
have relied on different means to record and analyze trajectories of
charged particles under study.  Over the years, diverse designs for
charged particle tracking detectors have emerged, and were put to use.
All of them operate on the principle that a charged particle moving
through matter (historically in a liquid or gas, more recently also
including solids, primarily semiconductor crystals) ionizes the atoms
near its trajectory.  The ionization pattern is recorded with the aim
to reconstruct the charged particle's trajectory for further event
analysis.  Current advancement on the high energy and precision
frontiers requires ever more sophisticated tracking detectors and
techniques.  The simple cloud chambers of the 1930s, crucial for the
discovery of the muon and positron, are no longer adequate for present
day experiments in nuclear and particle physics.

In the late 1970s, David Nygren\,\cite{Nygren} developed the time
projection chamber (TPC), a sophisticated gas-filled tracking chamber
capable of reconstructing a charged particle's trajectory in all three
dimensions, by combining the features of a multiwire proportional
chamber and a drift chamber.  In its many design variations, the TPC has
become a reliable and precise charged particle tracking device,
sufficiently fast for many applications, often used in magnetic
spectrometers.

TPCs have been used in numerous experimental projects over the past more
than forty years.  Typically, one coordinate is determined from the hit
anode wire, another from the signal attenuation along the resistive
wires, and the third from the drift time of electrons in the gas.  In
early applications, TPC sense wire output signals have been fed to
charged coupled devices (CCDs) for storage prior to being read by ADCs,
or directly to flash ADCs.  In pursuit of minimizing dead time and
increasing the acquired information, more recent experiments have chosen
to digitize TPC electrode waveforms.  Time projection chambers vary in
size, from small table-top devices to large detectors of many meters
across, used in high energy experiments.

The international PEN collaboration, led by the University of Virginia
group, has designed and built a small, low-mass, mini time projection
chamber (mTPC), for use within a much larger, complex detector system.
The PEN mTPC is the primary subject of this paper.

The Standard Model (SM) provides an exceptionally precise calculation of
the pion electronic decay branching ratio,
$\Rpiemu=\Gamma(\pi\to\text{e}\bar{\nu}
(\gamma))/\Gamma(\pi\to\mu\bar{\nu}(\gamma))$:
\begin{linenomath}
\begin{equation}
  \RpiSMemu= 
    \begin{cases}
      (1.2352\pm 0.0005)\times 10^{-4}\ \quad \text{\cite{Mar93}}, \\
      (1.2354\pm 0.0002)\times 10^{-4}\ \quad \text{\cite{Fin96}}, \\
      (1.2352\pm 0.0001)\times 10^{-4}\ \quad \text{\cite{Cir07}}.
    \end{cases}
    \label{eq:sm_br}
\end{equation}
\end{linenomath}
This level of theoretical precision, unmatched among decays of other
mesons, provides unique opportunities for tests of SM predictions.  The
$\pi_{\ell2}$ decay, $\pi^- \to \ell\bar{\nu}_\ell(\gamma)$,  connects the
pion pseudoscalar $0^-$ state to the $0^+$ vacuum.  The strong helicity
suppression of the $\pi_{e2}$ decay makes this process uniquely
sensitive to a class of pseudoscalar ($P$), or $P$-loop coupled,
non-$(V$$-$$A)$ contributions, arising from new, ``beyond Standard
Model'' (BSM) physics, undetectable in analogous, helicity-unsuppressed
leptonic decays, such as the $\pi \to \mu \bar{\nu}$, or $\pi_{\mu2}$.
At the precision of $10^{-3}$, \RpiEXPemu probes the $P$, $A$, and $S$
BSM mass scales up to 1000, 20, and 60\,TeV, respectively, sets
competitive limits on the violation of electron-muon universality, on
certain SUSY partners, neutrino sector anomalies and massive sterile
neutrinos (for a survey of rare pion decays see, e.g.,
Ref.~\cite{Poc14}).  We note that lepton universality, axiomatic in the
SM, may be in question following recent LHCb results~\cite{LHCb21}.
Sadly, the current value $\RpiEXPemu = 1.2327(23)\times
10^{-4}$\,\cite{Agu15,PDG20} lags in precision behind the theoretical SM
evaluations by an order of magnitude.
The objective of the PEN experiment is to obtain the pion electronic
decay branching ratio, \Rpiemu with a relative uncertainty of $5\times
10^{-4}$ or better.

This paper discusses the design and performance of the mini time
projection chamber, as well as its use in the PEN analysis.
Section~\ref{sec:pen-det} reviews the PEN detector system.
Section~\ref{sec:mTPC-design} discusses the design and construction of
the mTPC.  Section~\ref{sec:mTPC-recon} explains and documents the mTPC
performance in extracting the trajectory coordinates from raw electrode
signals.  Section~\ref{sec:MC-simul} discusses the generation of
realistic mTPC synthetic data using Monte Carlo (MC) simulation methods.
Section~\ref{sec:mTPC-in-PEN-anal} demonstrates the use of the mTPC
information in the PEN analysis with a focus on ensuring reliable
calibration of the detection efficiency.

\section{The PEN experiment and  detector system \label{sec:pen-det}}
The PEN data were acquired in the $\pi$E1 beam area of the Paul Scherrer
Institute (PSI) Ring Cyclotron, Switzerland, during three runs, from
2008 through 2010, for a total of approximately 25 weeks of in-beam data
acquisition, using the apparatus described below.  The data analysis is
in an advanced stage for all relevant decay channels in parallel.  The
main \RpiEXPemu analysis is blinded.  The unblinding of the \RpiEXPemu
result is planned after the completion of a full set of papers
describing the analysis (of which this is one), and of the final
analysis pass with fine parameter/cut adjustments, minimizing the total
uncertainty, plus ensuring that all the analyzed decay observables are
mutually fully consistent in a comprehensive analysis with independent
checks.

Since $\Rpiemu \approx$\,$10^{-4}$, the goal precision of PEN, relative
to the observed number of pion decays, is of ${\mathcal O}(10^{-8})$.
Even though many uncertainties, such as those associated with the number
of stopped pions, and the solid angle acceptances for the two channels,
cancel when building the ratio \Rpiemu, the high level of precision
still results in stringent demands on the accurate treatment of various
subtle effects, such as pion and muon decays in flight and radiative
corrections.  One of the key supporting analysis tasks is beam particle
tracking, accomplished trough the use of the PEN mTPC, optimized for
small size, low mass, high double-pulse resolution, as well as easy
installation and alignment.

\begin{figure}[t]
\centering
\includegraphics[width=0.67\linewidth]{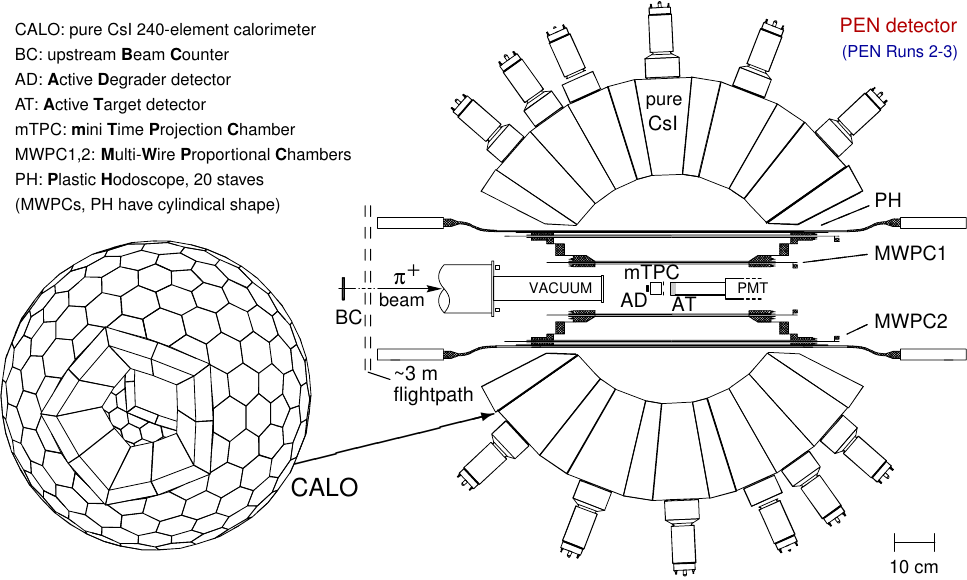}
\caption{Annotated schematic section through the PEN detector system for
  Runs\,2--3; inset: CsI calorimeter geometry.  }
\label{fig:pen_schem}
\end{figure}

The detector system used for the PEN experiment, schematically shown in
Fig.~\ref{fig:pen_schem}, is an upgraded version of the one previously
used in the PIBETA experiment~\cite{PiBeta_detector}.
The PIBETA project studied rare pion and muon decays, focusing on the
pion beta decay, $\pi^+\to\pi^0e^+\nu_e$, in measurement runs from 1999
to 2004~\cite{Poc04,Frl04,Byc09}.  Both PIBETA and PEN have studied
decays at rest, with the beam pions stopping in an active target.
During PEN data taking, the incoming pion beam momentum varied in the
range between 71.5\,MeV/$c$ and 83.5\,MeV/$c$, depending on the run
period.  The main PEN detector components used in Runs\,2 and 3 are
listed below.
\begin{itemize}
  \item A forward beam counter (BC) and active degrader (AD), both made
    of polyvinyltoluene (PVT), were used for beam particle
    discrimination.  The 3.67\,m separation between BC and AD was used
    to calculate the velocity of the particles and thus their mass,
    using the known beam momentum.
 \item A mini time projection chamber was used for the determination of
   the pion trajectory in all three dimensions.
 \item A cylindrical active target (AT) made of fast PVT, with both
   radius and length equal to 15\,mm, was used to measure the energy
   depositions of the incoming pion and its charged decay products.
 \item Two concentric cylindrical multi-wire proportional chambers
   (MWPC1, 2) tracked charged decay particles emerging from the target.
   The MWPC1, 2 anode wires were located at $r=60$ and 120\,mm,
   respectively.  Track polar angle, $\cos\theta$, is used to define the
   acceptance of the spectrometer for the signal \pieiisg\ and
   normalization \pimiisg\ decay channels, thus controlling a significant
   source of systematic between the two processes.
 \item Tightly surrounding MWPC2 was a hodoscope detector (PH)
   consisting of 20 identical PVT staves, each 4\,mm thick, used for
   particle identification and timing.  Thanks to strong separation
   between minimum ionizing positrons from weak decays, and highly
   ionizing protons from interactions of pions in the central detector
   region, the PH detector provided for efficient suppression of the
   copious hadronic background.
 \item A spherical, 240-module pure CsI electromagnetic calorimeter
   surrounded the target and tracking detectors, covering $\Delta\Omega
   \simeq3\pi$\,sr. The crystal shapes were hexagonal and pentagonal
   truncated pyramids.
\end{itemize}
The recorded detector signals are primarily used to identify,
characterize, and discriminate between the signal decay, $\pi \to
\text{e}\nu (\gamma)$ and the normalization process, $\pi \to \mu \nu
(\gamma)$.  In addition to their primary uses outlined above, individual
PEN detectors are used to check calibrations in the other detectors.

Main recorded event types were triggered with a signal in the plastic
hodoscope and a minimum energy in the CsI calorimeter.  Events were only
recorded within a window extending between $\sim$50\,ns before, and
$\sim$220\,ns after the time of the pion coming to rest in the active
target.  Events with low energy recorded in the CsI, overwhelmingly due
to positrons from muon decays, were prescaled by factors ranging from 8
to 256 depending on the positron-induced electromagnetic shower energy,
and on the run period.

The incoming beam contained pions, positrons and muons, most of which
can be separated by using the particle energy loss in the upstream beam
detector BC, and time of flight (TOF) between BC and AD.  The
reconstructed pion trajectory is used to determine the pion stop
location, which is in turn used to construct methods of cleanly
separating the main pion decay mode $\pi \to \mu \nu (\gamma)$ from the
rare decay mode of interest, $\pi \to \text{e}\nu (\gamma)$.

\section{The PEN mTPC design and construction \label{sec:mTPC-design}}
Beam particle tracking prior to the stopping of pions in the target was
performed in Run\,1 by a four-piece wedge degrader detector made of PVT.
To improve on the wedge degrader, for Runs\,2 and 3 the PEN collaboration
designed and built low-mass mini time projection chambers, discussed
below.

\begin{figure}[b!]
  \centering 
  \parbox{0.38\linewidth}{
    \includegraphics[width=\linewidth]{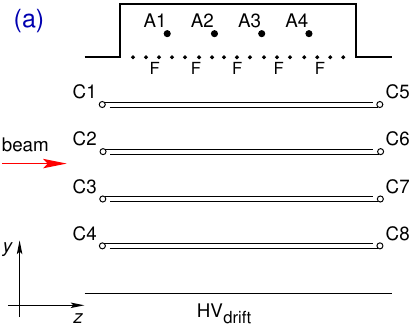}\\[4pt]
                         }
  \hspace*{0.05\linewidth}
  \parbox{0.42\linewidth}{
    \includegraphics[width=\linewidth]{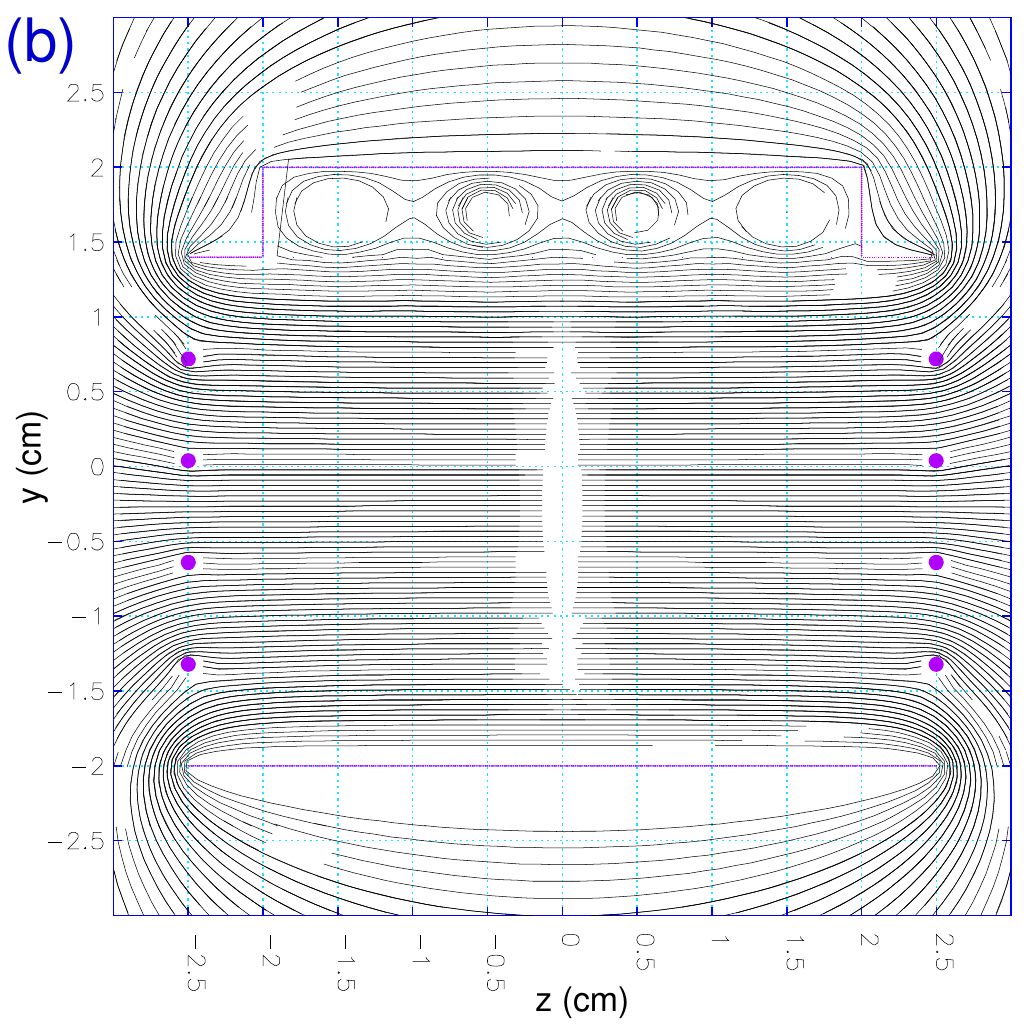}
                         }  
  \caption{(a) Schematic diagram of the mTPC design for Run\,2, with four
    anode wires, A1--A4, field shaping electrodes C1--C8, fixed on the
    inner surface of the $y$-$z$ chamber walls, the main (drift)
    cathode, HV$_{\textsf{drift}}$ at the bottom, and grid wires, F,
    separating the drift and proportional volumes.  C1--4 were mounted
    on the beam-right chamber wall (front in sketch view), and C5--8 on
    the beam-left wall (rear in sketch).  Electrode C$(i+4)$ was held at
    the same potential as C$(i)$.  Further design details are given in
    the text.  (b) The corresponding distribution of equipotential
    surfaces produced by the GARFIELD code.}
  \label{fig:mTPC_1_design}
 \end{figure}

\begin{figure}[t!]
  \centering 
  \parbox{0.38\linewidth}{
    \includegraphics[width=\linewidth]{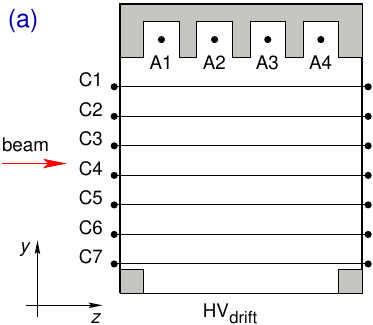}\\[11pt]
                         }
  \hspace*{0.05\linewidth}
  \parbox{0.42\linewidth}{
    \includegraphics[width=\linewidth]{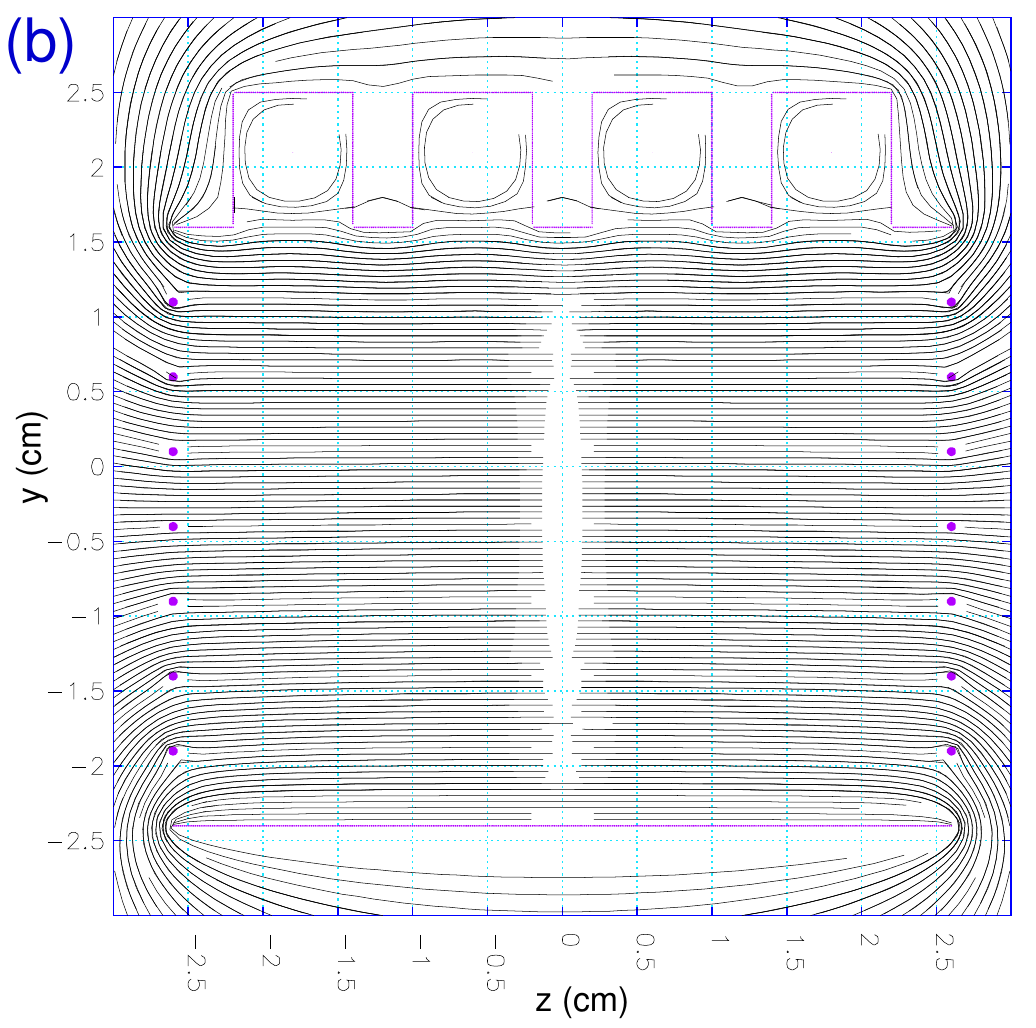}
                         }  
  \caption{(a) Schematic diagram of the mTPC design for Run\,3, with four
    anode wires, A1--A4, field shaping electrodes C1--C7, each wrapped
    fully around the light chamber frame, the main (drift) high voltage
    electrode at the bottom.  Unlike the Run\,2 mTPC, there were no grid
    wires separating the drift and proportional volumes.  The shaded
    areas depict the light styrofoam frame of the chamber.  (b) The
    corresponding distribution of equipotential surfaces produced by the
    GARFIELD code.}
  \label{fig:mTPC_2_design}

\end{figure}

The mTPC used in PEN Run\,2, shown in Fig.~\ref{fig:mTPC_1_design}, was
a low-mass cubic box with dimensions $50 \times 50\times 50$\,mm$^3$.
The geometry was studied using the CERN GARFIELD \cite{Garfield} drift
chamber simulation package with the goal to optimize the electric field
homogeneity in the drift region, and to equalize gas amplification for
the four anode wires.  Further modifications and improvements to the
field cage were implemented in the second generation mTPC for Run\,3,
shown in Fig.~\ref{fig:mTPC_2_design}.  We first focus on the Run\,2
device, and then highlight the differences in the Run\,3 version.

The Run\,2 chamber's active volume was subdivided by a grid into drift
and proportional regions.  The drift region dimensions were $40\times
40\times 40$\,mm$^3$.  The homogeneous electric field in the drift
region was shaped by the cathode electrode, the separating grid, and
eight field shaping electrodes affixed to the inner surfaces of the side
walls.  The drift plate was at $-4$\,kV with respect to ground (anode)
potential, and the drift velocity $\sim$\,1.8\,cm/$\mu$s.  The thickness
of the Mylar windows perpendicular to the beam was 20\,$\mu$m, and the
potential electrodes were made of copper wire, 0.1\,mm in diameter.  The
grid was made of 0.1\,mm bronze wires with 2\,mm spacing.  Above the
separating grid was the 7\,mm wide amplification region with the four
anode wires arranged perpendicular to the beam direction, with a spacing
of 10\,mm.  The 40\,mm long anodes were made of Nichrome (NiCr) wire,
12\,$\mu$m in diameter, with the resistance of 60\,$\Omega$/cm.  The
mTPC was filled with a gas mixture of 90\% argon and 10\% methane at
atmospheric pressure (slow bubblers).  No connectors were placed
directly on the mTPC.  All electric connections for signal, power, and
HV cables were made by soldering.  The voltage dividers for the field
shaping electrodes were placed outside the detector.  The signals from
the two sides of all four anode wires were fed through to a CAEN V1720
unit where they were digitized into waveforms at the sampling rate of
250\,MS/s.

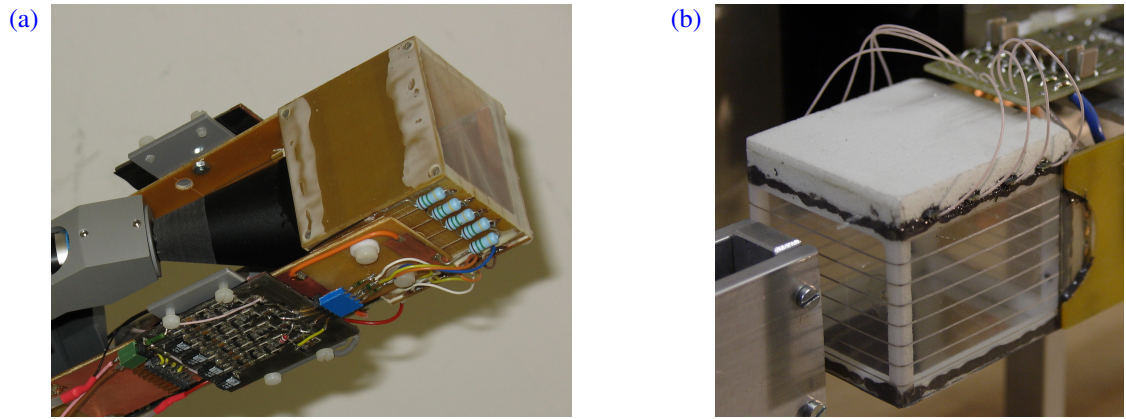
\begin{figure}[t!]
  \parbox{0.5\linewidth}{\hspace*{\fill}
    \includegraphics[height=0.25\textheight]{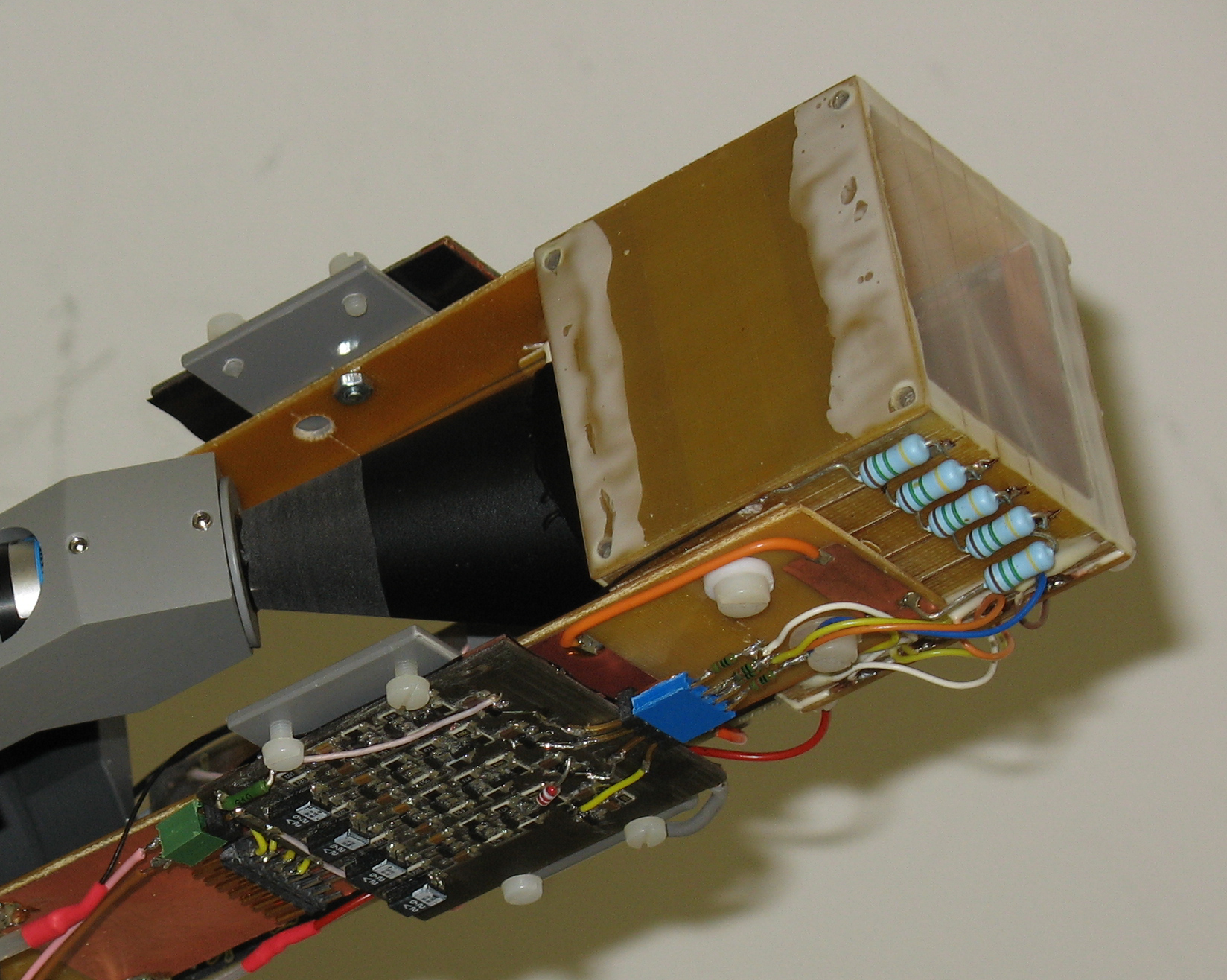}
    \begin{picture}(0,0)
      \put(-213,147){\color{blue}(a)}
    \end{picture}
                         }
  \hspace*{0.1\linewidth}
  \parbox{0.39\linewidth}{
   \includegraphics[height=0.25\textheight]{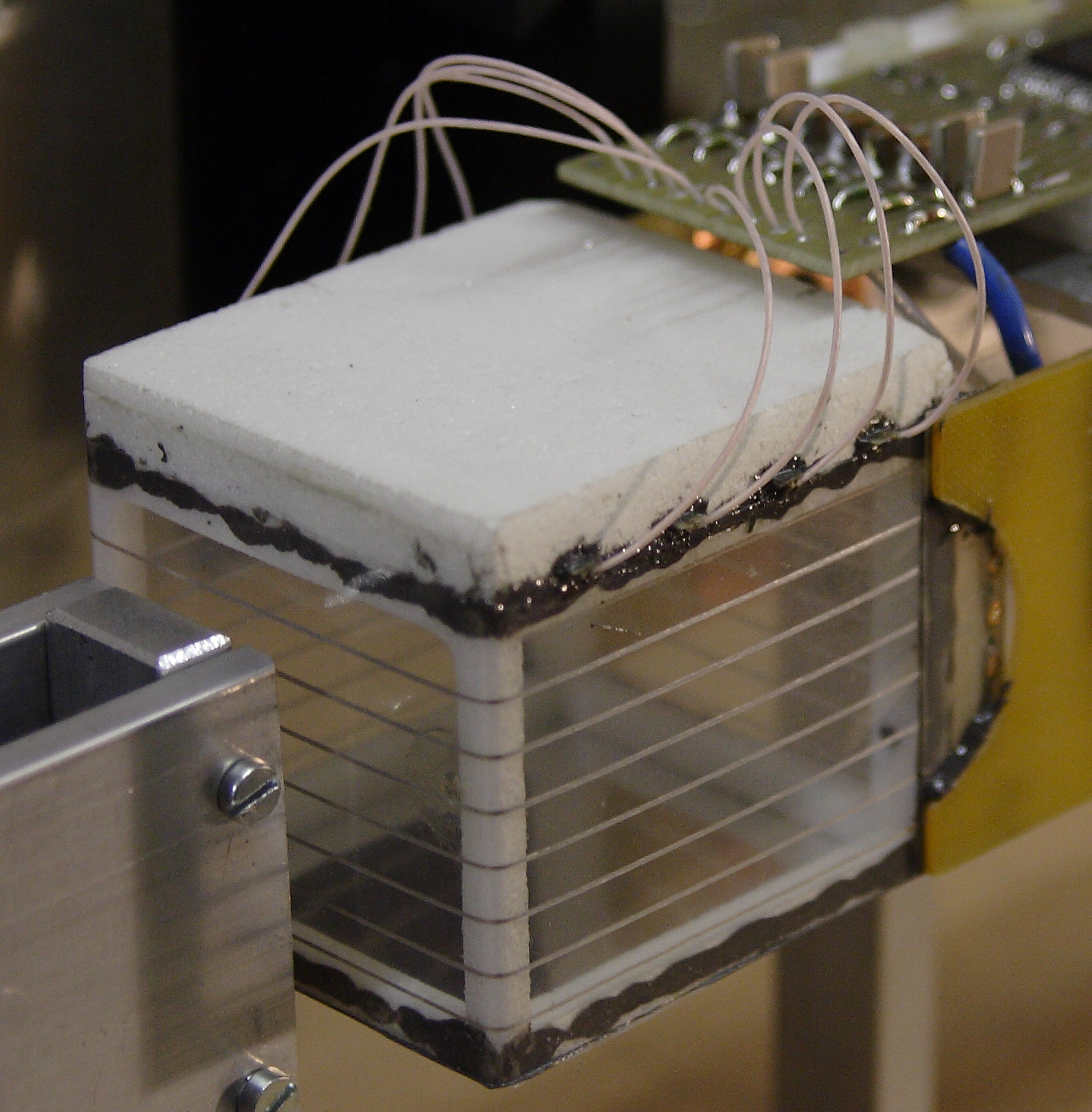}
    \begin{picture}(0,0)
      \put(-172,147){\color{blue}(b)}
    \end{picture}
                         }

  \caption{Photographs of the mTPC detectors used in PEN Run\,2 (a) and
    Run\,3 (b).  The significantly lighter construction of the latter,
    clearly evident in the image, allowed its placement closer to the
    active target, resulting in an improved resolution of the decay
    vertex.}
  \label{fig:mTPC_photos}
\end{figure}

A modified version of the detector was used in Run\,3 \cite{Baranov}.
Notable differences include significantly reduced mass, 12.0\,mm wire
spacing instead of 10.0\,mm, and elimination of the grid wires.  Its
lower mass allowed the placement of the detector further downstream,
closer to the active target.  Photographs of the two mTPCs are shown in
Fig.~\ref{fig:mTPC_photos}.

The basic mechanisms of signal development in a TPC are well understood.
In a first stage the electrons from the track of some ionising particle
drift in a more or less constant drift field towards the anode plane.
The observed drift time allows one to calculate the position coordinate
in the direction of the drift field (vertical, or $y$, for PEN mTPCs).
In most cases the anode consists of thin sense wires where the
avalanches, developing in the $E \propto 1/r$ field, result in the
necessary signal gain.  After reaching a wire the secondary electrons
create electric signals which propagate towards both wire ends.  The
signal exponential fall time constant reflects the velocity at which the
cloud of gas ions moves away from the anode.  The electron signals are
split between the two anode ends with a ratio inversely proportional to
the wire resistances between the ends and the charge arrival location on
the wire.  Thus, the observed signal amplitude ratio may allow
reconstruction of that location along the anode coordinate (horizontal,
$x$, for PEN mTPCs).

\begin{figure}[h!]
  \centering
  \renewcommand{\fwfr}{0.40}
  \includegraphics[width=\fwfr\linewidth]{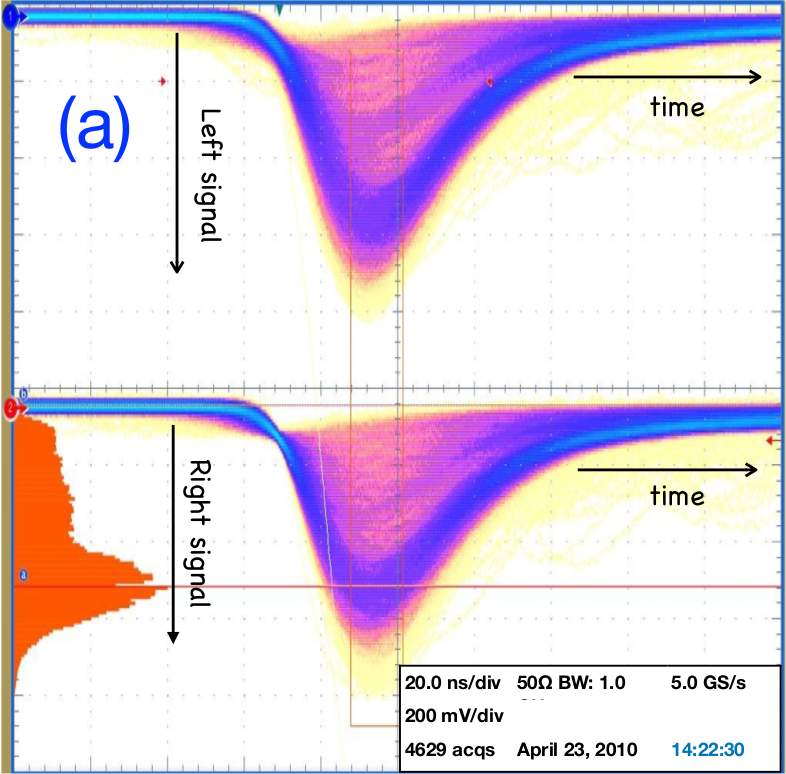}
    \qquad
    \includegraphics[width=\fwfr\linewidth]{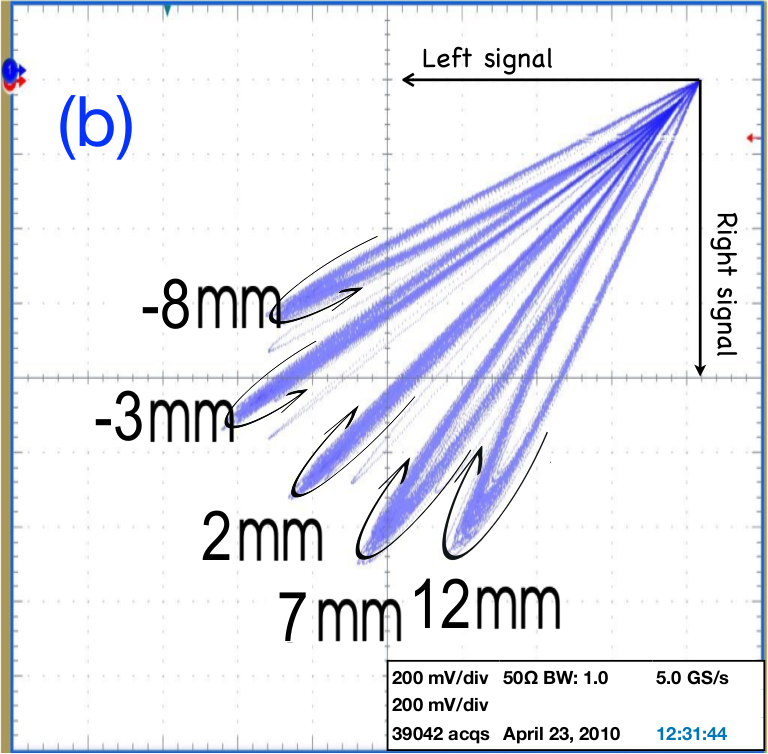}
   \caption{Oscilloscope images of mTPC anode signals observed at both
     ends of the resistive wire. The detector was irradiated with a
     collimated $^{55}$Fe X-ray source. Both the individual signals (a),
     and the left-right signal correlation (b) are shown. See text for
     further discussion.}
   \label{fig:Fe-55_traces}
\end{figure}
Signal shapes are distorted by reflections at the amplifier inputs and
as a result the tails of the two signals gradually lose their position
dependence, as illustrated in Fig.~\ref{fig:Fe-55_traces}.  The figure
shows oscilloscope pictures of the left and right signals of an anode
wire of the Run\,2 mTPC recorded as the detector was irradiated with a
collimated $^{55}$Fe X-ray source at different locations along the
anode. The energy distribution peaks around 5.9\,keV, about double the
energy loss of our beam pions, a Landau distribution with mean of
$\sim$\,3.2\,keV and peak probability of $\sim$\,2.2\,keV per wire.
Part\,(a) shows the individual left and right signals for central source
location. The D0M Ampl-8.3 amplifier \cite{Alexeev2001} has
leading/trailing edge of 7\,ns. The histogram on the left shows the
energy distribution obtained from the signal integral in the indicated
gate around the peak. Part\,(b) shows the correlation between left and
right signal, varying the source position along the wire in steps of
5\,mm. Note that the signals are separated by ten or more standard
deviations, corresponding to a position resolution well below a mm,
i.e., $\mathcal{O}$(1\%) of the wire length. This value includes the
unknown contribution of the $^{55}$Fe source collimator.  We also note
that charge division does not only affect the amplitudes, but also the
signal shapes.  Signals follow a loop starting and ending at the origin
at the top right.  The direction of time evolution through the loops is
indicated by the curved arrows at the signal peaks, clockwise on the
right side and counter-clock on the left.  Central hits result in
identical signals which land on the diagonal (not shown in the figure).
For off-center hits the amplitude ratio deviates most from unity during
the rise time.  At later times the position dependence gradually fades
away, so the loops return to the diagonal.  For this reason the tracking
algorithm optimized for off-line analysis only uses the first
$\sim$\,30\,ns of the recorded wave forms.

Signal development in TPCs with meters long sense wires, used in
high-energy experiments, is much more affected by reflections and
cross-talk.  In those situations numerical calculations have been quite
successful in describing the experimental observations \cite{Bock12}. In
this analysis we have followed a more phenomenological approach based on
accurate calibrations of the measured data, and adjusting the PEN
experiment Monte Carlo simulation to those.

\begin{figure}[t!]
  \renewcommand{\fwfr}{0.47}
  \parbox{\fwfr\linewidth}{
    \includegraphics[width=\linewidth]{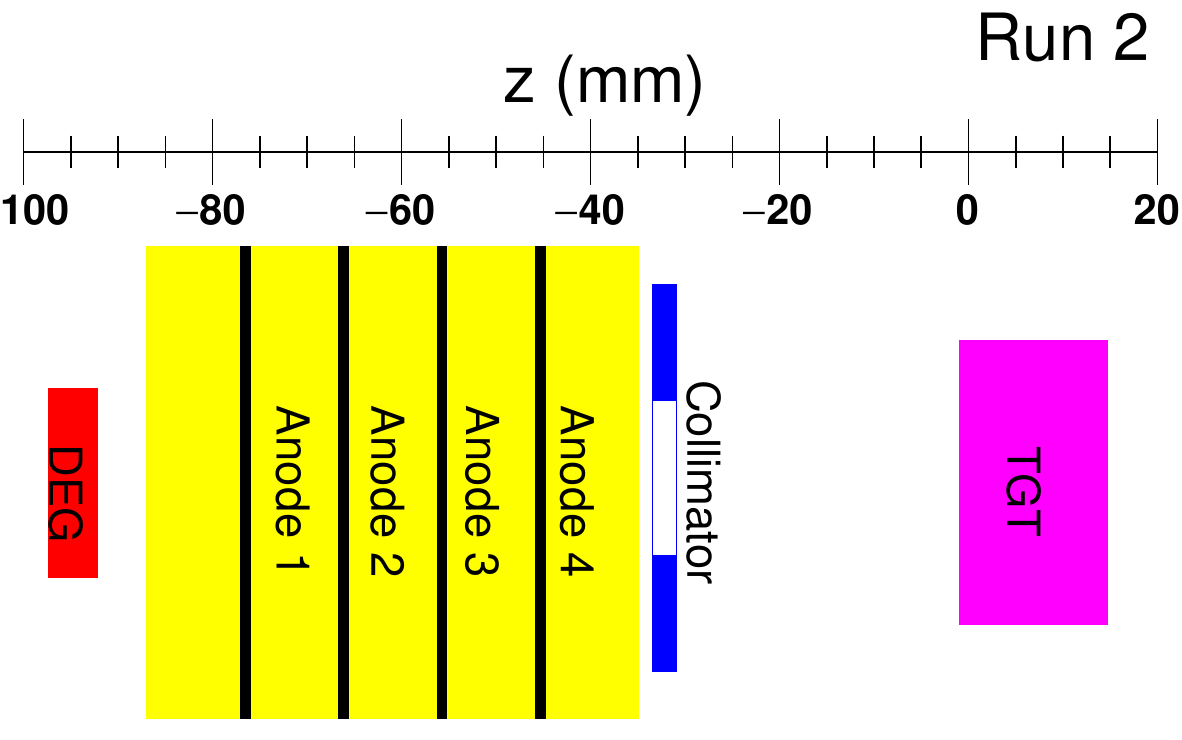}
                         }
  \hspace*{\fill}
  \parbox{\fwfr\linewidth}{
    \includegraphics[width=\linewidth]{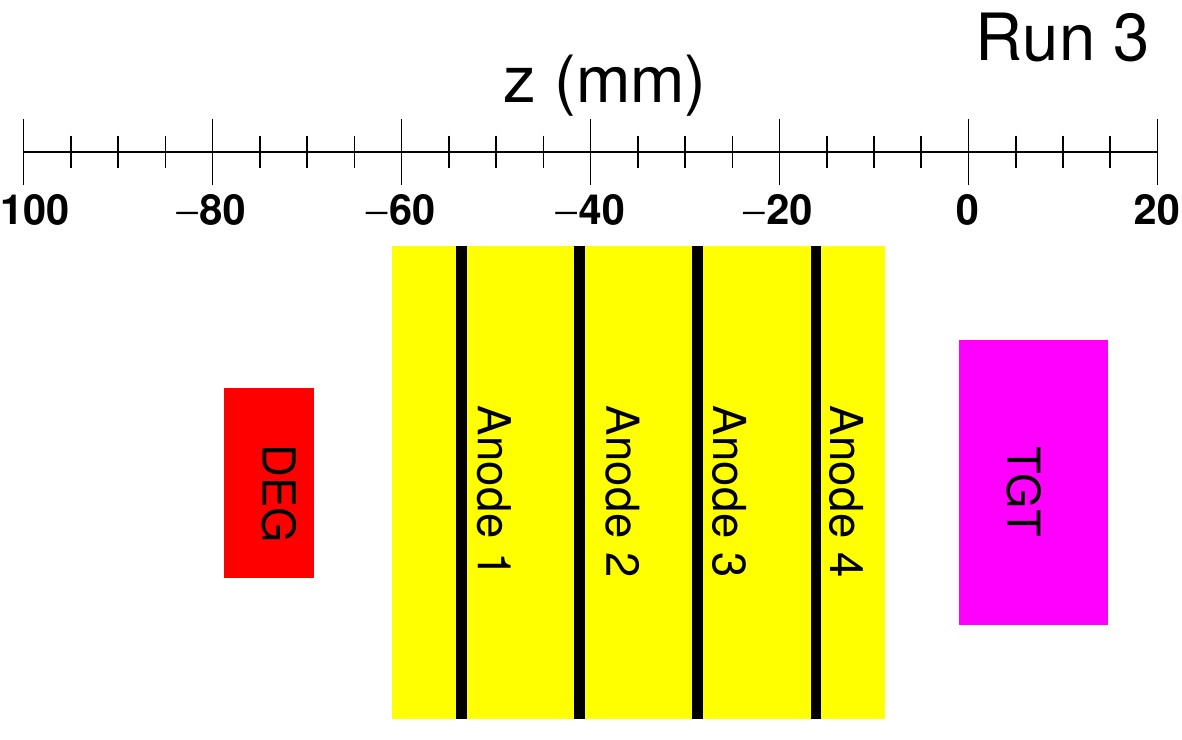}
                         }
  \caption{Schematic drawings of the mTPC detector placement in PEN
    Runs\,2 and 3, as labeled.  The Run\,3 detector geometry, with no
    collimator between the mTPC and the target, was significantly more
    compact than that used in Run\,2.}
  \label{fig:run_geom}
\end{figure}
The geometrical arrangements deployed in PEN Runs\,2 and 3,
respectively, are shown in Fig.~\ref{fig:run_geom}.  In Run\,2, the mTPC
was separated from the target by about 37\,mm, with a beam collimator in
between.  In Run\,3, the mTPC-target separation was about 10\,mm, with
no collimator between them, leading to an improved resolution of the
decay vertex.

\section{Particle tracking and event reconstruction with the PEN mTPC 
  \label{sec:mTPC-recon}}
\begin{figure}[t!]
  \renewcommand{\fwfr}{0.49}
  \centering
  \includegraphics[width=\fwfr\linewidth]{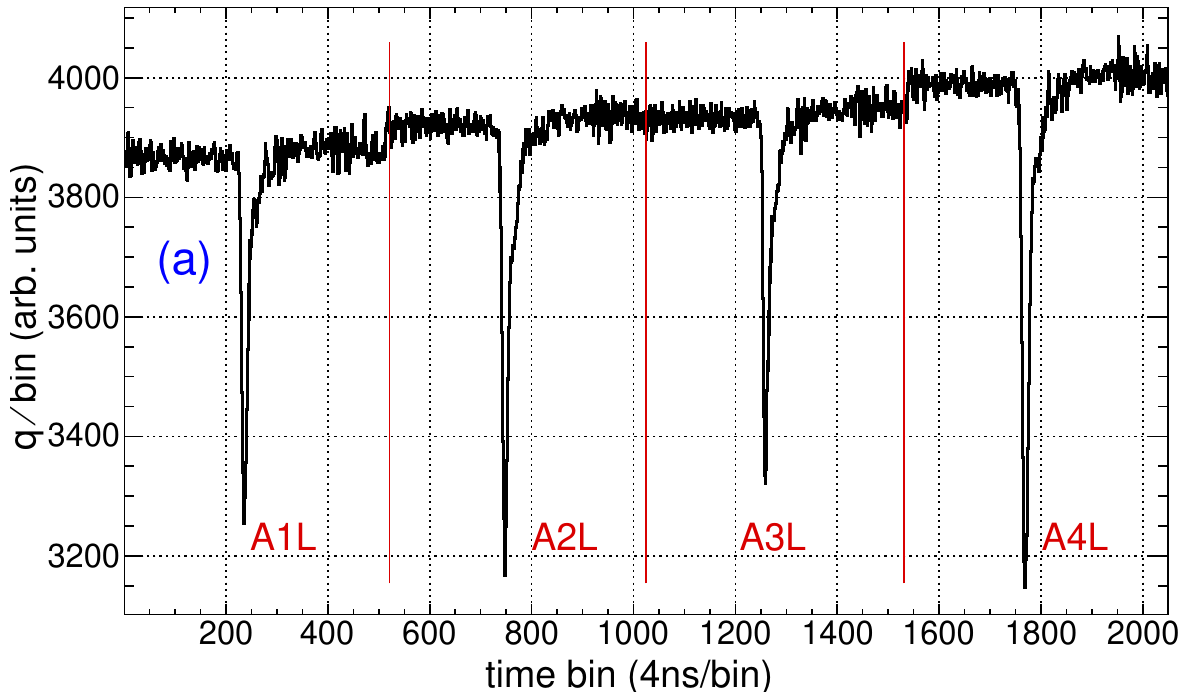}
  \hspace*{\fill}
  \includegraphics[width=\fwfr\linewidth]{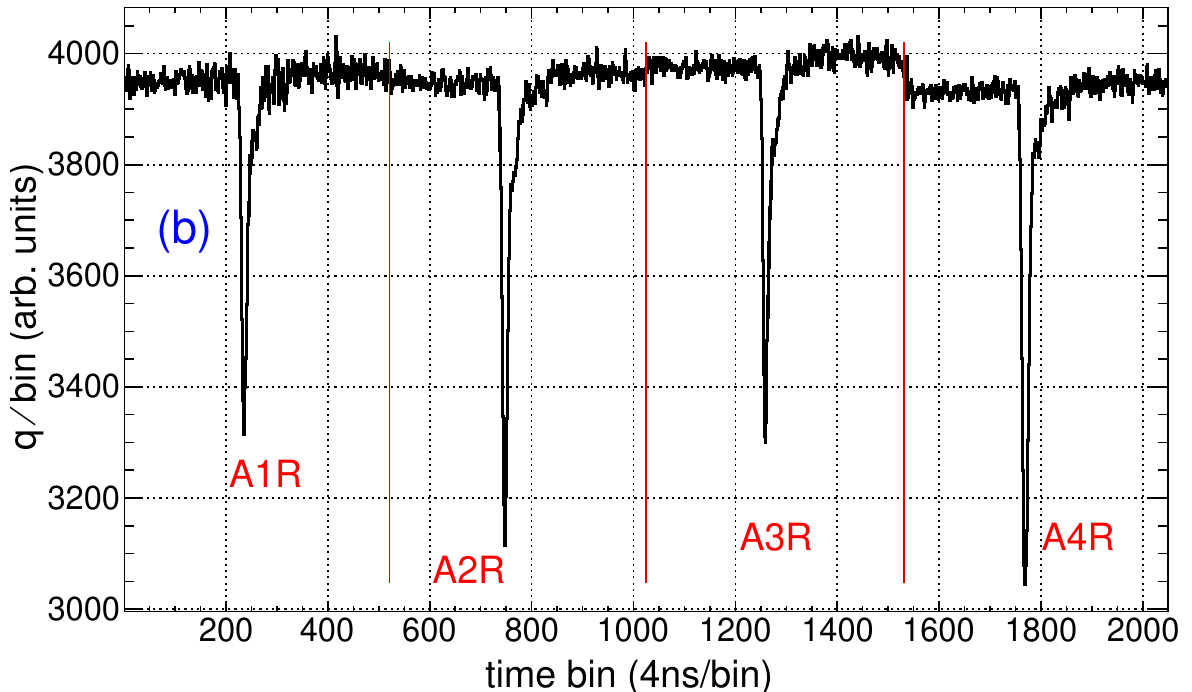}
  \caption{Typical single event raw mTPC anode charge waveforms for the
    left (a), and the right (b) ends of the resistive wires, multiplexed
    by means of adding a 500 bin ($\hat{=}\,2\,\mu$s) offset to
    each successive anode wire waveform, as indicated by red vertical
    delimiters.
    \label{fig:raw_wfs}}

  \vspace*{12pt}

  \includegraphics[width=\fwfr\linewidth]{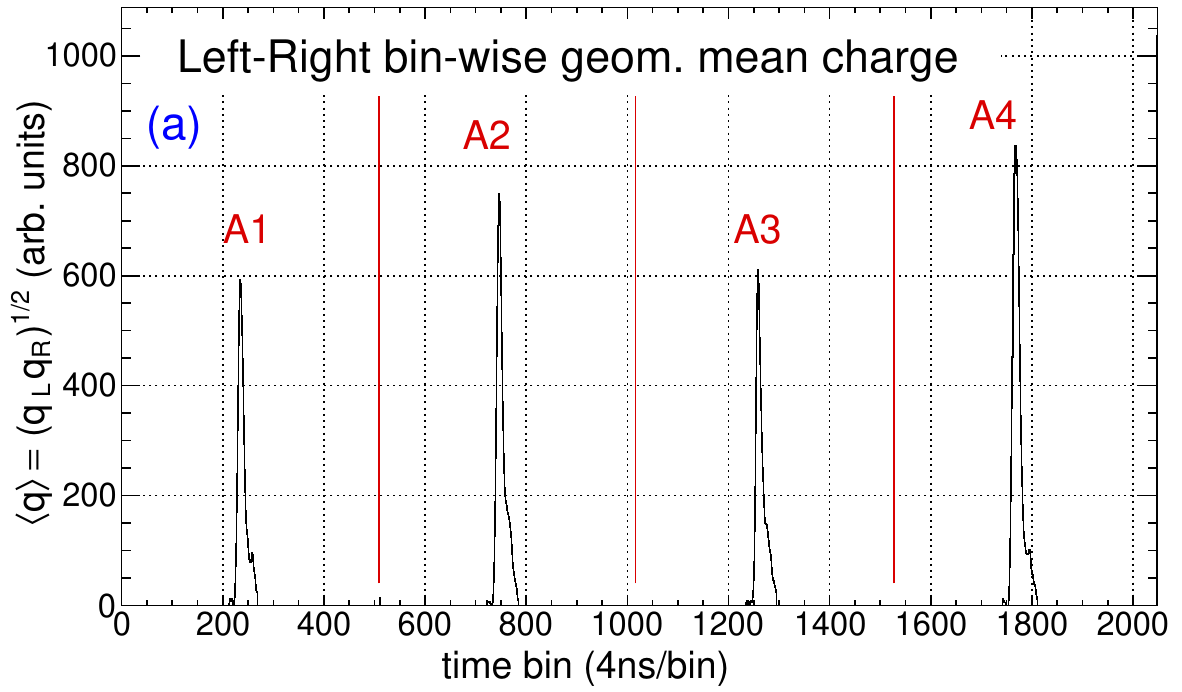}
  \hspace*{\fill}
  \includegraphics[width=\fwfr\linewidth]{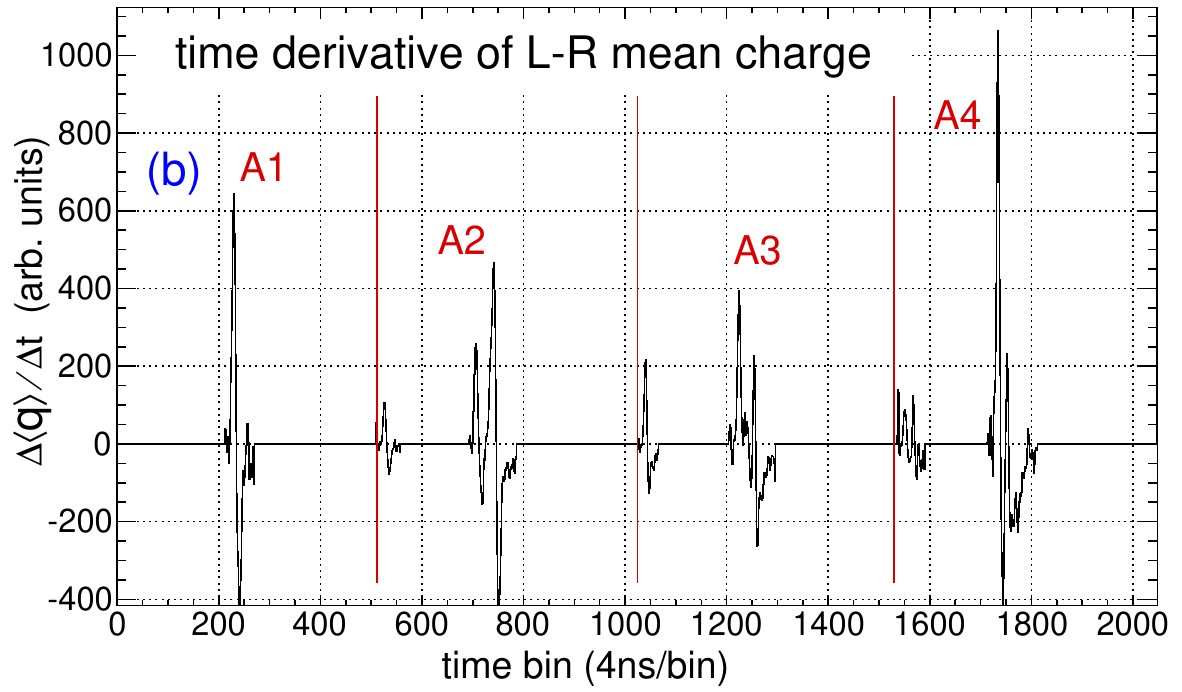}
  \caption{ (a) Waveforms of the binwise left-right geometric mean
    charge for the four anode wires, corresponding to the event shown in
    Fig.~\ref{fig:raw_wfs}. (b) Results of the time differentiation of
    the L-R mean waveforms in (a).
    \label{fig:wf_means}}
\end{figure}

As indicated in the preceding section, the relevant event-level mTPC
information is stored in the form of waveform arrays (aka records,
traces), one for each end of the four anode wires.  Waveform traces for
a typical event are shown in Fig.~\ref{fig:raw_wfs}.
This section presents the strategies used to search for signals in the
anode waveforms, their combination into tracks, the determination of the
associated trajectory coordinates, and relation to beam particle
identification.

\begin{figure}[t!]
  \centering
  \includegraphics[height=0.3\textheight]{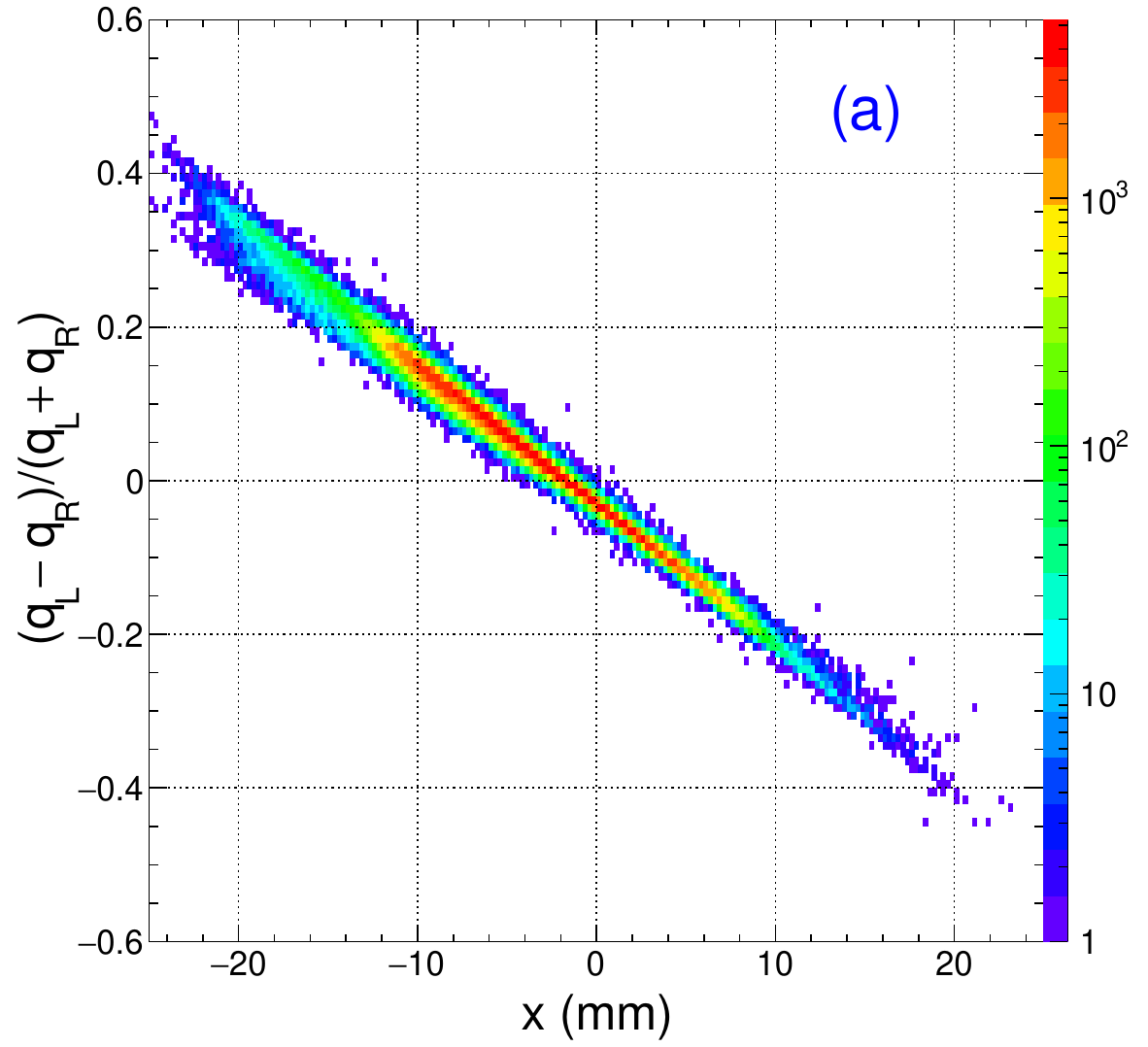}
  \hspace{3mm}
  \includegraphics[height=0.3\textheight]{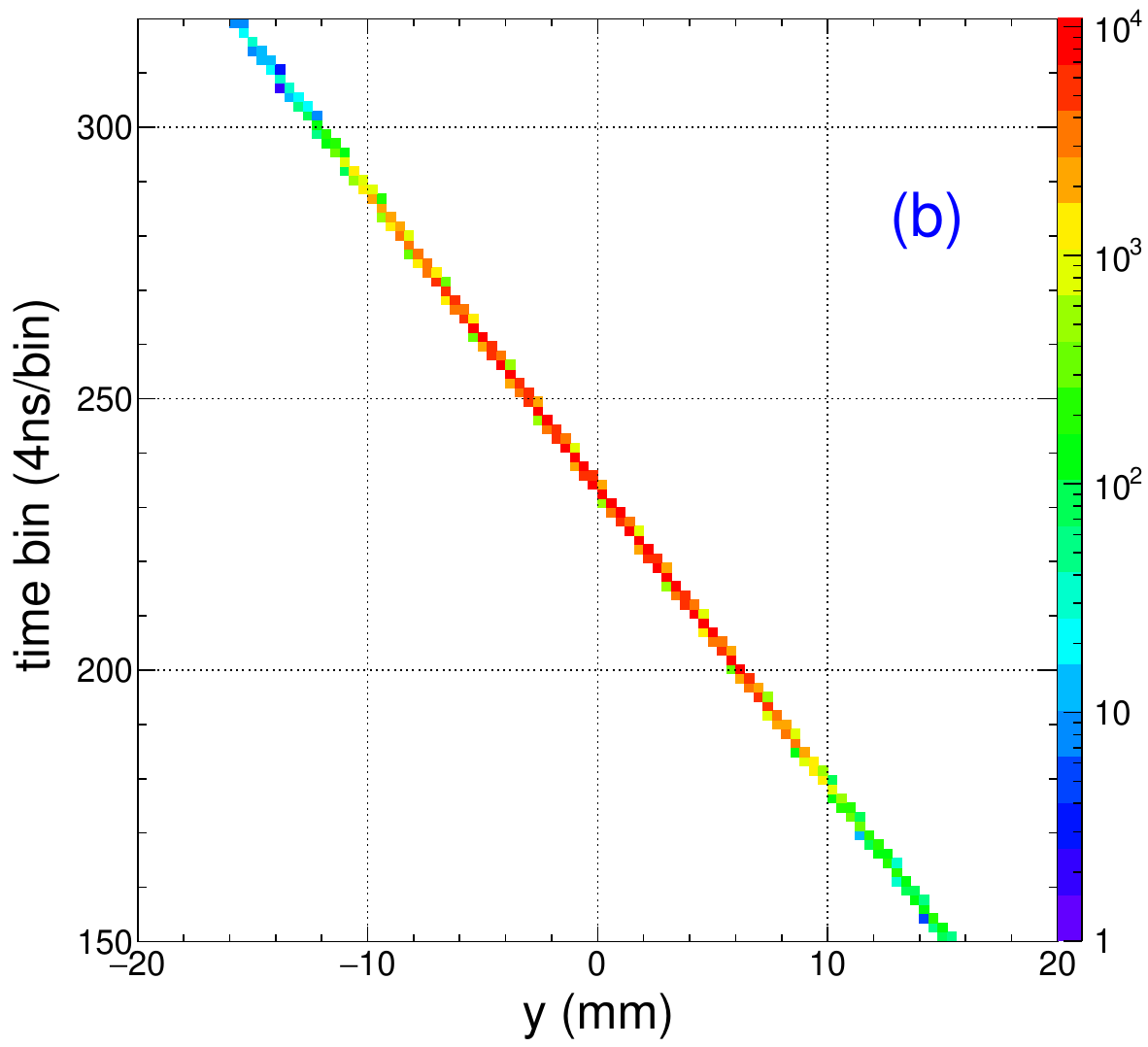}
  \caption{Coordinate calibration plots for the Run\,2 version of the
    PEN mTPC: (a) Charge fraction as a function of the $x$ coordinate of
    the signal on wire 4, after calibration.  (b) Reconstructed event
    signal time bin plotted against $y$ of the wire 4 signal, after
    calibration.  Note: fiducial volume for tracks of interest is $x,y
    \in (-15,15)$\,mm.
   \label{fig:coord_calib}} 
\end{figure}

After subtraction of appropriately averaged baseline levels from the raw
waveforms, the geometric means of the left and right waveform bins are
constructed bin-wise for each anode wire, and the result is
differentiated as shown in Fig.~\ref{fig:wf_means}.  The peaks in the
resulting distributions are used as a measure of the anode arrival
times.  The signal amplitudes at both ends of a wire reflect the energy
deposition and their ratio is a measure of the $x$ coordinate of a hit
(Fig.~\ref{fig:coord_calib}a).
The anode wire number defines the $z$ coordinate along the pion
trajectory.  The wire spacing was measured with an accuracy of 0.1 mm.
The anode arrival times are used to calculate the drift time and thus
the $y$ coordinate.  A valid hit in the degrader detector, which was
separated by 6\,mm (Run\,2) and 8\,mm (Run\,3) of air from the mTPC's
front face, provided the start signal for the drift time.  The
correlation between the calibrated mTPC signal drift times and the track
$y$ coordinates for anode wire~4 in Run\,2 is shown in
Fig.~\ref{fig:coord_calib}b.

\begin{figure}[H]
  \centering
  \includegraphics[height=0.3\textheight]{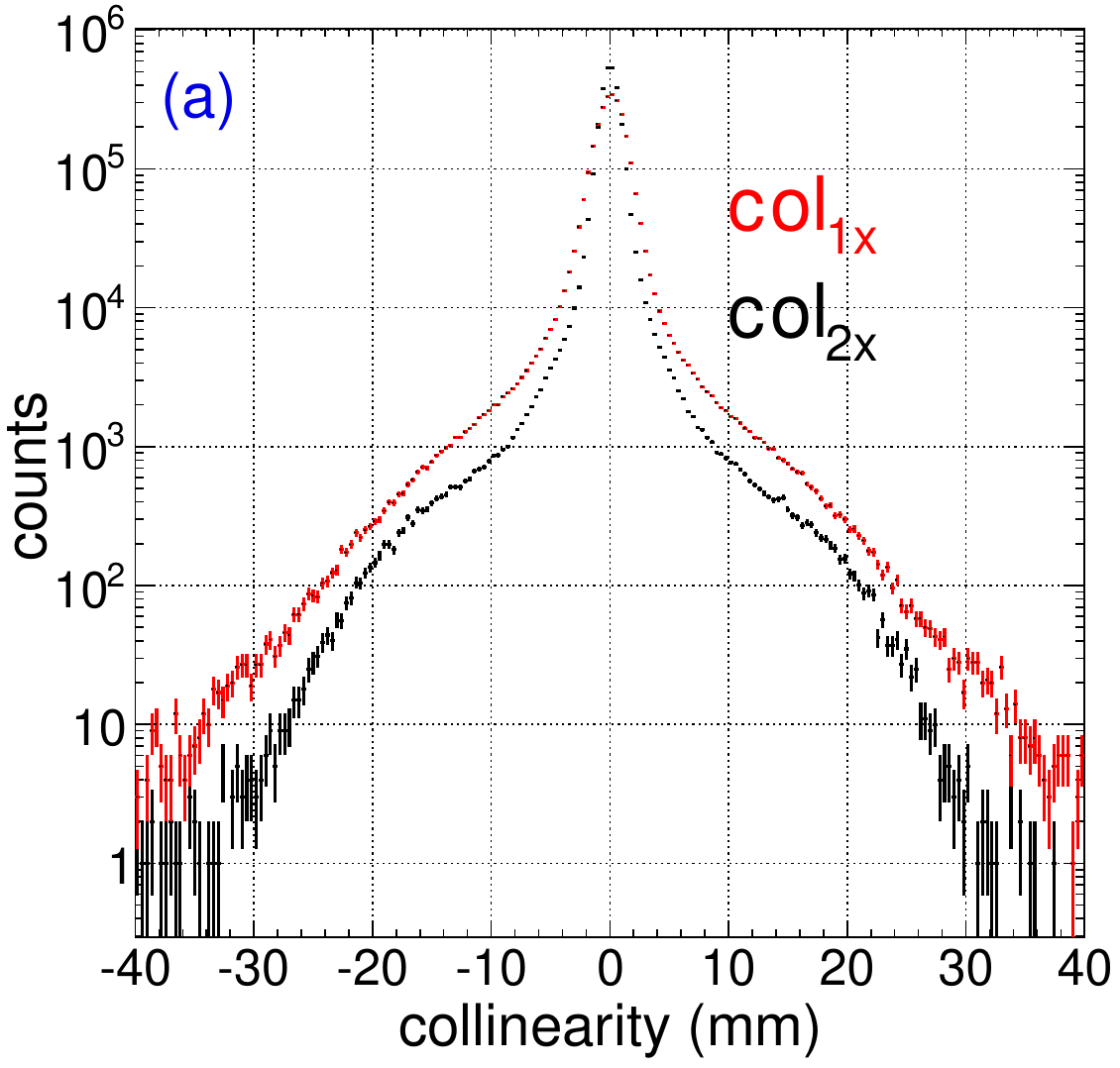} 
  \includegraphics[height=0.3\textheight]{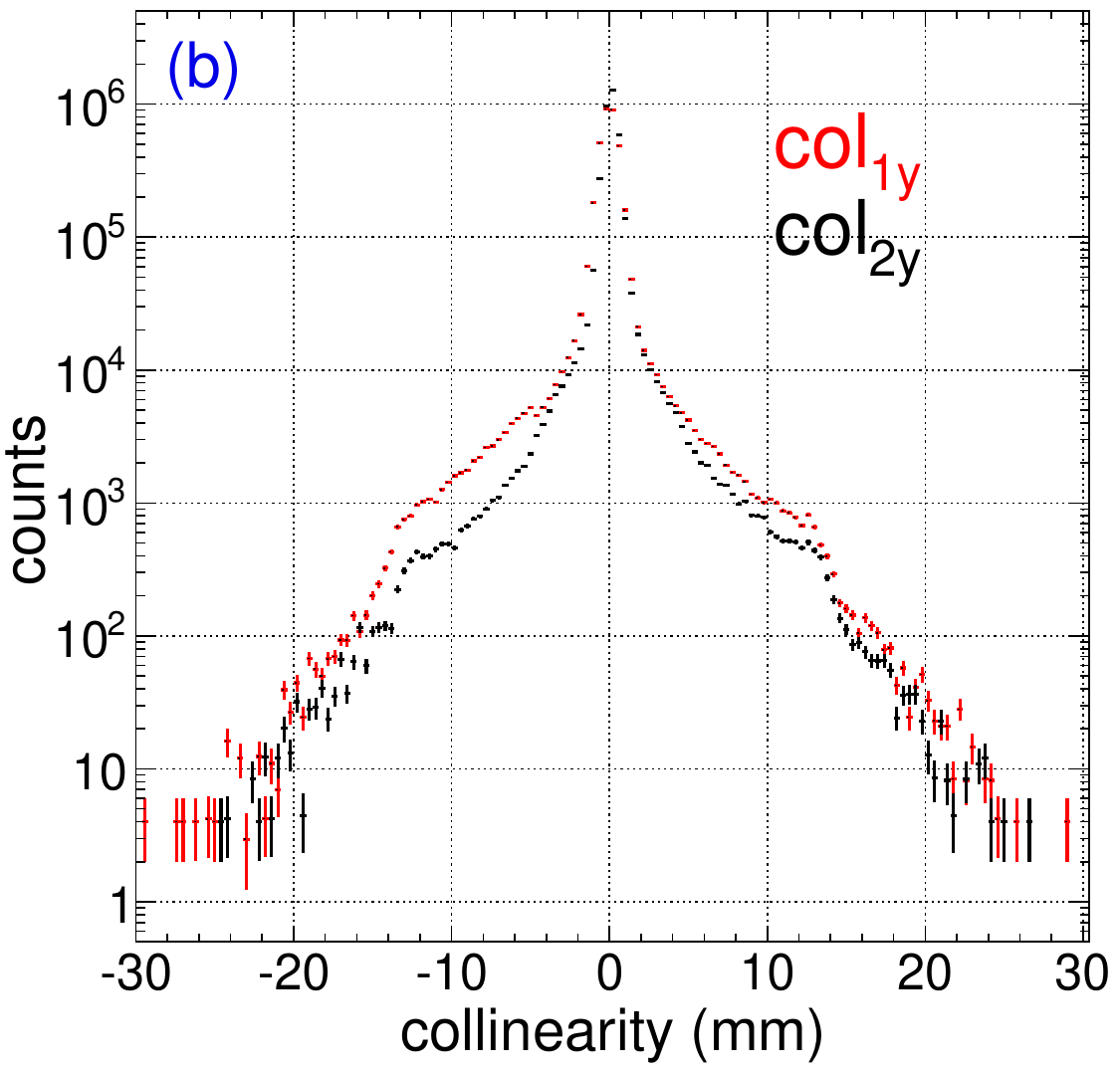}
  \caption{(a) Collinearity values,  col$_{1x}= x_1+ x_2 -x_0-x_3$
    (red), and col$_{2x}$=$\left(x_1-x_2\right)-(x_0-x_3)/3$ (black).
    (b) Same for the $y$ collinearities. 
    \label{fig:coll_tests-1d}  }
\end{figure}

Once $x,y$ anode hits have been found, straight tracks are
reconstructed, based on the following principles.  Each hit can be used
for one track only.  Tracks are searched in order of the number of hits
per track, beginning with fully efficient tracks.  Collinearity tests
are used to remove hit combinations that do not belong to a single beam
particle.  In the case of fully efficient tracks the straight trajectory
defined by $(x_2,y_2)$ and $(x_3,y_3)$ must agree with the one defined
by $(x_1,y_1)$ and $(x_4,y_4)$.  Collinearity is tested by comparing the
$(x,y)$ locations at mean $z$ and the slopes in $x$ and $y$, so for $x$:
\begin{linenomath}
\begin{align}
  \text{col}_{1x} \equiv
    x_2+ x_3 -x_1-x_4 &=0\,,\qquad \text{and} \label{eq:col_x1} \\
  \text{col}_{2x} \equiv
    \left(x_2-x_3\right)-(x_1-x_4)/3 & =0\,,   \label{eq:col_x2}
\end{align}
\end{linenomath}
where indices $i$ denote anode wires.  Because of multiple small-angle
scattering and intrinsic detector resolution, the above collinearity
tests are not exactly satisfied in practice.  Nevertheless, the two
expressions center at or close to zero for valid tracks.  Analogous
expressions are constructed for the deduced $y$ values of the track.
Figure~\ref{fig:coll_tests-1d} shows distributions of the two
collinearity tests for both $x$ and $y$.

\begin{figure}[H]
  \centering
\includegraphics[height=0.25\textheight]{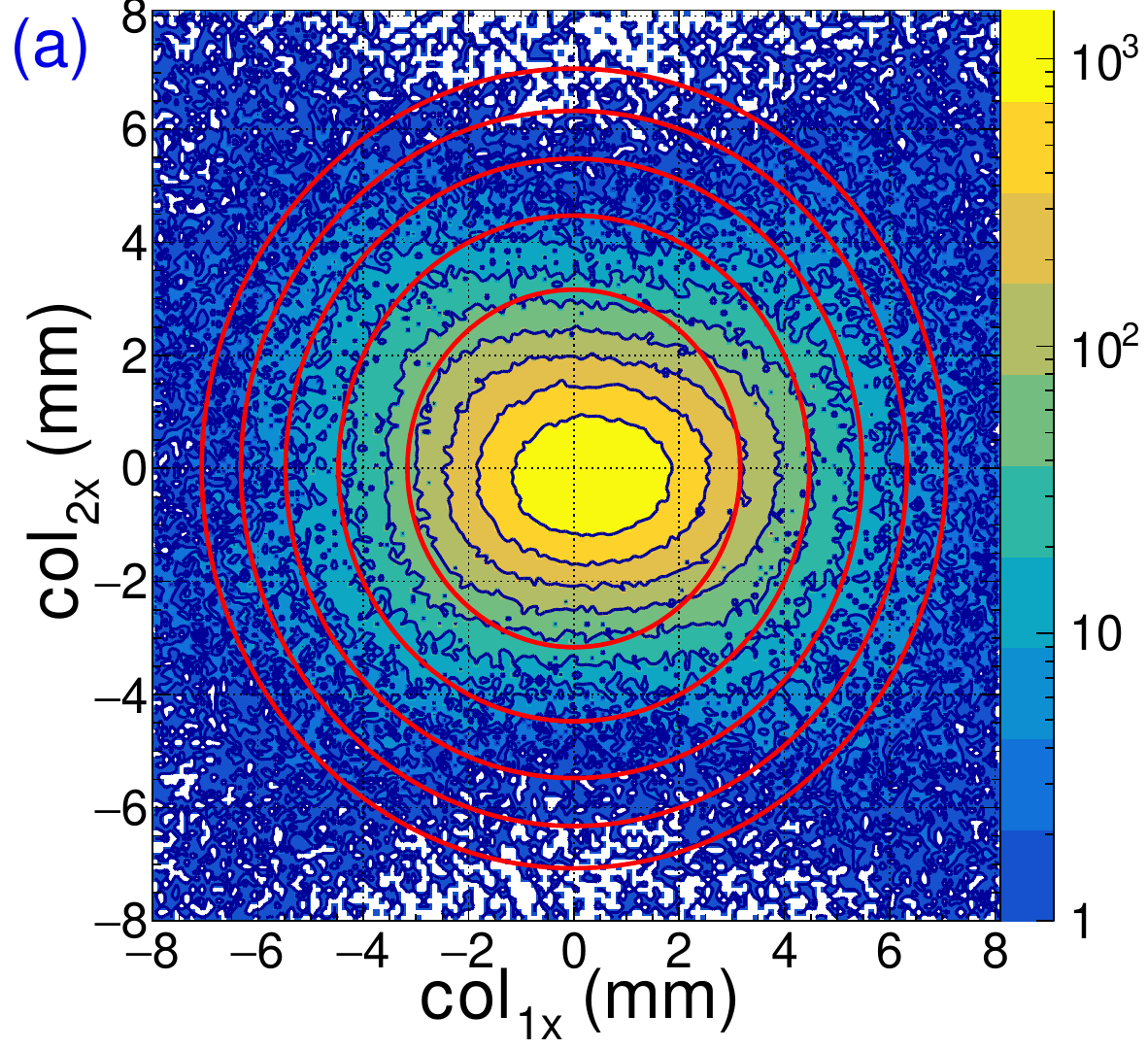} 
  \includegraphics[height=0.25\textheight]{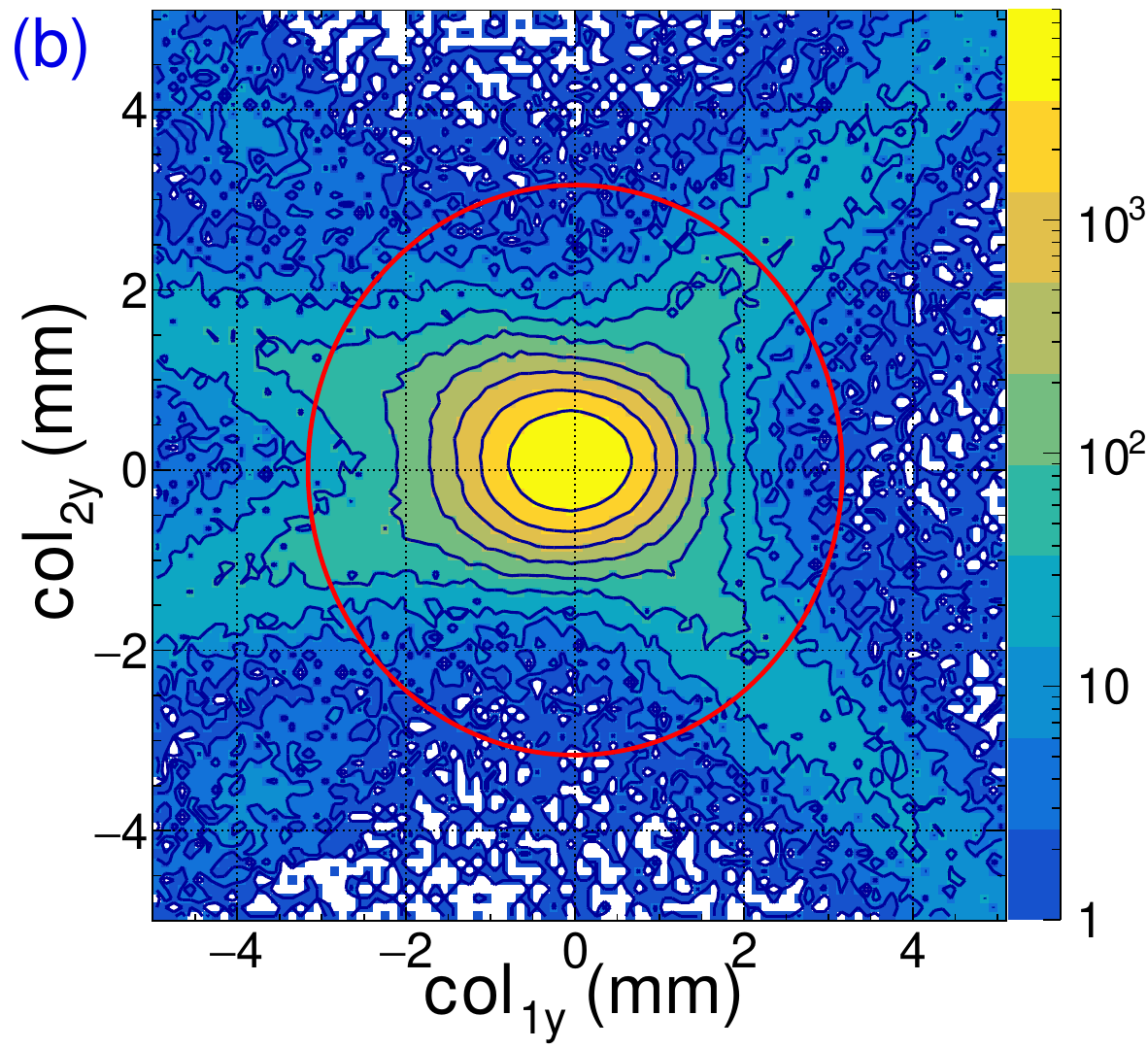} \\
  \includegraphics[height=0.23\textheight]{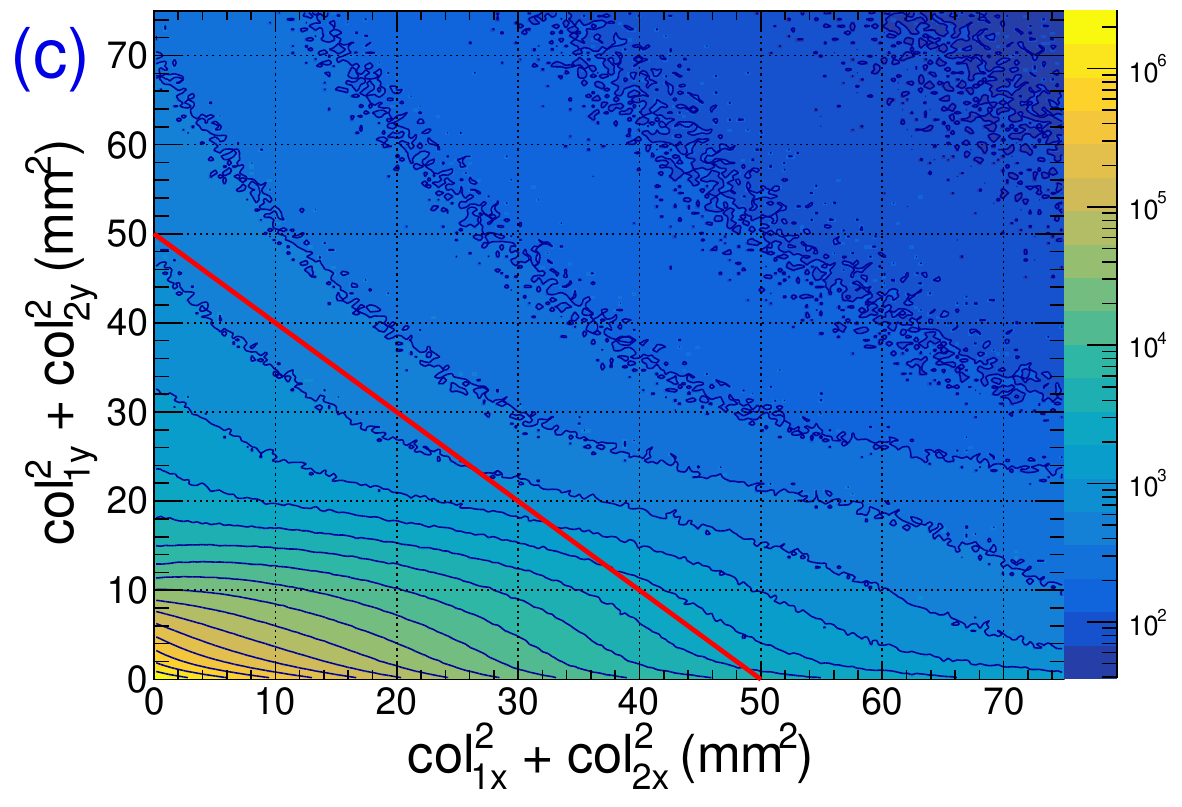}
  \caption{(a) Contour plot of $\text{col}_{2x}$ against
    $\text{col}_{1x}$.  The red circles corresponds to the col$_x^2$
    cuts referenced in the text.  (b) The corresponding plot in $y$. Red
    circle: the col$_y^2 \leq 10\,\text{mm}^2$ cut.  (c) Contour plot of
    col$_y^2$ against col$_x^2$.  Red line: final track acceptance cut
    on collinearity, $\text{col}_x^2 + \text{col}_y^2 \leq
    50\,\text{mm}^2$.  See text for further details. }
  \label{fig:coll_tests-2d}
\end{figure}

\begin{figure}[b!]
  \centering
  \includegraphics[height=0.18\textheight]
                {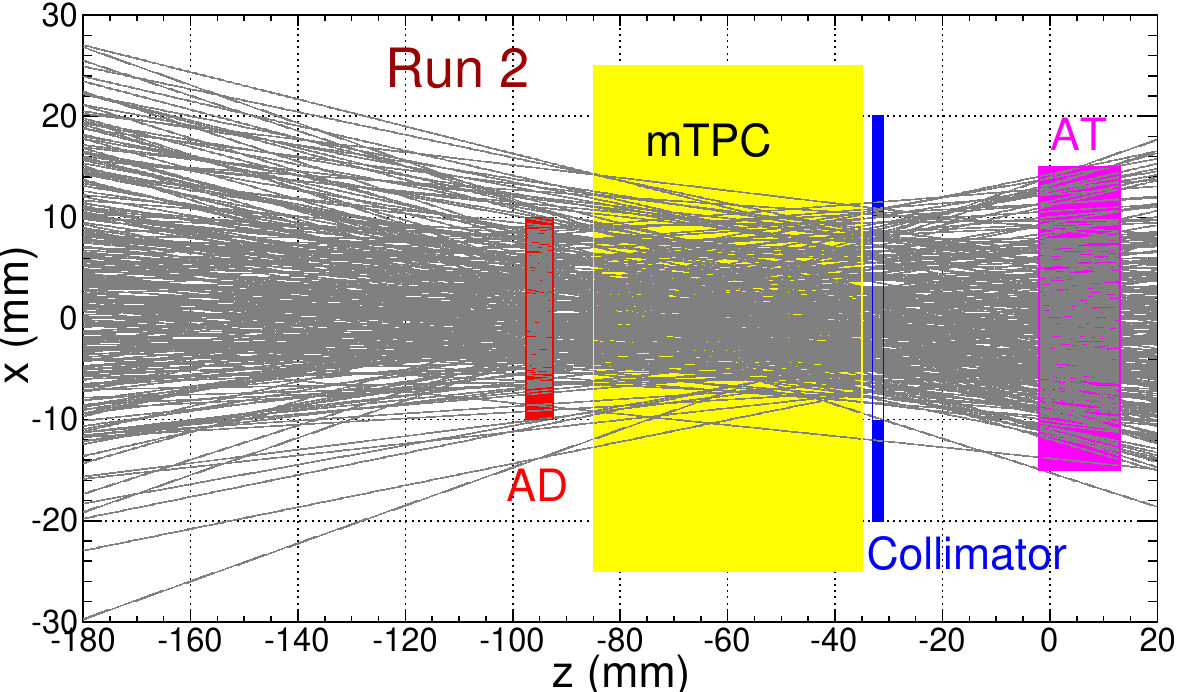}
  \includegraphics[height=0.18\textheight]
                {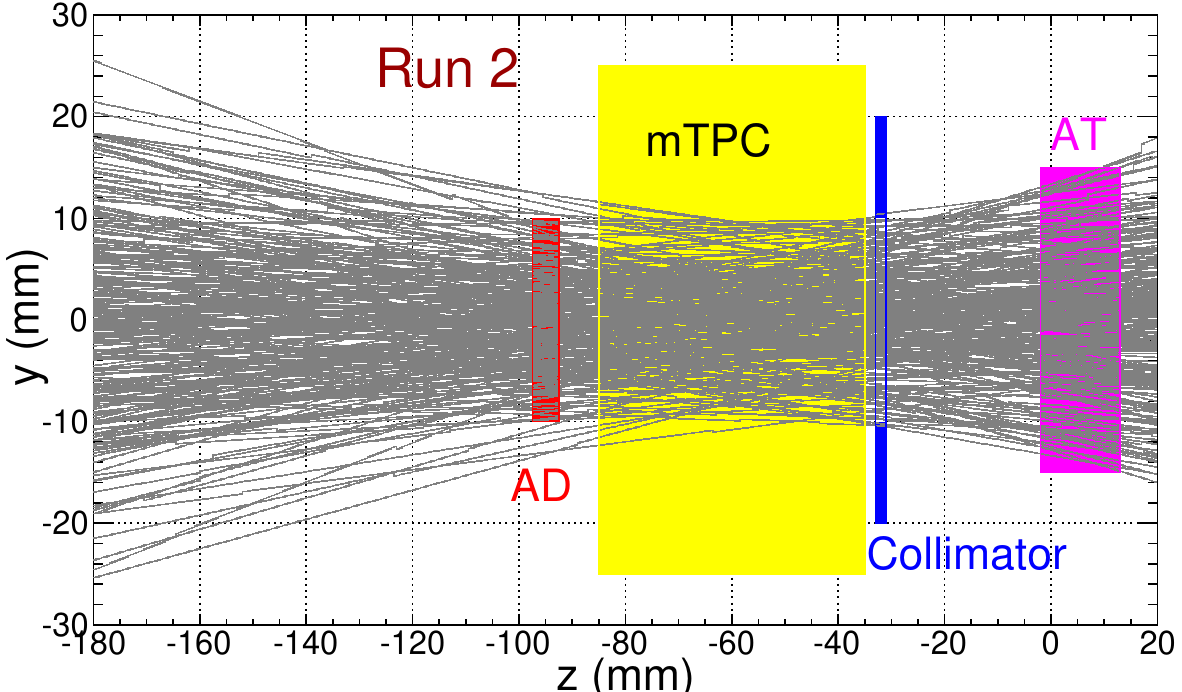}\\
  \vspace{3pt}
  \includegraphics[height=0.18\textheight]
                {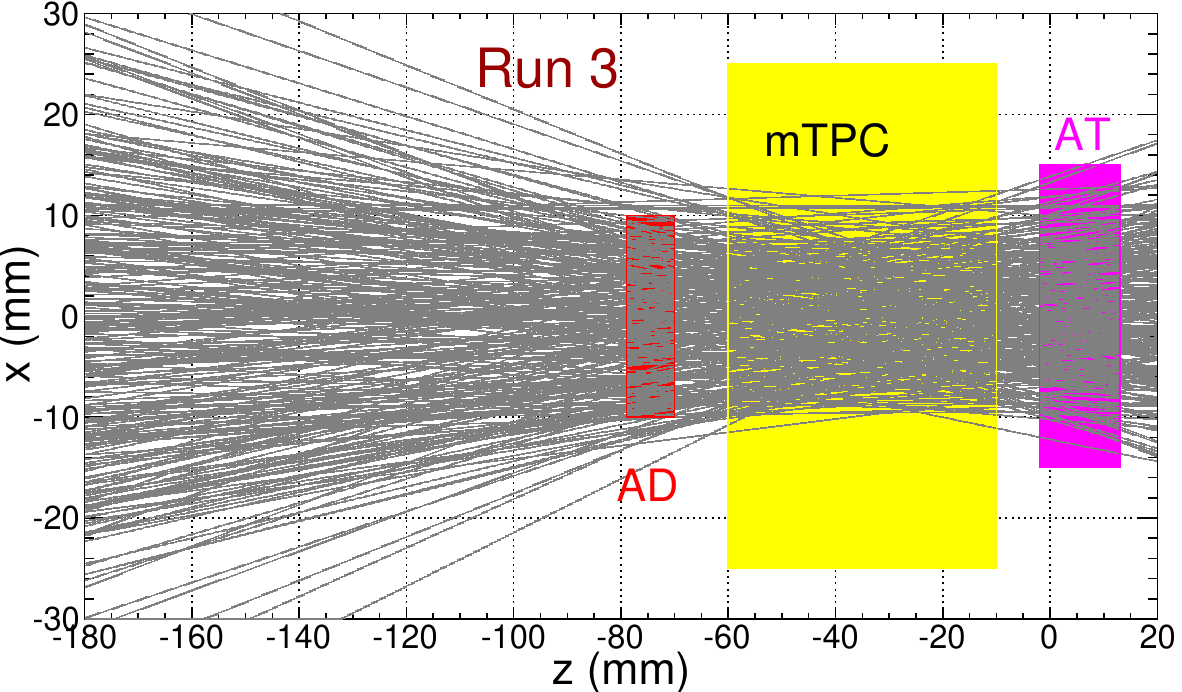}
  \includegraphics[height=0.18\textheight]
                {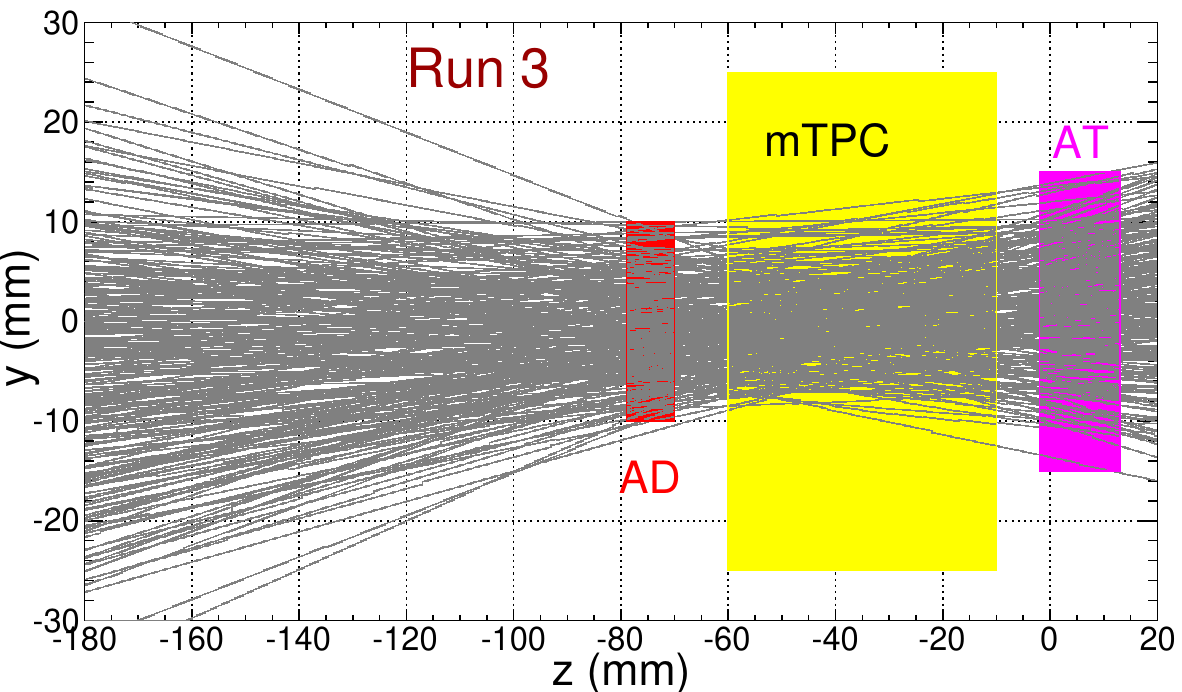}
  \caption{Representative pion beam trajectories reconstructed by the
    mTPC: $x$ vs.\ $z$ (left) and $y$ vs.\ $z$ (right) in Run\,2 (top)
    and Run\,3 (bottom), with outlines of the central region beam
    detectors, AD, mTPC, and AT, following Fig.~\ref{fig:run_geom}.  }
  \label{fig:trac_profiles}
\end{figure}

The linear independence of the two collinearity observables is
illustrated in Fig.~\ref{fig:coll_tests-2d}.
Correlations do appear for large deviations from 0, which are relatively
rare, as would be expected in case of a kink in the trajectory.  For
small deviations there is no such correlation, consistent with just the
broadening due to intrinsic detector resolution.
Collinearity tests may fail when there are two particles in the beam,
when one of the wires doesn't fire, if the tracked particle scatters, or
in rare cases through waveform distortions due to $\delta$-electron
presence.  Thus, a series of tests on the detected wire hit coordinates
are applied in order to identify and select valid tracks, applying the
following procedure:

\begin{enumerate}

  \item The full set of observed $x,y$ space points is scanned for a
    combination that satisfies the condition
    $\text{col}_{1x}^2+\text{col}_{2x}^2 \equiv \text{col}_x^2 \leq c_x$
    where $c_x$ is widened from 10 to 50\,mm$^2$ in steps of 10\,mm$^2$
    until a solution is found.  The corresponding cut circles are drawn
    in red in Fig.~\ref{fig:coll_tests-2d}(a).  Once a combination of
    $x_i$ values is found to satisfy one of the above $\text{col}_x^2$
    cuts, it is saved and removed from the set.  Search for tracks
    continues among the remaining wire hits until all are sorted into
    track candidates, or have failed the col$_x^2\leq 50\,\text{mm}^2$
    test.

  \item The $y_i$ values corresponding to the saved track candidates in
    step 1 are analyzed next using analogous criteria:
    $\text{col}_{1y}^2+\text{col}_{2y}^2 = \text{col}_y^2 \leq c_y$,
    again with $c_y$ from 10 to 50\,mm$^2$ in steps of 10\,mm$^2$.  The
    circle corresponding to the first of these tests, $\text{col}_y^2
    \leq 10\,\text{mm}^2$, is shown in Fig.~\ref{fig:coll_tests-2d}(b).
    At this stage, a track candidate consists of a set of four wire hit
    $(x,y)$ coordinate pairs that independently satisfy a $c_x$ and a
    $c_y$ cut.

  \item Sets of track candidate $x_i$ and $y_i$ hit coordinates found in
    the above two steps are further subjected to the final condition for
    a valid track: $\text{col}_x^2 + \text{col}_y^2 \leq
    50\,\text{mm}^2$.  This last test is indicated by the red line in
    Fig.~\ref{fig:coll_tests-2d}(c).
\end{enumerate}
If no track with signals from all four wires is found, the beam
trajectory may still be reconstructed.  In such situations, new
collinearity observables are evaluated based on the signals of three
wires only.  For instance, in the absence of $x_1$, a new collinearity
test can be constructed:
\begin{linenomath}
\begin{equation} 
    x_4-2x_3+x_2= 0\,.  \label{eq:col_x_no_x1}
\end{equation}
\end{linenomath}
Similar tests can be made in case a different wire didn't report.
Finally, if still no tracks are found, combinations of just two hits are
searched, with the condition that the trajectory crosses AD and AT.
However, for events with fewer than four anode wires reporting
analyzable signals, the ability to identify trajectory kinks is reduced.
Once tracks have been found, the corresponding trajectories are
evaluated by linear interpolation.  Figure~\ref{fig:trac_profiles} shows
a representative set of reconstructed pion trajectories in horizontal
and vertical projections.

\begin{figure}[t]
 \centering
 \includegraphics[height=0.32\textheight]{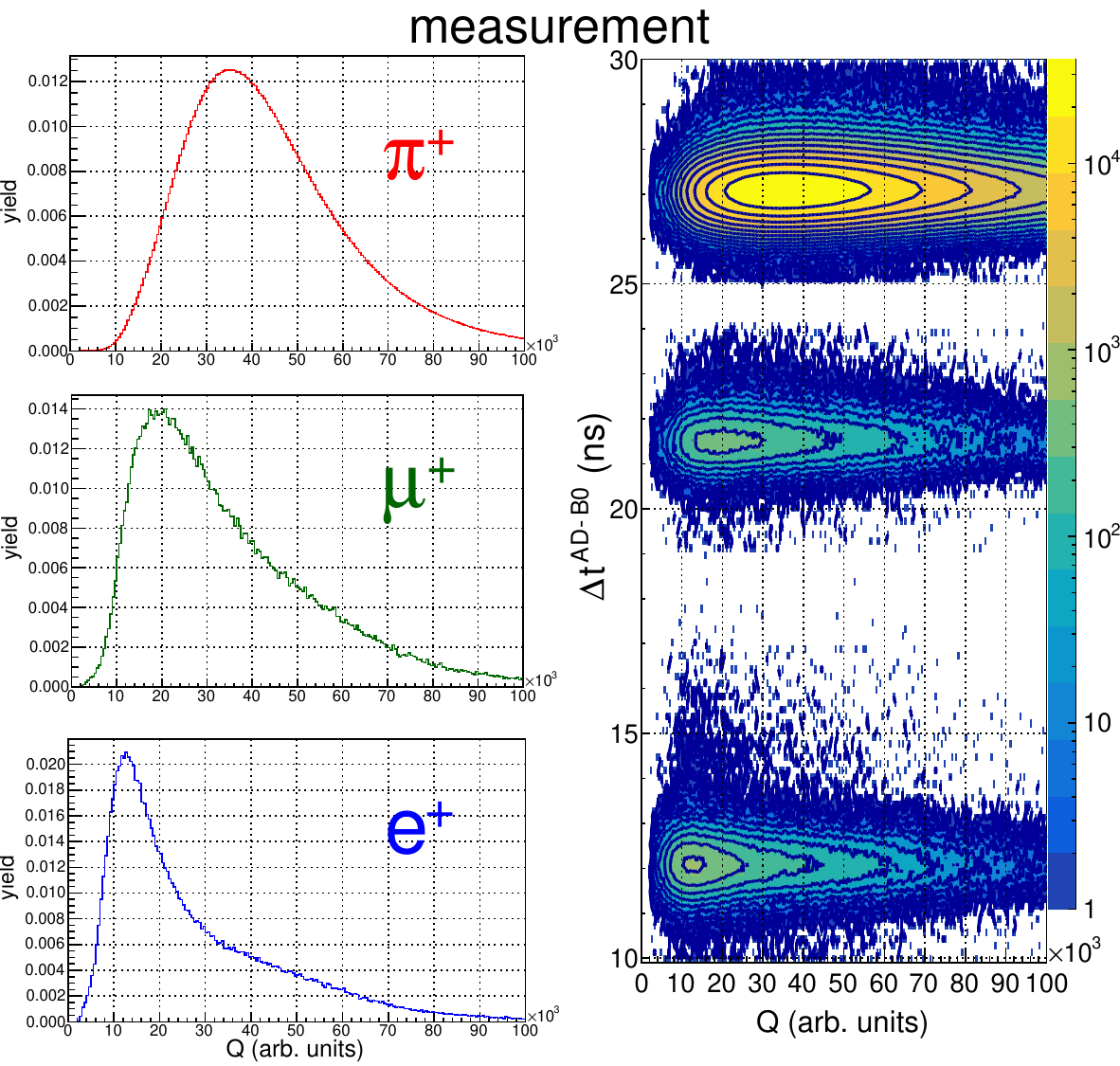}
 \hspace*{12pt}
 \includegraphics[height=0.32\textheight]{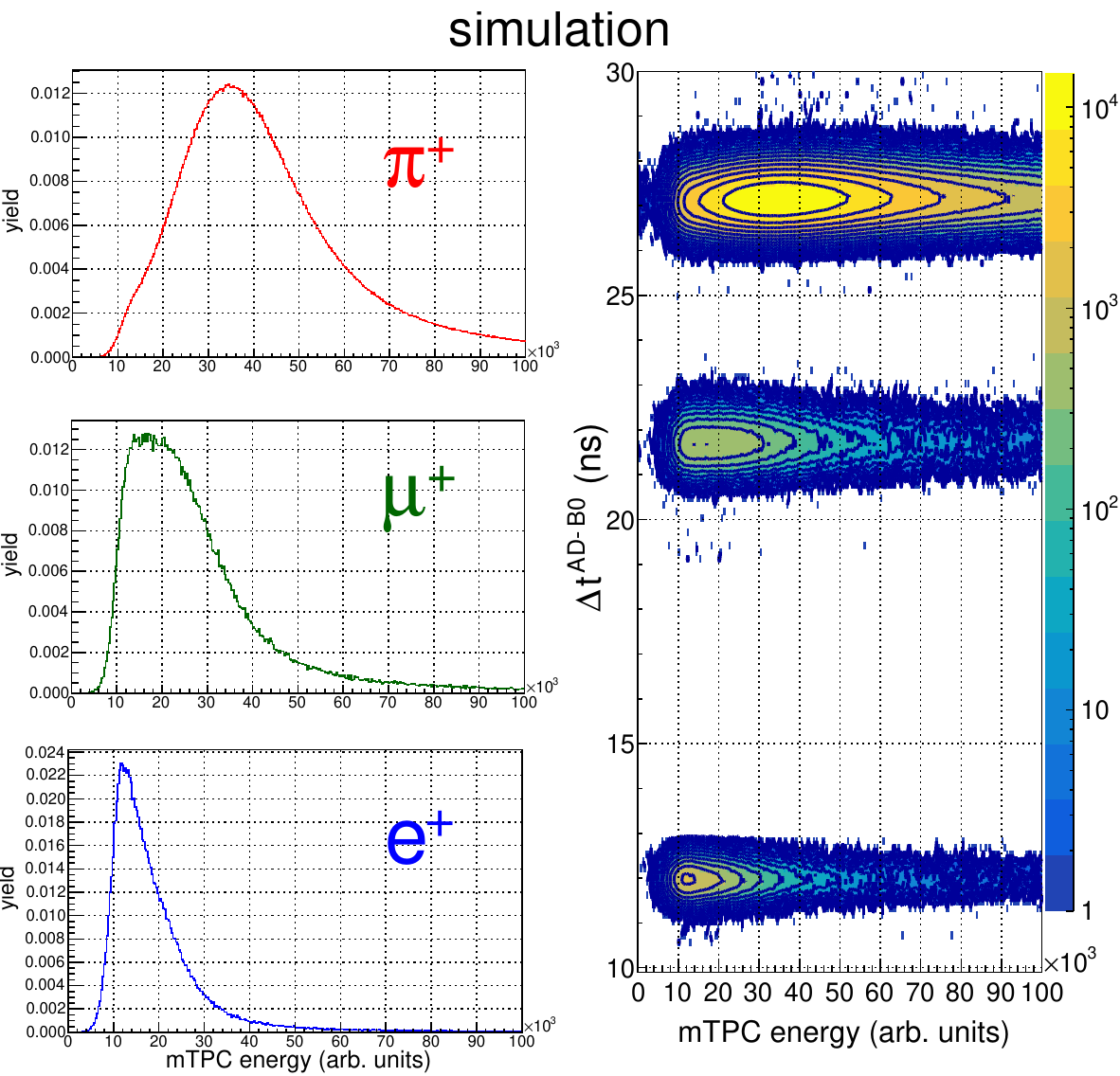}
 
 \caption{Beam PID from BC--AD TOF, and associated energy loss in the
   mTPC gas.  Left: sum of measured anode geometric mean charges $Q =
   \sum_{i=1}^4\langle q_i\rangle$.  Right: simulated energy loss in
   mTPC gas, scaled to anode charge units.  PID for the measured events
   (left plots) was made by imposing appropriate cuts on the $\Delta
   t^{\text{AD$-$BC}}$ TOF variable (2-dim.\ contour plots).  The
   disparate velocities of $e$, $\mu$ and $\pi$ sharing the same
   $\sim$73\,MeV/$c$ momentum, induce different energy losses in the
   chamber gas, producing distinct amplitude spectra, here compared
   qualitatively.  Energy loss distributions are normalized to unit
   integral for ease of comparison; hence, ``yield'' is relative.  See
   text for additional discussion.  \label{fig:beam_e-mu-pi} }
\end{figure}

In addition to reconstructing the pion beam trajectories, discussed
above, the mTPC (along with the AD) provides independent confirmation of
correct identification of the three beam particle types.  The dissimilar
velocities of the $e$, $\mu$, and $\pi$ sharing the same momentum,
result in different energy losses in the chamber gas, thus producing
distinct amplitude spectra, as illustrated in
Fig.~\ref{fig:beam_e-mu-pi}.
 Hence, beam particle identification (PID), accomplished mainly through
 the BC--AD TOF observable, is independently validated by the
 qualitative agreement between the measured and Monte Carlo predicted
 energy deposited in mTPC gas by the three particle types.  We note that
 a detailed simulation of the anode charge collection, e.g., accounting
 for delta electrons, was not implemented as the effects cancel (apart
 from an inconsequential rescaling) in the anode charge asymmetries
 (Fig.~\ref{fig:coord_calib}), and ultimately do not affect the
 evaluation of \RpiEXPemu.  The relevant instrumental effects are
 incorporated through smearing, discussed in Sec.~\ref{sec:MC-simul}.

As seen above, mTPC beam tracking provides the projected stopping
$(x,y)$ coordinates for each beam pion that stops and decays in the AT.
The $z$-stop coordinate is determined as follows.  The energy of a beam
pion as it reaches the active degrader, $E_{\text{AD}}^{\pi}$, is
calculated from the BC--AD time of flight.  To obtain the pion's energy
as it reaches the target, and thus its range in the target,
$E_{\text{AD}}^{\pi}$ is reduced by the observed energy deposited by the
pion in the degrader.  In this way the pion stop position is predicted
in three dimensions.  Accurate prediction of the location of the stopped
pion in the target is critical to controlling the PEN analysis
systematics, primarily because it allows accurate attribution of
fractions of observed target energy to the pion and positron,
respectively, thus increasing the reliability of detection of the
intermediate muon in the decay chain $\pi \to \mu \to e$.  Reliable
sorting of decay types, \pieii\ or \pimii, is the main systematic
challenge in the measurement.  This point is further revisited in
Sec.~\ref{sec:mTPC-in-PEN-anal}.

\begin{figure}[h!]
\centering
 \includegraphics[height=0.28\textheight]{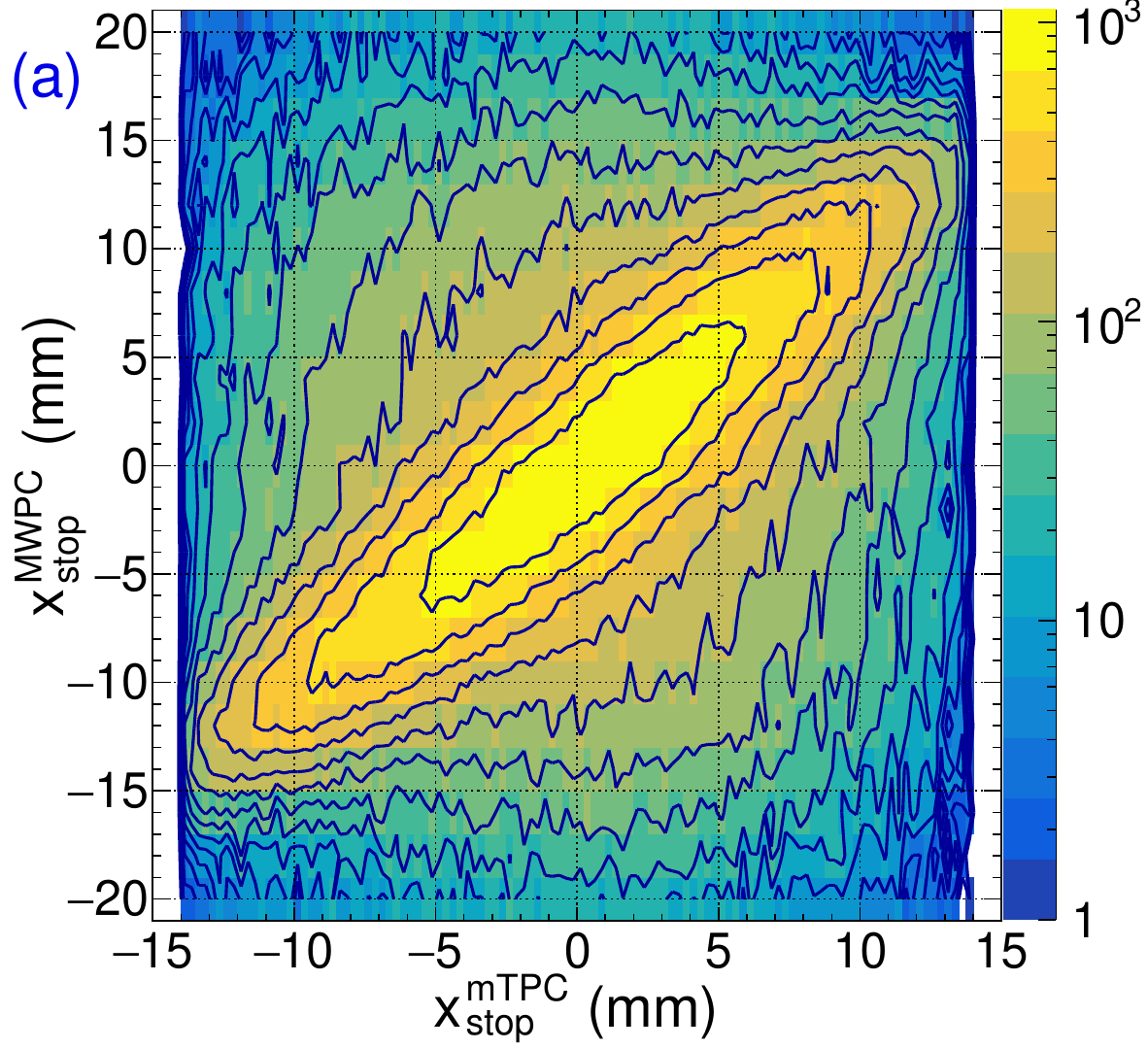}
 \includegraphics[height=0.28\textheight]{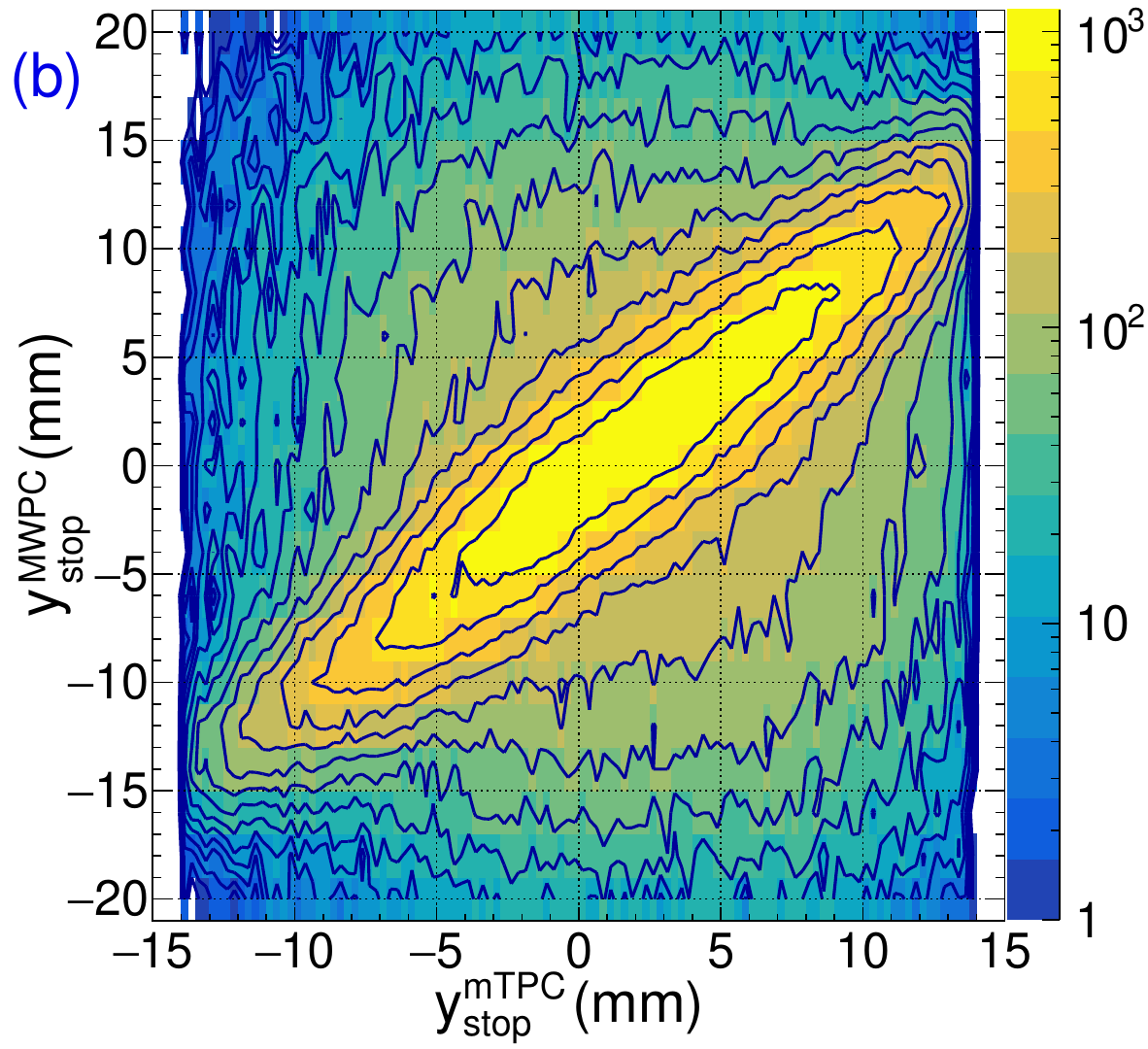}
 \caption{(a) $x$ coordinate of the pion stopping location
   reconstructed from MWPC vertical tracks of decay positrons from the
   target vs.\ the $x$ value reconstructed from the mTPC and beam TOF
   information.  (b) corresponding distribution for $y$ from the analysis
   of horizontal positron trajectories. \label{fig:mwpc_stop_check} }
\end{figure}

PEN analysis allows for independent calibration of mTPC tracking using
the two MWPCs.  Specifically, the distribution of pion stopping
$(x,y,z)$ coordinates predicted by mTPC, BC and AD data, has been
cross-calibrated using decay positron trajectories observed with the
MWPCs.  Horizontal (constant $y$) and vertical (constant $x$) tracks
through the MWPCs provide an independent calibration of the
mTPC/TOF-predicted stopping $(x,y)$ coordinates, as shown in
Fig.~\ref{fig:mwpc_stop_check}.

Detection efficiency for a given mTPC anode wire has been determined by
observing the fraction of valid tracks for which that wire registers a
signal when the remaining three anode wires have fired.  The four mTPC
wire detection efficiencies are tabulated in Tab.~\ref{tab:eff_mtpc}
for PEN Runs\,2 and 3.  We note that the A1 wire failed completely
through the second half of Run\,3, and the A2 wire failed near the end
of Run\,3.  For this reason the mTPC wire efficiencies for Run\,3 are
calculated and listed only for the first half of that run.  We also note
that the two wires that eventually failed were less efficient from the
start; the actual cause of the problem has not been identified.
\begin{table}[t!]
  \caption{Pion detection efficiencies extracted for each mTPC anode
    wire during the two PEN run periods, as indicated.  Physically
    different mTPCs were used in Runs\,2 and 3, as discussed in
    Sec.~\ref{sec:mTPC-design}.  Comparison of Run\,2 and 3 wire
    efficiencies indicates that the Run\,2 mTPC design with the grid
    wires (labeled ``F'' in Fig.~\ref{fig:mTPC_1_design}) worked better.}
  \label{tab:eff_mtpc}
  \begin{center}
    \begin{tabular}{ccc}
      \hline
      Anode wire & Run\,2 & Run\,3\\
      \hline
      1 & 99.590 $\pm 0$.004\% &95.84 $\pm 0$.04\% \\
      2 & 99.288 $\pm 0$.006\% &96.52 $\pm 0$.02\% \\
      3 & 99.095 $\pm 0$.007\% &98.76 $\pm 0$.01\%\\
      4 & 99.794 $\pm 0$.003\% &99.43 $\pm 0$.01\% \\
      \hline
    \end{tabular}
  \end{center}
\end{table}
\noindent

\section{Monte Carlo simulation of the mTPC response \label{sec:MC-simul}}
\begin{figure}[t!]
  \centering	
  \includegraphics[height=0.25\textheight]{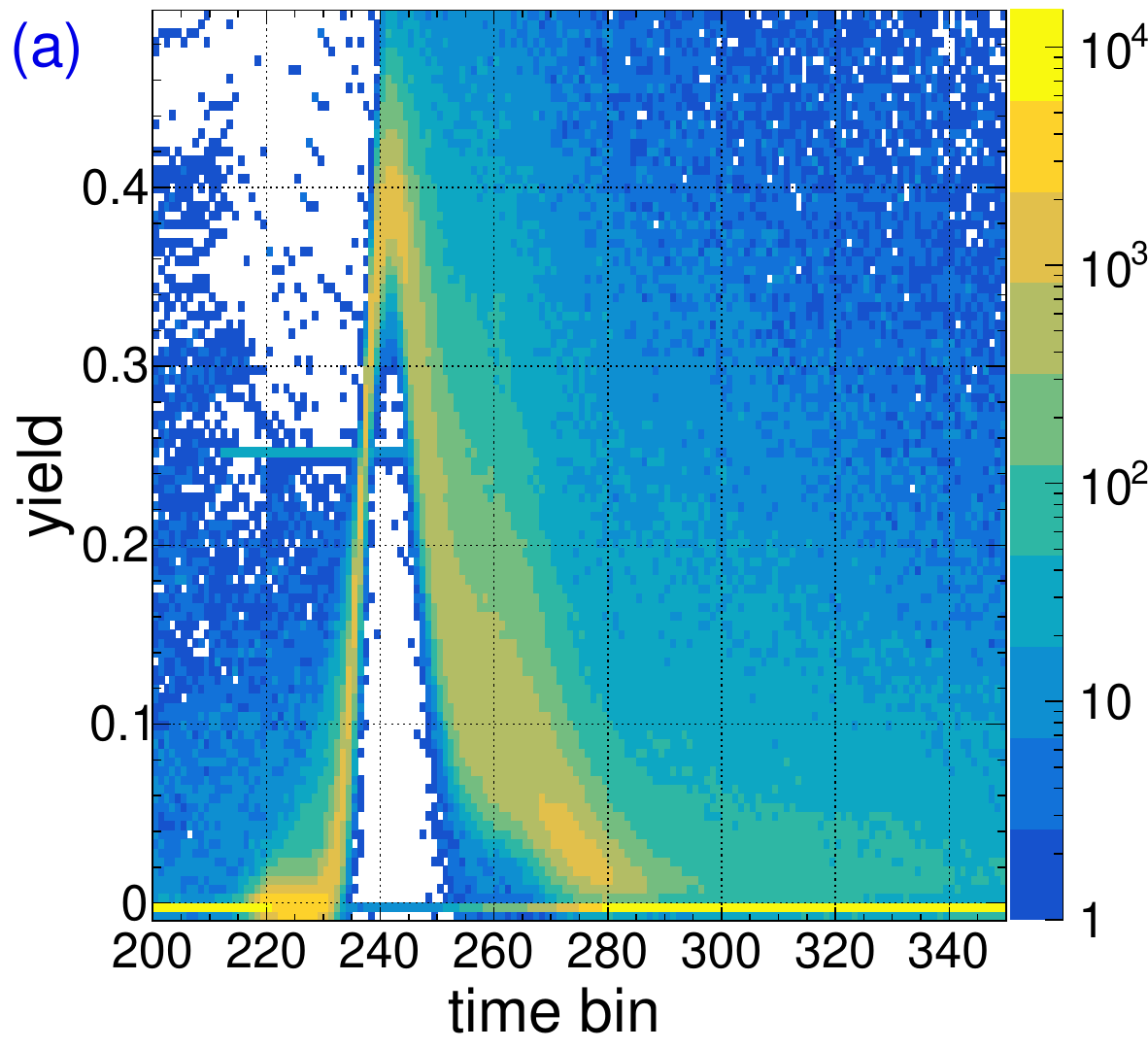}
  \includegraphics[height=0.25\textheight]{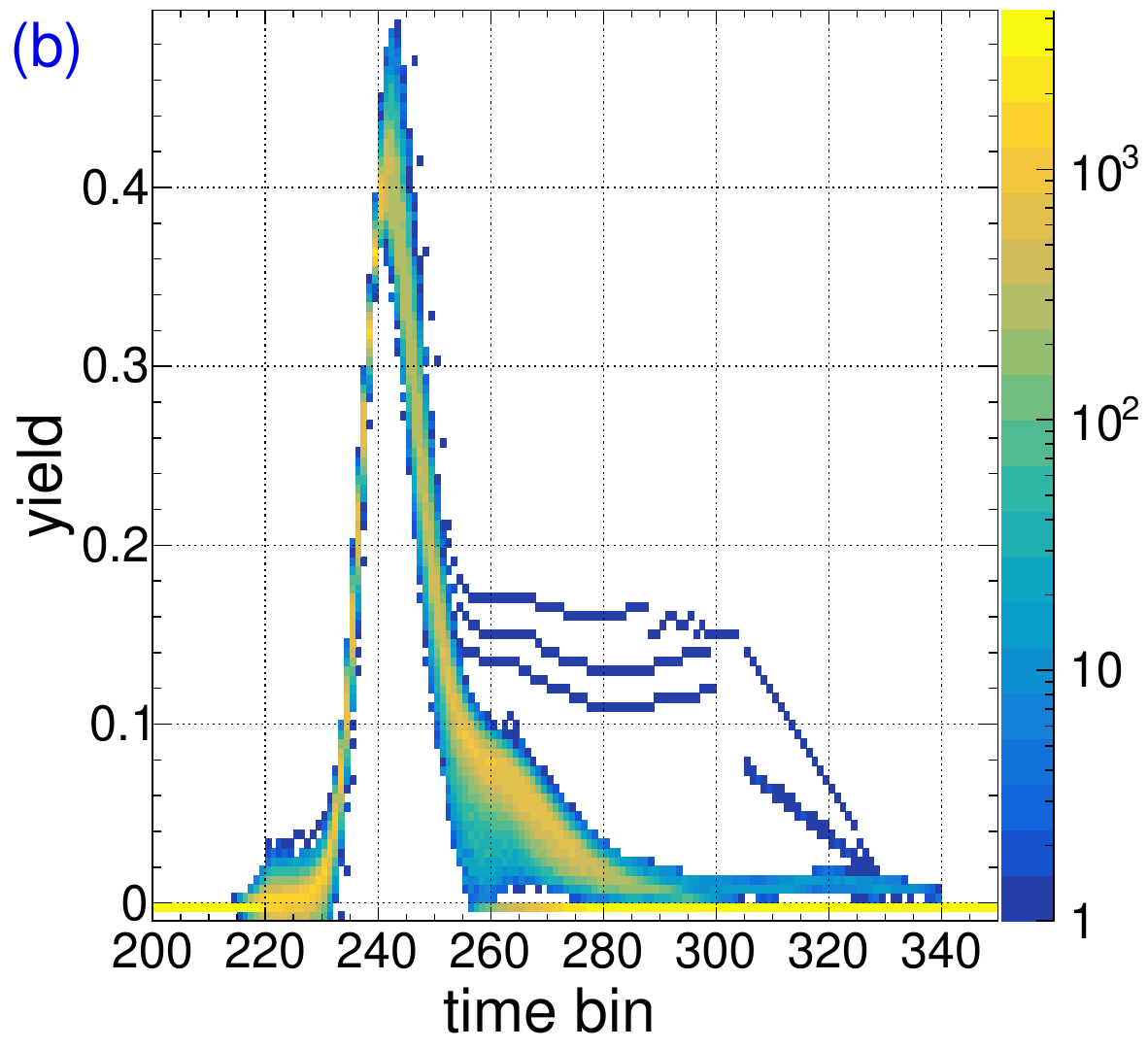}
  \caption{(a) Superimposed mTPC digitized waveform response from
    measured data. 
    (b) Synthetic mTPC waveforms constructed by waveform averaging of
    measured response (a), for use in the Monte Carlo simulation.  
   \label{fig:sim_wf}}
\end{figure}

\begin{figure}[b!]
  \centering
  \includegraphics[height=0.25\textheight]{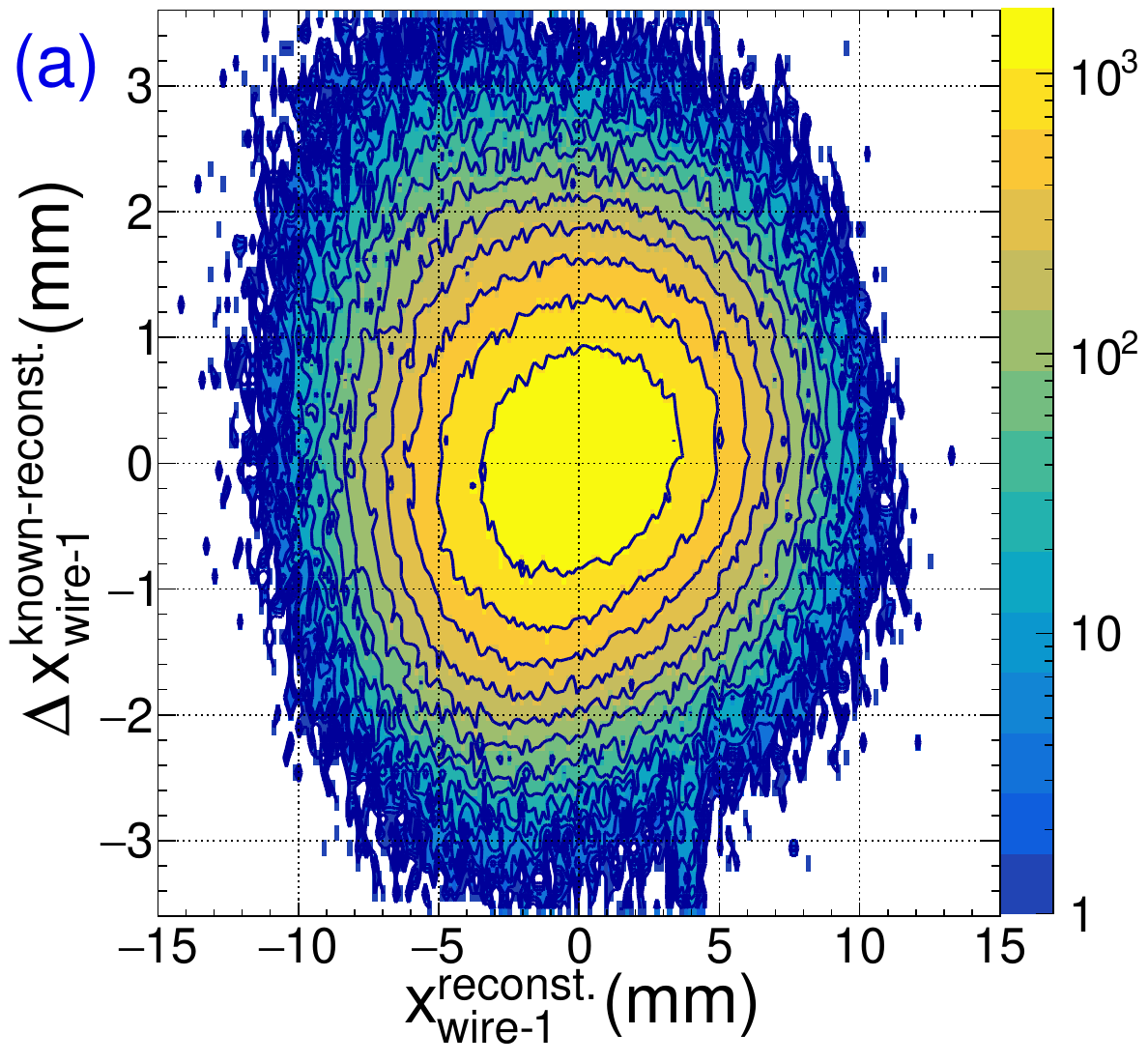}
  \includegraphics[height=0.25\textheight]{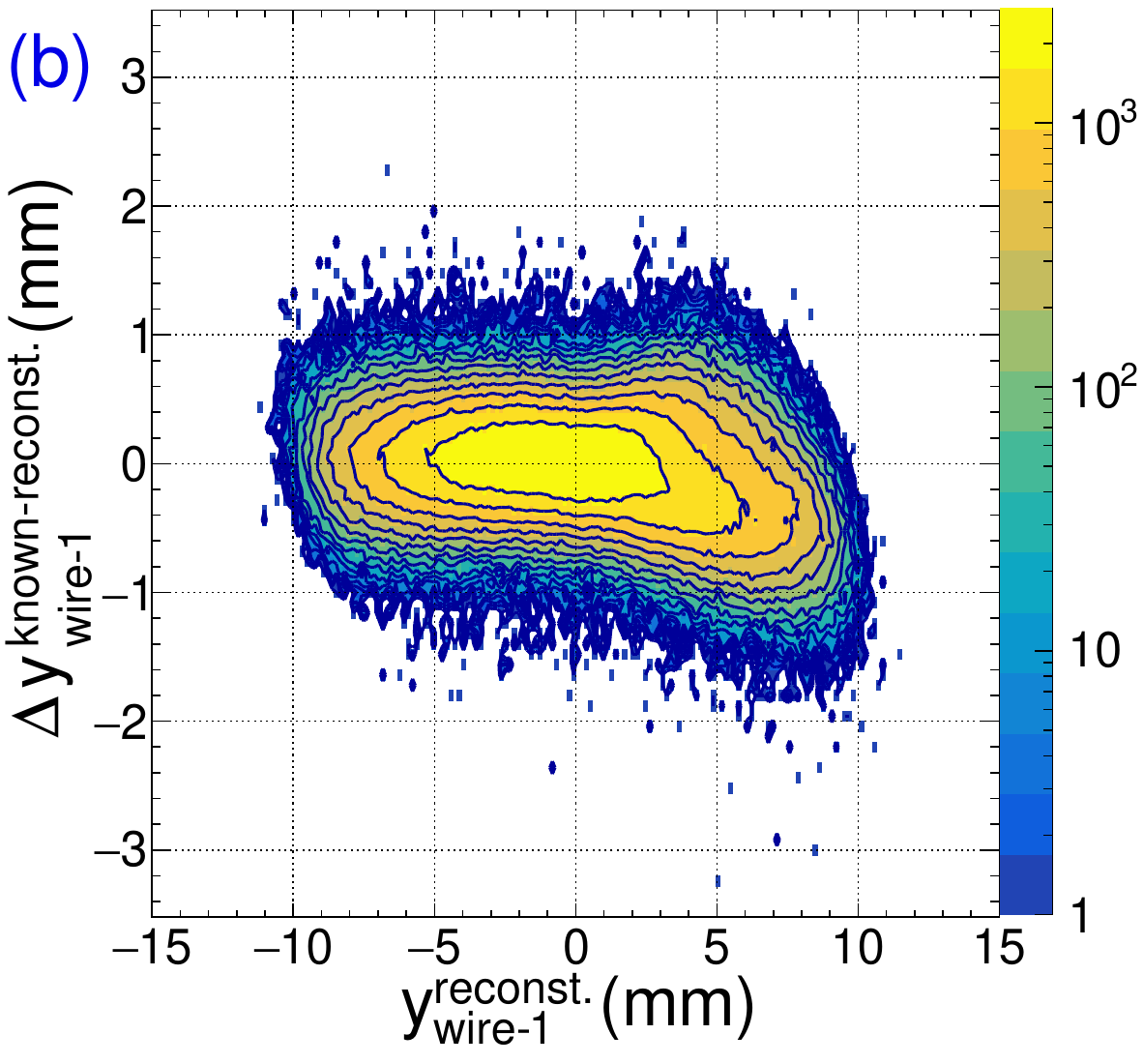}
  \caption{(a) Difference between known and reconstructed pion beam $x$
    coordinates plotted against the reconstructed $x$ at first wire.
    (b) Same for $y$.  The narrower width of the $\Delta y$ distribution
    reflects the superior position resolution obtained from the drift
    time as compared to charge division.
    \label{fig:sim_x_y-reconst} } 

\end{figure}

\begin{figure}[t!]
  \centering	
  \includegraphics[height=0.25\textheight]{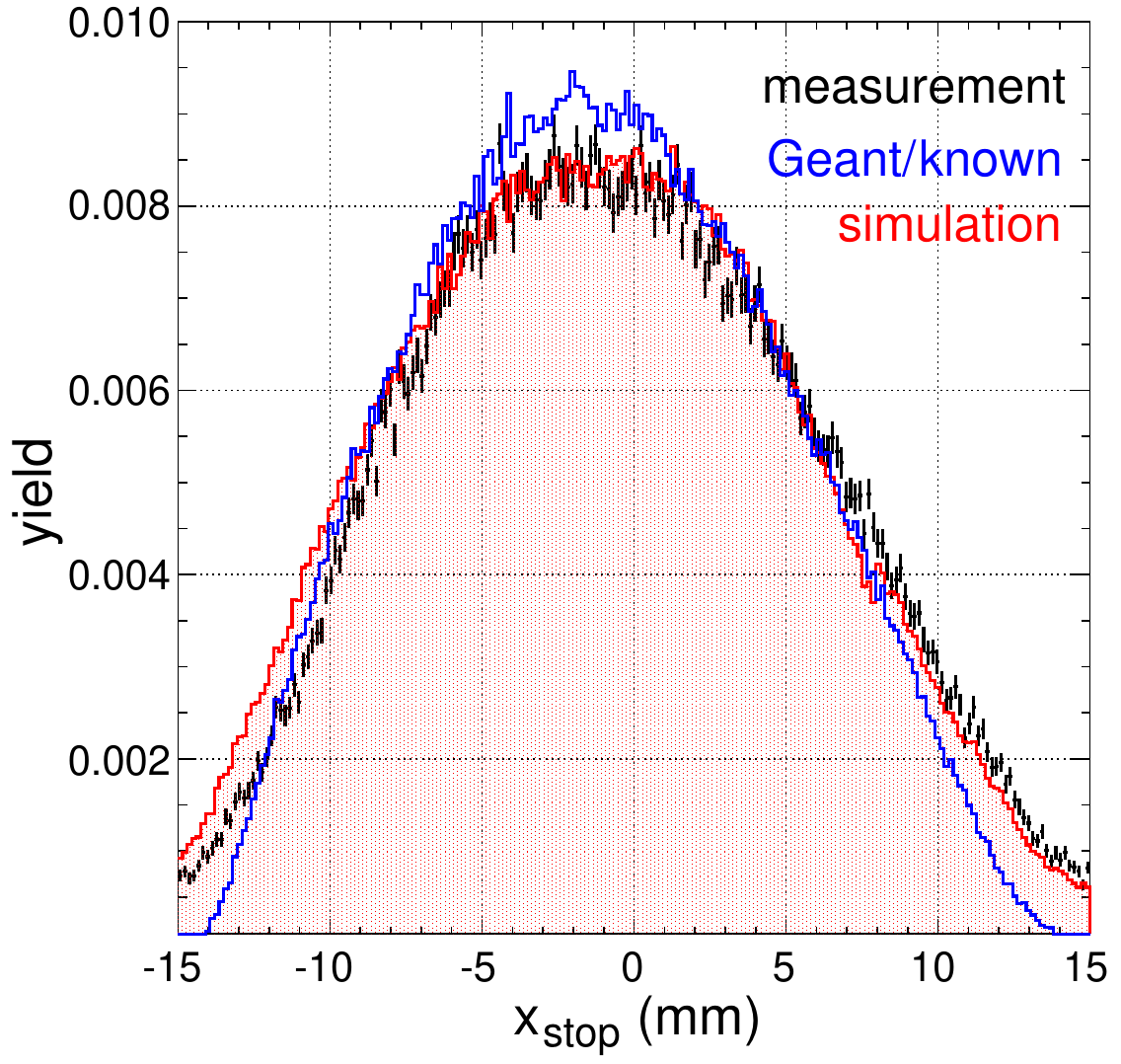}
  \includegraphics[height=0.25\textheight]{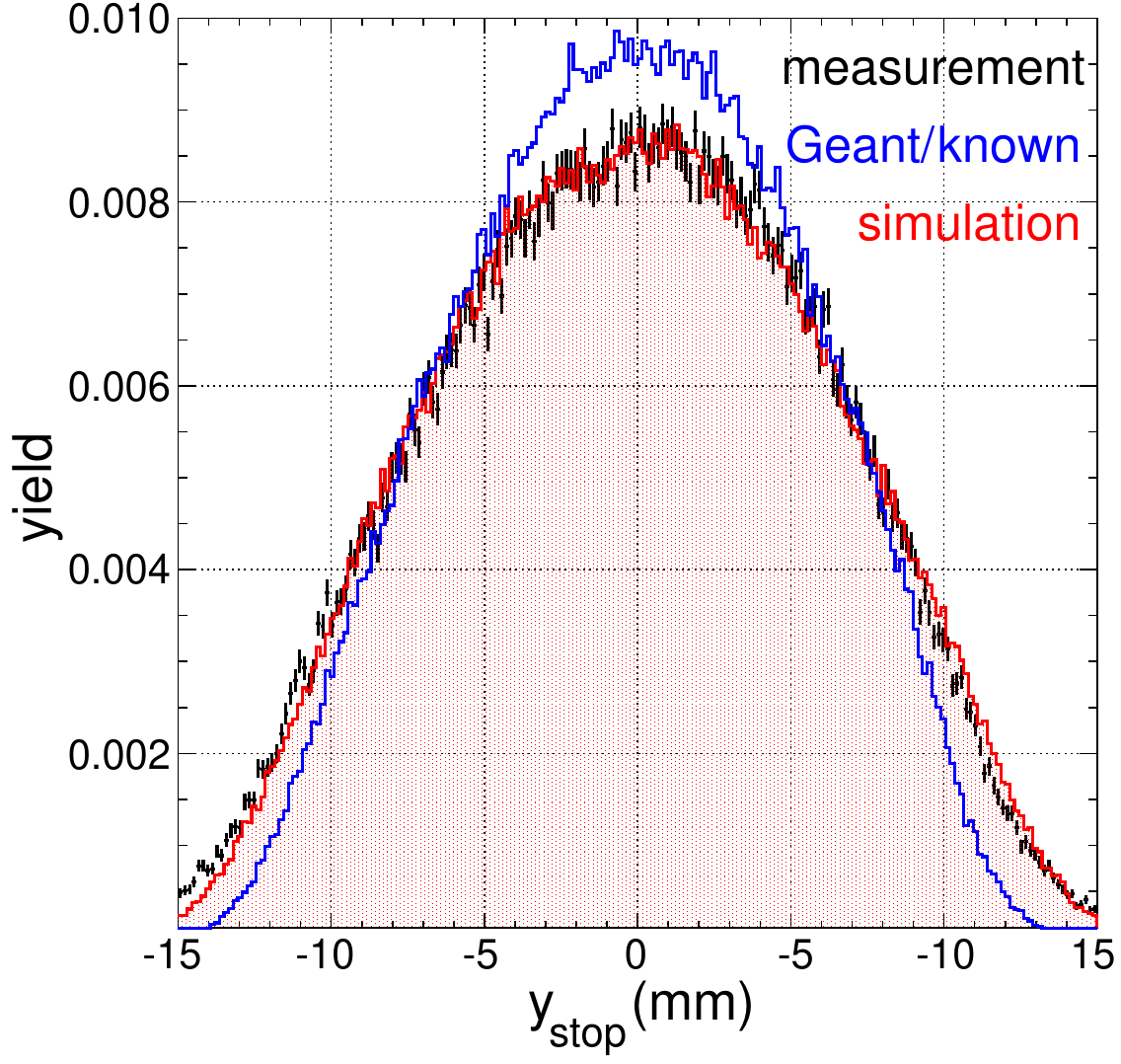} \\
  \hspace*{8pt}
  \includegraphics[height=0.20\textheight]{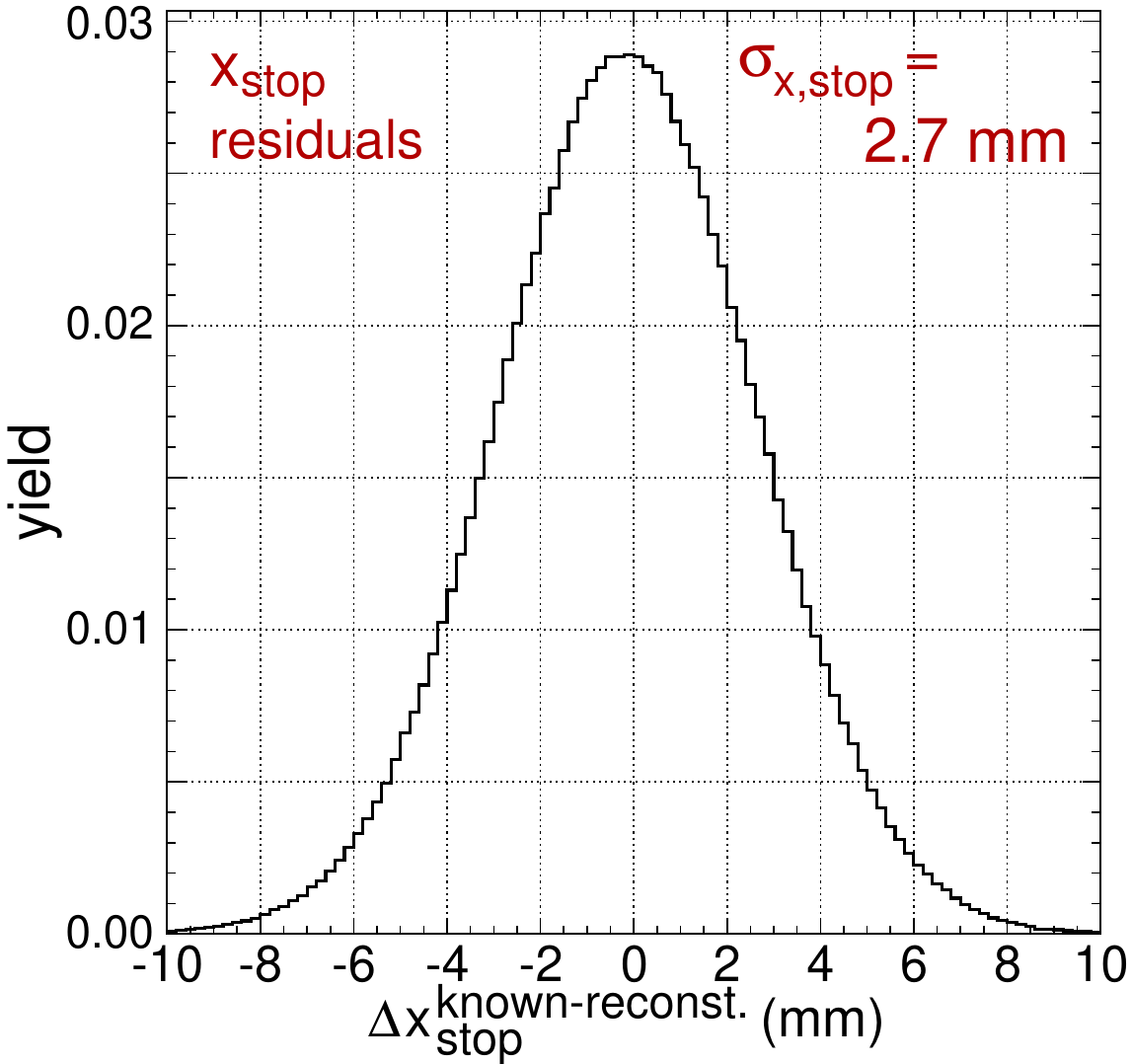}
  \hspace{28pt}
  \includegraphics[height=0.20\textheight]{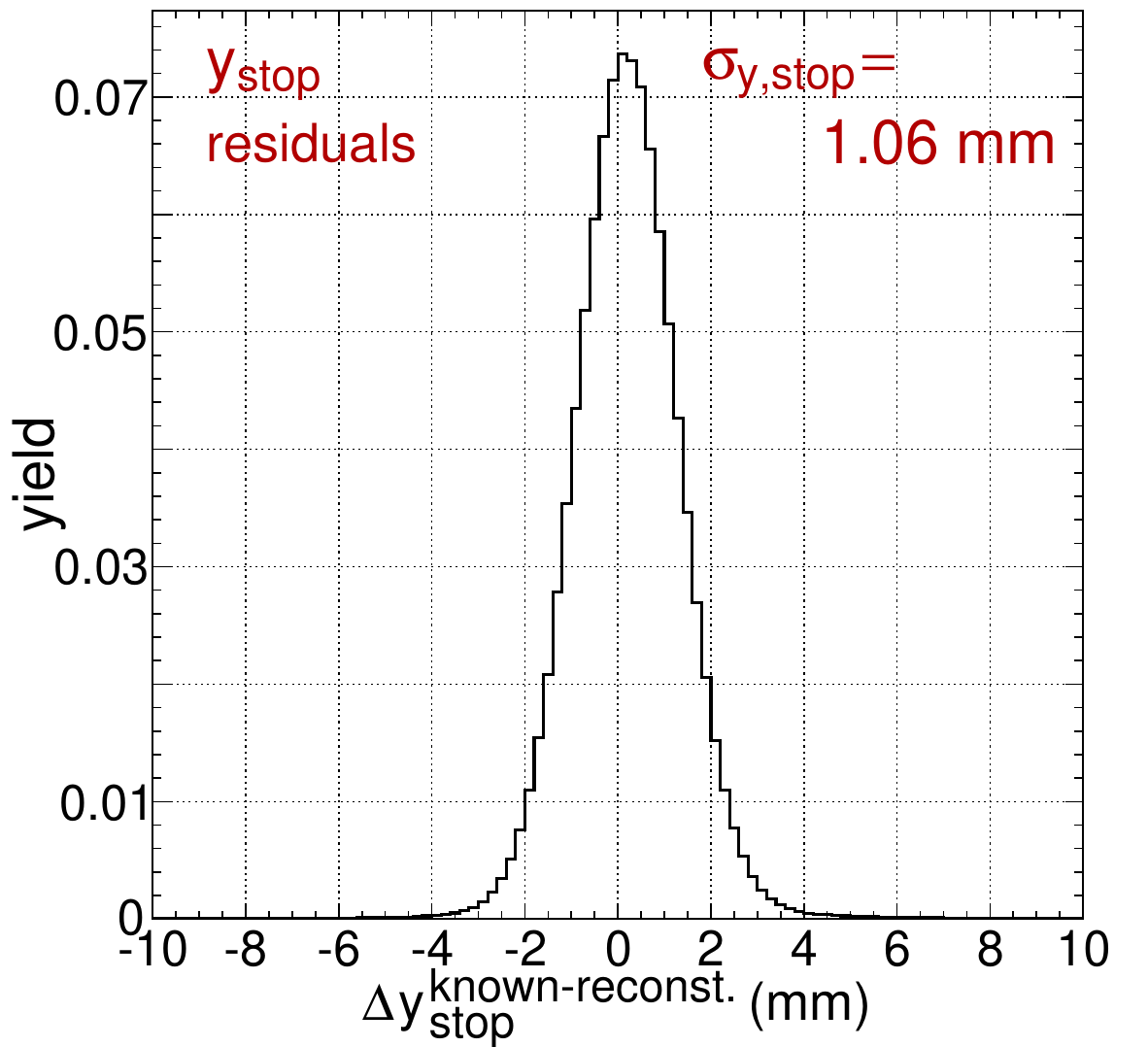}
  
    \caption{Upper left: distribution of pion stopping $x$ coordinates
      in the target detector.  Upper right: stopping $y$ distribution.
      The simulation or ``reconstructed'' histogram (red) reflects the
      full detector response (mTPC waveforms and smearing); it is in
      excellent agreement with measurement histogram (black).  Geant
      (blue) histogram plots the ``known'' stopping coordinate values in
      the simulation.  Lower left and right: corresponding
      known\,$-$\,reconstructed residuals for $x$ and $y$, respectively,
      with standard deviations $\sigma_{x,\text{stop}}$ and
      $\sigma_{y,\text{stop}}$ as indicated.  All data shown are from
      Run\,2. 
      \label{fig:xst_yst_meas-sim_comp}}
\end{figure}

Monte Carlo simulations producing highly realistic synthetic events are
needed in order to reach the precision goal of the PEN data analysis.
Detailed detector geometries and responses are implemented using the
Geant4 toolkit\cite{Geant4}, supplemented by the PEN C++ code.  The
simulation uses waveform kernels obtained by taking an average of
numerous observed detector responses, properly normalized and shifted,
as illustrated in Fig.~\ref{fig:sim_wf}, to faithfully simulate details
of the detector response.
Synthetic waveforms are constructed using the $x$ and $y$ trajectory
coordinates, along with the energy deposited in the $z$ regions
instrumented by the individual wires.  In the process, Monte Carlo
values of the $x$ and $y$ coordinates are appropriately smeared to
account for the resolution of the detector.  The waveforms are
constructed by using a linear function in $y$ for the time bins, and two
separate linear functions in $x$ for the anode left and right signal
amplitudes, with opposite signs of their respective slopes.  Simulated
synthetic waveforms, which include appropriate smearing due to the
measured detector response, are used to obtain the simulated pion
stopping distribution, which, in turn, must agree with the measured
data.  Unlike the measured, simulated waveforms are accompanied by
information on the actual (``known'' or ``true'') $x$, $y$, and $z$
coordinate values of the simulated pion as it passes by each mTPC wire.
Of course, the reconstructed values are not quite the same as those of
the ``known'' observables due to the statistical nature of detector
smearing.  Results of this procedure are illustrated in
Figs.~\ref{fig:sim_x_y-reconst} and \ref{fig:xst_yst_meas-sim_comp}.
Fig.~\ref{fig:sim_x_y-reconst} demonstrates the absence of correlation
between the smeared (reconstructed) and known values of an anode wire
$(x,y)$ coordinates.  It also affirms the superior resolution in $y$,
derived from charge drift time in gas, compared to $x$, derived from
charge splitting of the anode signals.  There is a caveat associated
with the smearing of the stopping distribution, as follows.  The
simulation was constructed so as to match the shape and rms of the
stopping position in both the $x$ and $y$ coordinates (as well as $z$).
This can be achieved in two ways.  The first is to generate the pion in
the simulation with wider $p_x$ and $p_y$ momentum component
distributions.  The second is to smear the simulated detector response
of the mTPC.  Both methods will have the same effect of broadening the
stopping distribution.  However, the two methods will produce different
spectra in the downstream detectors, specifically in vertex quality
which also uses the information from the MWPC.  The best way to produce
highly realistic stopping distributions is to construct the simulated
waveforms not with the known positions of $x$ and $y$, but rather using
the known values plus a random variable representing the smearing due to
the detector resolution.  The known, reconstructed, and measured pion
stopping $x$ and $y$ distributions are compared for a set of Run\,2 data
in Fig.~\ref{fig:xst_yst_meas-sim_comp}.  Comparing the smeared and
reconstructed to the known Geant stopping $x$, $y$ values provides a
full measure of the mTPC tracking resolution at the pion stop location
in the AT.  The standard deviation values, displayed in the residuals
plots in Fig.~\ref{fig:xst_yst_meas-sim_comp}, were obtained by Gaussian
fits.  Even though the track position resolution at the measurement
(wire) location is much better, as seen in a closer look at the track
collinearity tests below, the pion stopping position affects the
branching ratio analysis more directly.

\begin{figure}[t!]
  \centering
  \includegraphics[height=0.22\textheight]{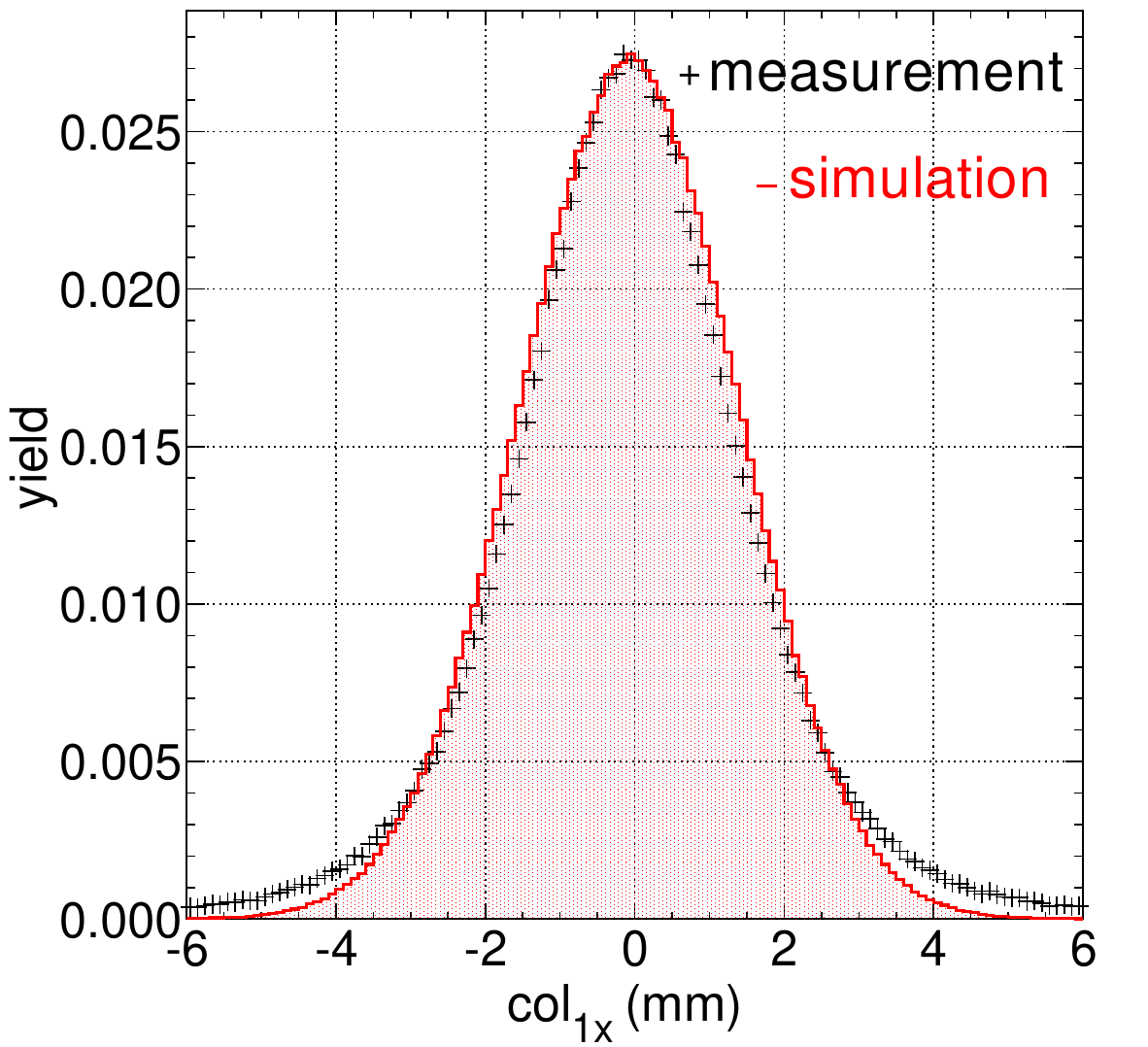}
  \includegraphics[height=0.22\textheight]{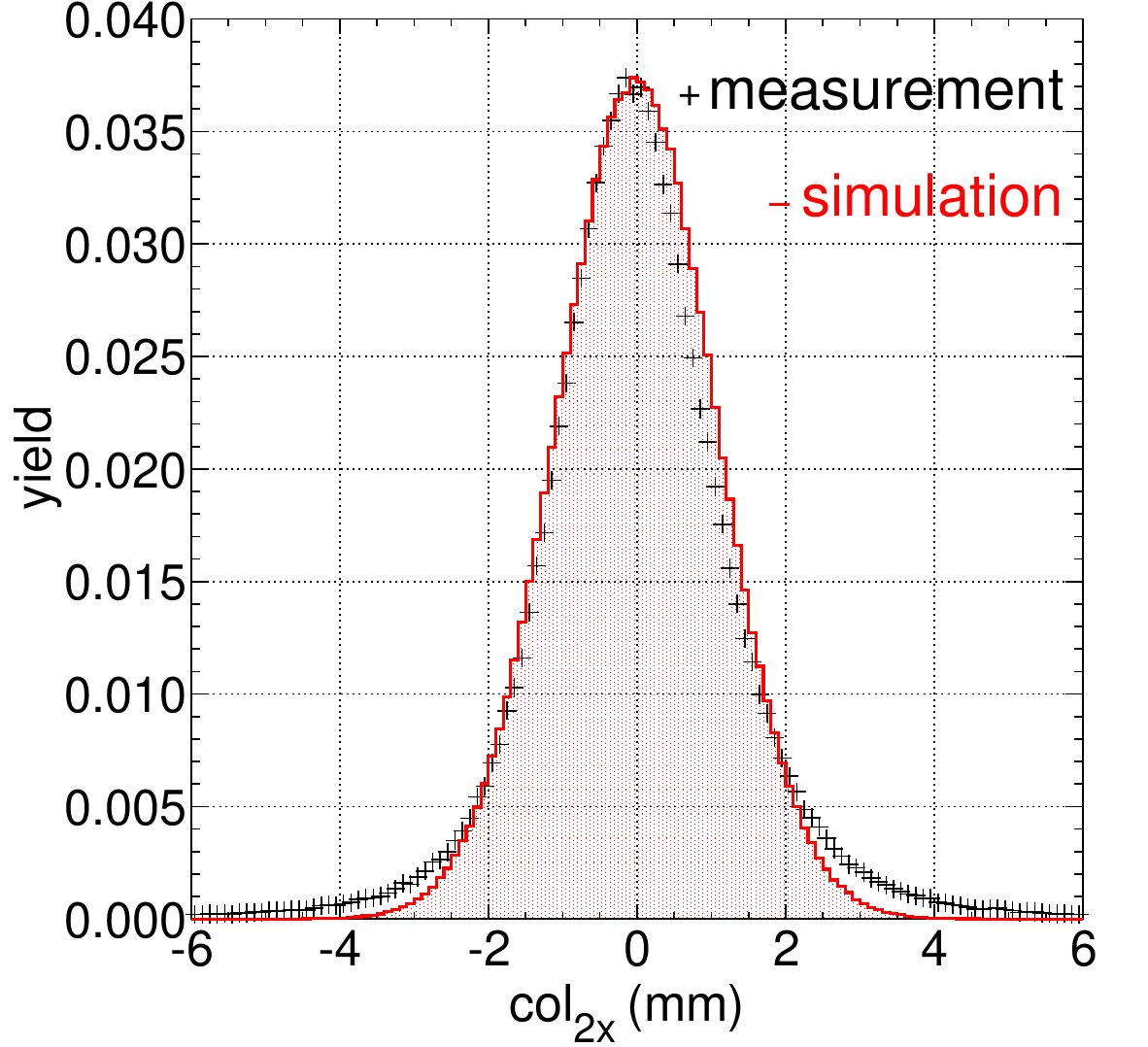}\\
  \includegraphics[height=0.22\textheight]{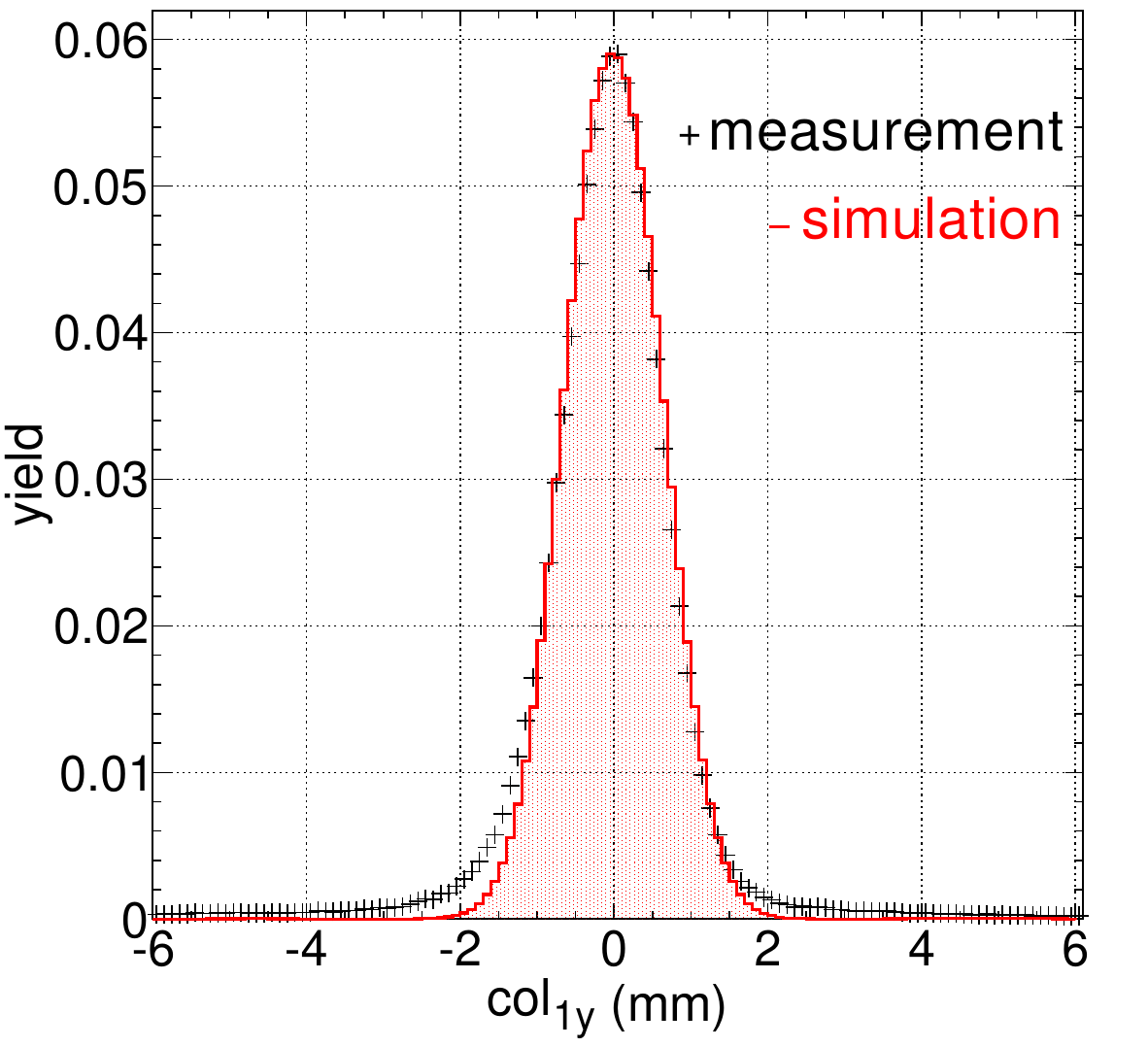}
  \includegraphics[height=0.22\textheight]{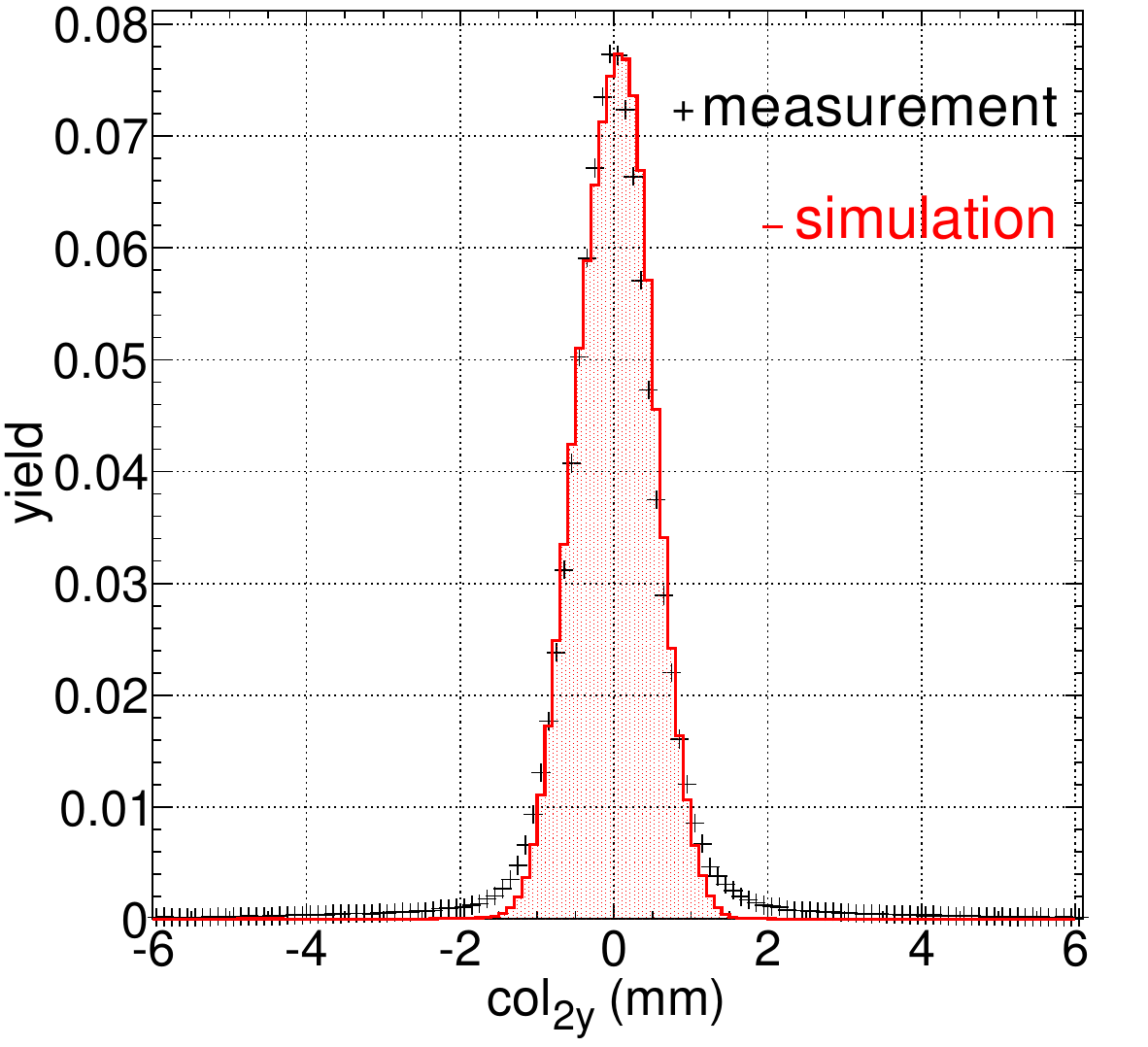}
  \caption{Comparison of two different collinearity tests measured in
    Run\,2 for $x$ (top row) and $y$ (bottom row) with corresponding
    Monte Carlo simulation results.  Measured (black) data are compared
    to the known (blue) values in the simulation, and to waveforms with
    realistic detector resolution (red).  Note that the same measured
    data (black) are plotted on logarithmic scale in
    Fig.~\ref{fig:coll_tests-1d}.  See text for discussion of
    resolution.
    \label{fig:coll_simul_x-y}}
\end{figure}

After smearing, the simulated mTPC $(x,y)$ space points fully reflect
the observed track coordinate resolutions.  This is confirmed in
Fig.~\ref{fig:coll_simul_x-y} by the good agreement between the typical
measured and simulated distributions of the four collinearity tests.
This figure therefore best illustrates the intrinsic tracking resolution
of the mTPC, averaged per anode wire.  The definitions of
$\text{col}_{1x}$ and $\text{col}_{2x}$, given in \eqref{eq:col_x1} and
\eqref{eq:col_x2}, respectively, yield
\begin{linenomath}
\begin{equation}
  \sigma(\text{col}_{1x}) 
                         = 2\langle\sigma_x\rangle_{\text{wire}}\,,
  \qquad \text{and} \qquad
  \sigma(\text{col}_{2x}) 
                         = 1.49\langle\sigma_x\rangle_{\text{wire}}\,,
  \label{eq:coll-sigmas}
\end{equation}
\end{linenomath}
where $\langle\sigma_x\rangle_{\text{wire}}$ is the average wire
tracking resolution in $x$, per wire.  Analogous expressions hold for
$y$.  By averaging $\langle\sigma_x\rangle_{\text{wire}}$ values derived
from col$_{1x}$ and col$_{2x}$, and doing the same for $y$, we arrive at
the best estimates for the average tracking resolution per wire in $x$
and $y$, respectively.  Values for the experimental standard deviations
of the four collinearities were determined by fitting the measured
distributions with Gaussian functions for the two PEN run periods
separately.  Thus obtained standard deviations, and the derived per-wire
track position resolutions, are summarized in Tab.~\ref{tab:track_resol},
\begin{table}[b!]
  \caption{Summary of the observed mTPC tracking resolutions, in mm:
    $\langle\sigma_x\rangle_{\text{wire}}$ and
    $\langle\sigma_y\rangle_{\text{wire}}$, average tracking resolutions
    per wire, evaluated from standard deviations of collinearity tests
    following \eqref{eq:coll-sigmas}; $\sigma_{x,{\text{stop}}}$ and
    $\sigma_{y,{\text{stop}}}$ for the projected $x$ and $y$ pion
    stopping coordinates in the AT, evaluated as shown in
    Fig.~\ref{fig:xst_yst_meas-sim_comp}.  All values are rounded off to
    the nearest 10\,$\mu$m. }
  \label{tab:track_resol} 
  \begin{center}
    \begin{tabular}{cccccc}
      \hline
        & $i$
          & $\sigma(\text{col}_{1i})$
            & $\sigma(\text{col}_{2i})$
              & $\langle\sigma_i\rangle_{\text{wire}}$
                & $\sigma_{i,\text{stop}}$ \\       
     \hline
     \multirow{2}{*}{Run\,2}
        & $x$ & 1.15 & 0.74 & 0.53 & 2.70 \\
        & $y$ & 0.63 & 0.48 & 0.32 & 1.06 \\
      \hline
     \multirow{2}{*}{Run\,3}
        & $x$ & 1.27 & 0.82 & 0.59 & 2.42 \\
        & $y$ & 0.81 & 0.71 & 0.44 & 0.95 \\
      \hline
    \end{tabular}
  \end{center}
\end{table}
along with the observed resolutions for the predicted pion stopping
coordinates in the AT, deduced from the $\Delta x_{\text{stop}}$ and
$\Delta y_{\text{stop}}$ residuals as shown above
(Fig.~\ref{fig:xst_yst_meas-sim_comp}).  These results are consistent
with the earlier observation of superior tracking resolution in $y$
(based on drift time) than in $x$ (based on charge division).  In all,
the obtained resolution values are very good, slightly above 0.5\,mm in
$x$, and below 0.5\,mm in $y$.  We note that the Run\,3 mTPC is somewhat
worse than its predecessor in intrinsic tracking resolution at the wire
locations.  In spite of that, it is better than the Run\,2 mTPC in terms
$\sigma_{x,{\text{stop}}}$ and $\sigma_{y,{\text{stop}}}$, resolution of
the pion stopping coordinates in the AT, thanks to the more compact
AD--AT geometry (Figs.~\ref{fig:run_geom} and \ref{fig:trac_profiles}),
made possible by its significantly lower mass.  Again, accurate
knowledge of the pion stop coordinates, and therefore of the decay
vertex, affects the branching ratio systematics more strongly, as we
discuss next.

\section{mTPC in the PEN analysis \label{sec:mTPC-in-PEN-anal}}
\begin{figure}[t!]
  \parbox{0.49\linewidth}{
    \includegraphics[width=0.94\linewidth]{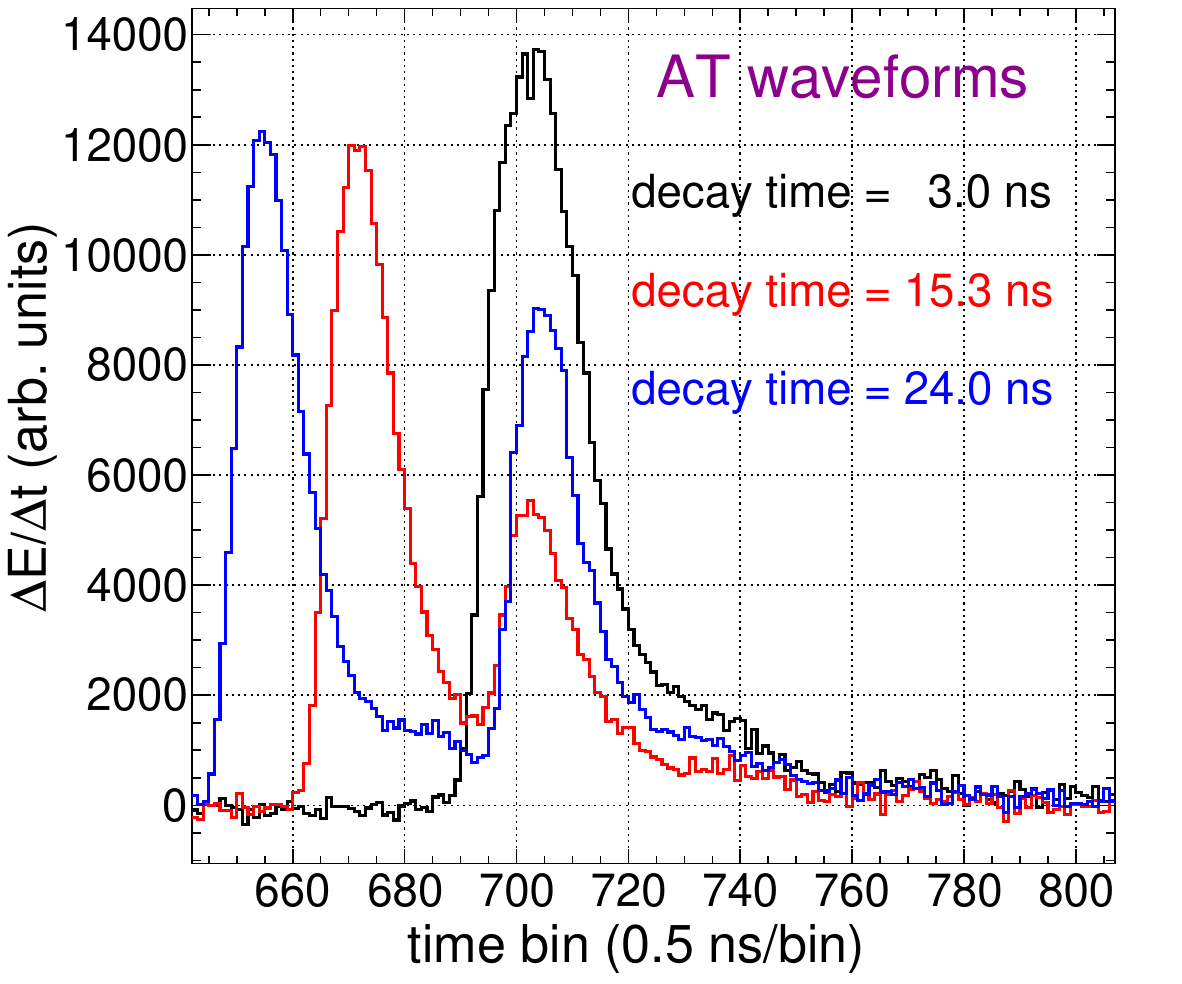}
    \caption{Illustration of the effect of AT waveform pileups.  A
      shorter delay between peaks represents a faster decay, which in
      turn complicates target energy reconstruction for the outgoing
      decay positron.  By subtracting the predicted $\pi$ and $e$
      energies, a ``rest energy'' is evaluated for every event, which
      may or may not show an intermediate muon.  \label{fig:tgt_pileup}
  } }
  \hspace*{\fill}
  \parbox{0.49\linewidth}{
    \centering
    \includegraphics[width=0.92\linewidth]  
                    {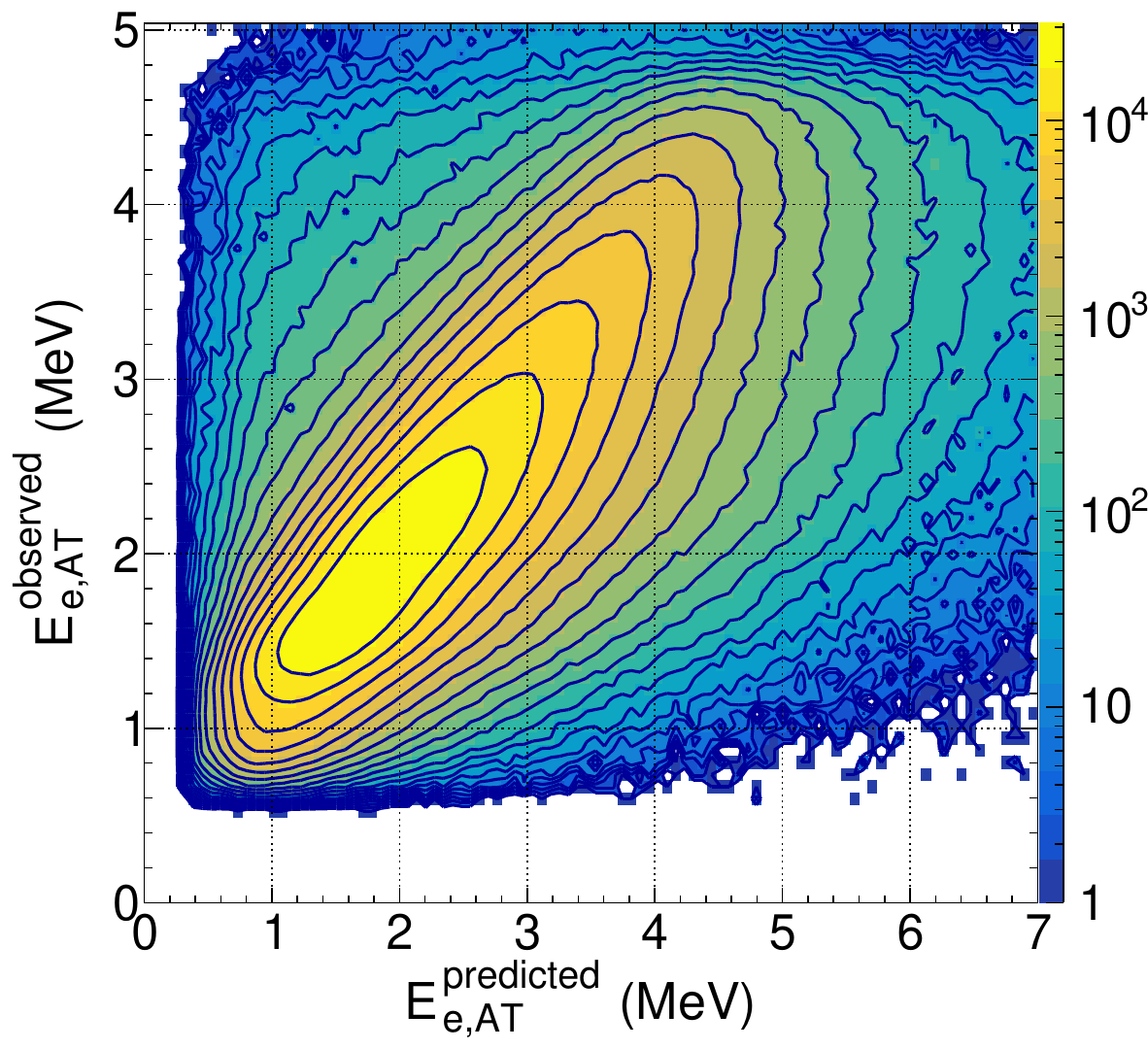}
    \caption{Observed versus predicted positron energy in the active
      target for well separated (in time) ``Michel'' decays, i.e.,
      positrons that emerge from the decay chain $\pi \to \mu \to e$.
      \label{fig:Ee_tgt_pred_vs_meas}  }
                        }
\end{figure}

The ability to reproduce the values of several beam-related observables
is key to the reliable reconstruction and interpretation of the PEN
measured events in the following analysis steps.
\begin{itemize}
  
  \item Predicted stopping position $(x,y,z)_{\pi,\text{stop}}$ of the
    pion in the target (mTPC plus BC--AD TOF), yields
    $E_{\pi,\text{AT}}^{\text{predicted}}$, the predicted energy deposited
    by the pion in the AT .  Combined with the $e^+$ trajectory (MWPCs),
    $(x,y,z)_{\pi,\text{stop}}$ enables a geometrical estimate of
    $\lambda_{e,\text{AT}}$, the decay positron's pathlength in the AT.

  \item Combining $\lambda_{e,\text{AT}}^{\text{predicted}}$ with the
    known stopping power d$E$/d$x$ of $e^+$'s in PVT, yields
    $E_{e,\text{AT}}^{\text{predicted}}$, the predicted energy deposition
    by the decay positron in the target.

  \item Evaluating the target rest energy, $E_{\text{AT}}^{\text{rest}}
    = E_{\text{AT}}^{\text{total}} -
    (E_{\pi,\text{AT}}^{\text{predicted}} +
    E_{e,\text{AT}}^{\text{predicted}})$, enables the discrimination of
    the two main decay channels of interest to PEN: $\pi^+ \to
    \mu^+\nu_{\mu}(\gamma)$ and $\pi^+ \to e^+\nu_{e}(\gamma)$, without
    relying on information from the CsI calorimeter, as discussed below.
    Distinguishing these two processes in the measured data with high
    accuracy is the greatest challenge of the experiment.
\end{itemize}
\begin{figure}[b!]
  \centering
  \includegraphics[height=0.27\textheight]{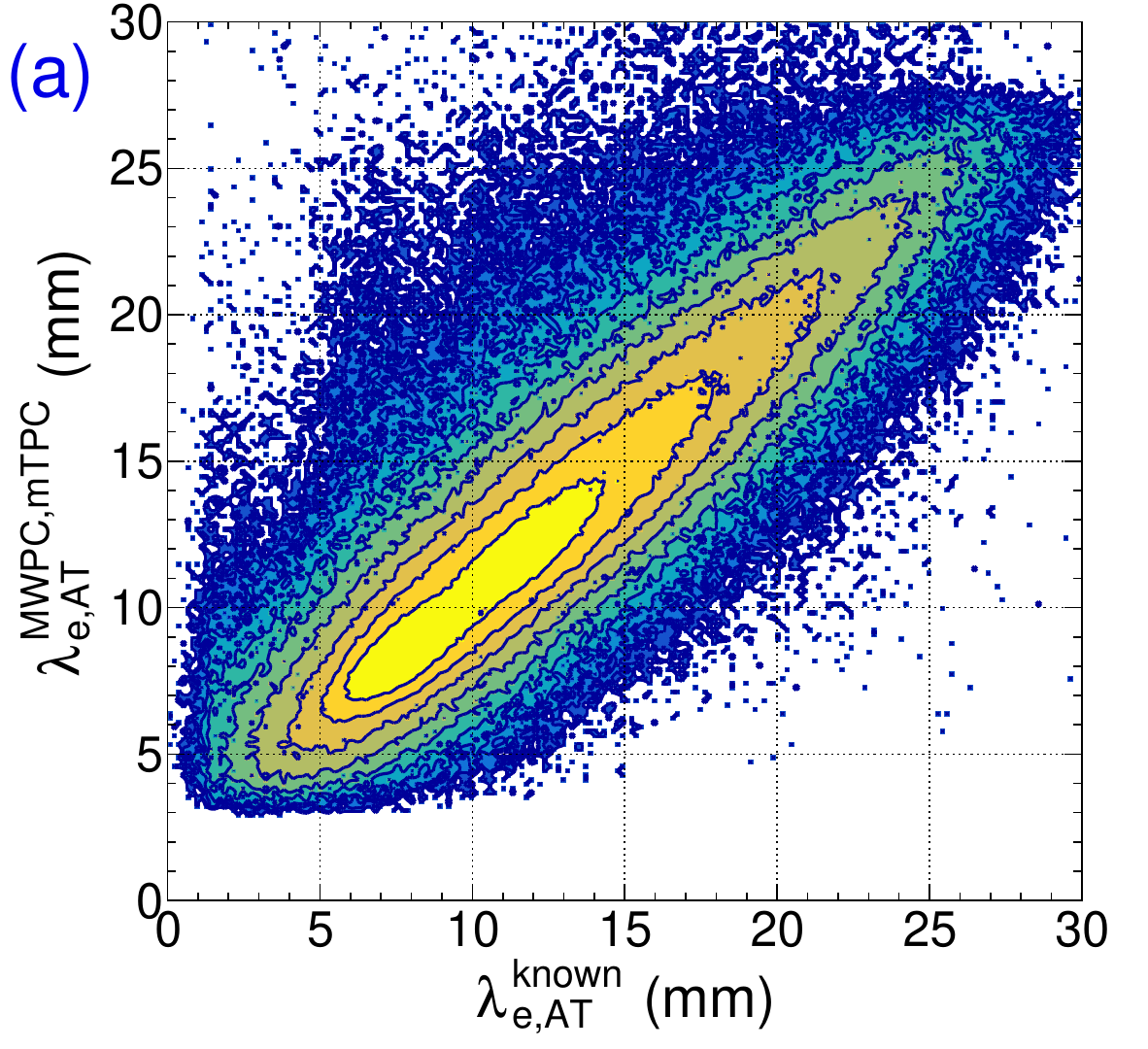}
  \includegraphics[height=0.27\textheight]{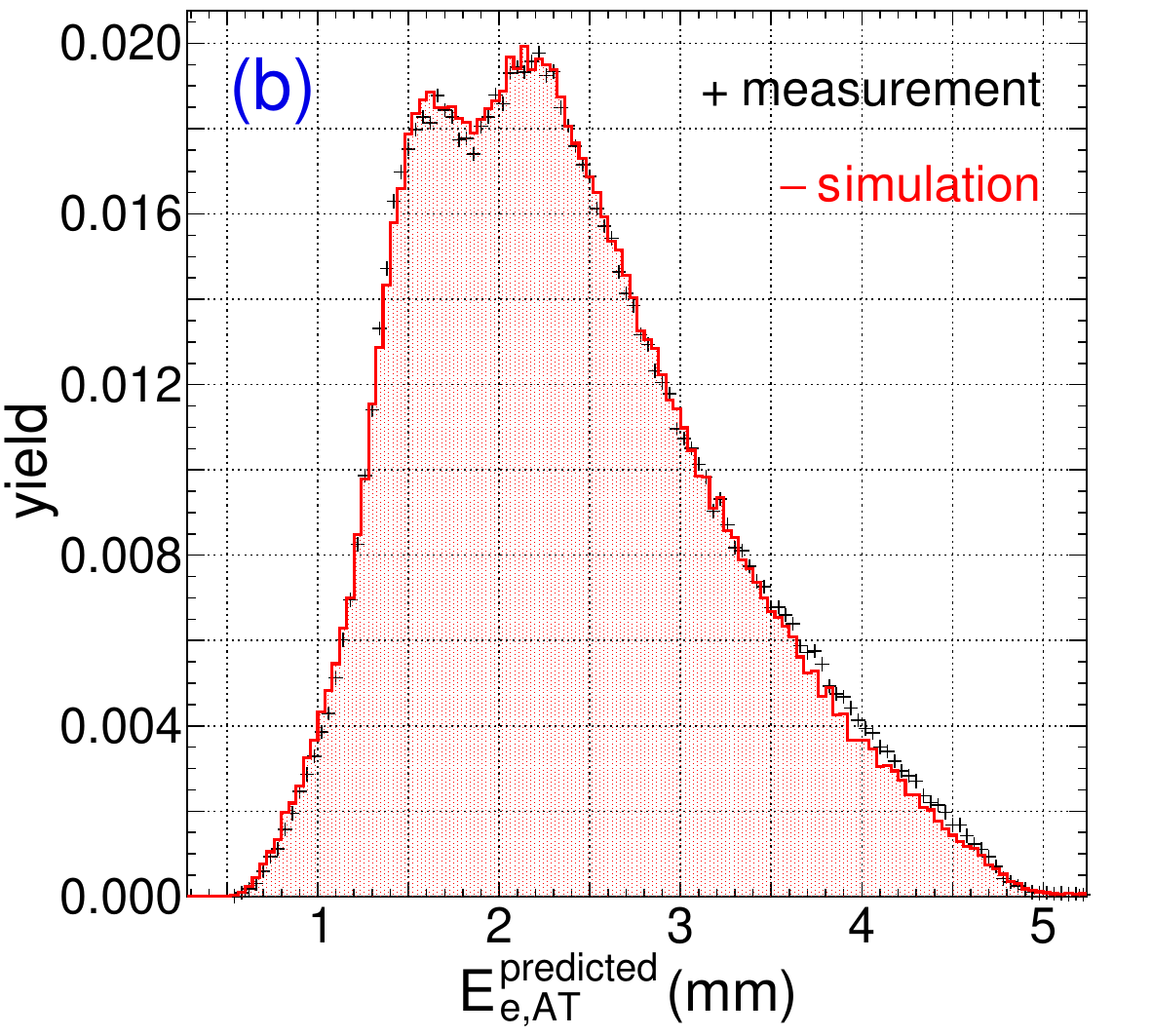}
  \caption{(a) MC simulation: reconstructed $e^+$ pathlength in AT using
    the MWPC/mTPC combination algorithm for PEN data vs.\ known
    pathlength.  (b) Predicted positron energy in the active target,
    calculated using reconstructed pathlength.  The two-peak structure
    is a consequence of the positron leaving through the curvilinear
    side, or one of the flat surfaces of the AT (pion stops were not
    centered at the $z$ midpoint of the AT).  \label{fig:pred_Ee_tgt} }
\end{figure}
The accuracy of the reconstruction of the positron trajectory, and thus
its energy deposition inside AT, is illustrated in
Fig.~\ref{fig:pred_Ee_tgt}.  Of crucial importance is the fact that the
method doesn't make use of the AT waveform which otherwise might
introduce a dependence on pion decay time, a dangerous source of
systematic error in \RpiEXPemu.  In the cases where the decay occurs
relatively quickly, the resulting superposition (``pileup'') of the
stopping (pion) and decay particle (muon, positron) overlapping signals
in the target waveform makes it difficult to determine the energy
deposited in the target by the positron as opposed to the pion or muon.
The effect is illustrated with synthetic waveforms in
Fig.~\ref{fig:tgt_pileup}.  By relying on subtraction of predicted $\pi$
energy in the target, our method avoids pulse fitting, so there is no
minimum pulse separation or minimum pulse amplitude requirement for a
valid result.

For events with well separated stopping and decay particle signals in
the target waveforms, we observe a strong linear correlation between the
positron energy deposition from the AT signal, and its value from the
reconstructed e$^+$ path, as shown in Fig.~\ref{fig:Ee_tgt_pred_vs_meas}.

An astute reader will notice two minor anomalies in two figures above.
First, the shortest reconstructed pathlengths in
Fig.~\ref{fig:pred_Ee_tgt} depart from the diagonal, trending slightly
above the corresponding known values.  These events are decays occurring
close to the edge of the AT.  The required minimum AT positron energy
cut filters such reconstructed events asymmetrically, favoring longer
pathlengths in the AT, thus producing the observed effect.  The second
anomaly concerns a slight high-$E_{e,\text{AT}}^{\text{predicted}}$ tail
in Fig.~\ref{fig:Ee_tgt_pred_vs_meas}, another target edge effect.
These rare events correspond to pion stops predicted to lie slightly
outside the AT, or at its very edge, followed by a decay positron which
turns back and traverses the full extent of the AT on its way to the
calorimeter.  Meanwhile, the actual pion stop occurs inside the AT, and
the decay positron traverses a shorter pathlength through the AT.  As
seen in Fig.~\ref{fig:trac_profiles}, in recognition of the
$\mathcal{O}$(2\,mm) mTPC track resolution in the target, the PEN
analyzer accepts tracks that skirt or narrowly miss the AT.

\begin{figure}[t!]
  \centering
  \hspace*{6pt}
  \includegraphics[height=0.25\textheight]{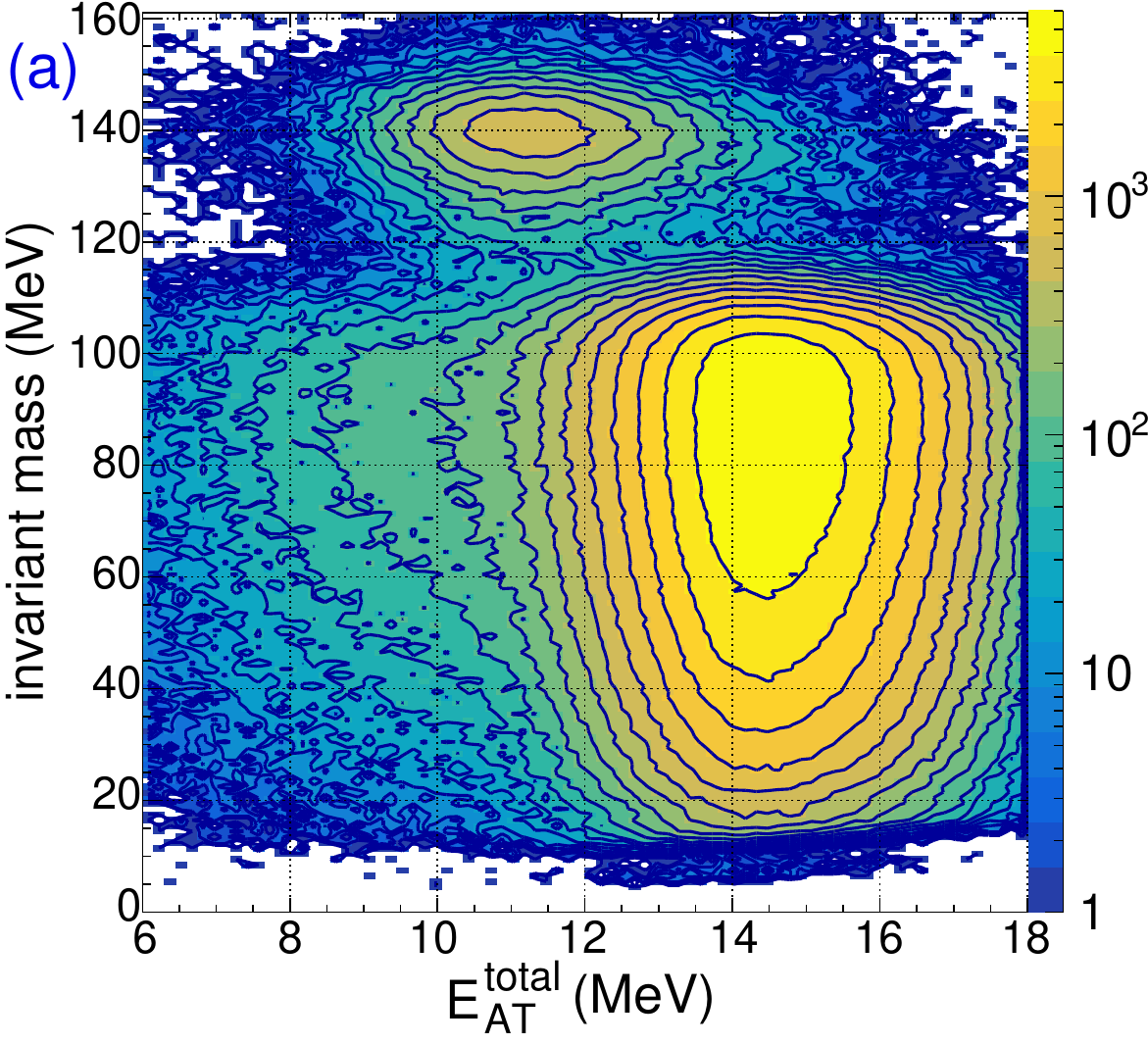}
  \hspace*{8pt}
  \includegraphics[height=0.25\textheight]{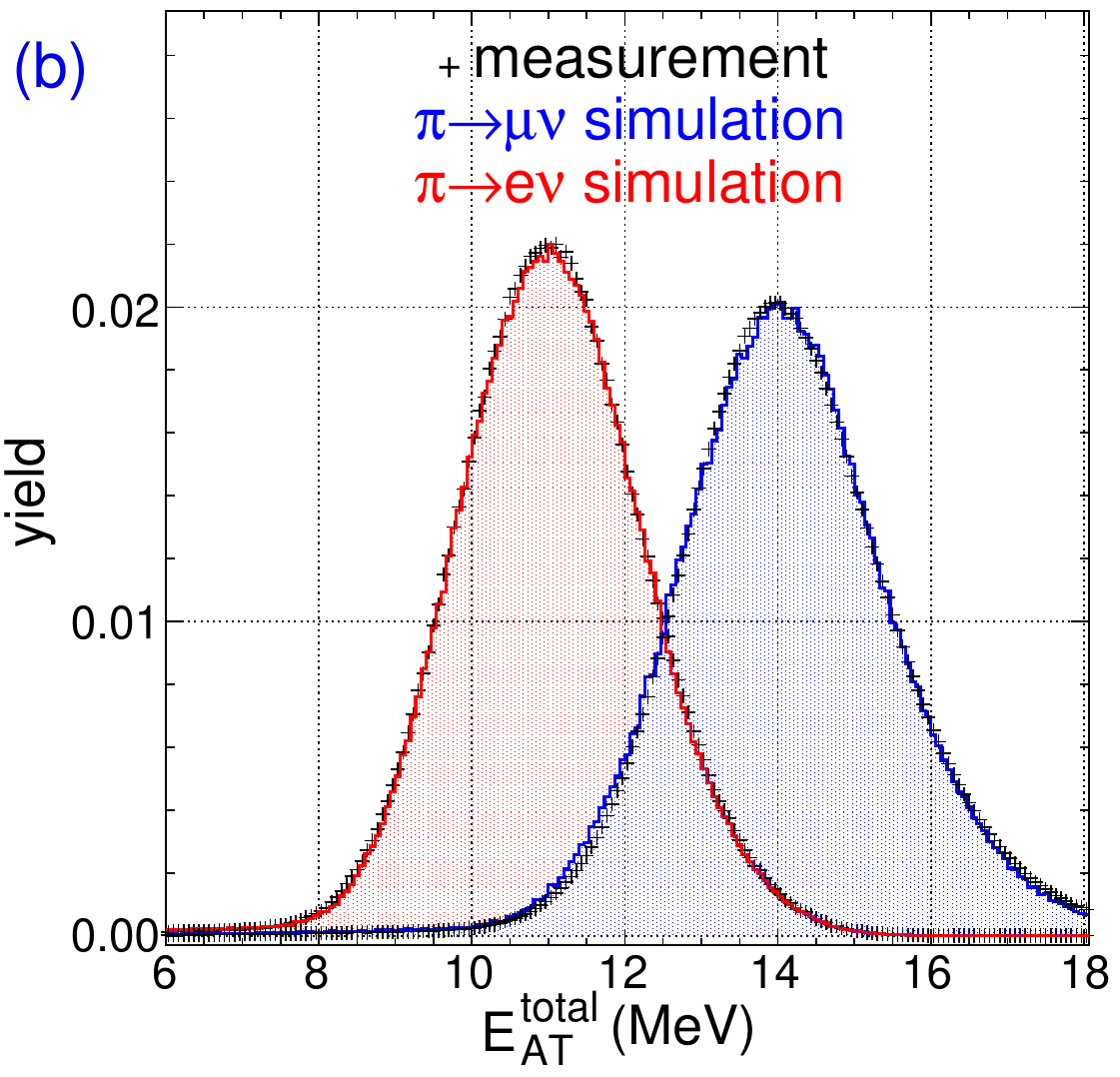}
  \includegraphics[height=0.25\textheight]{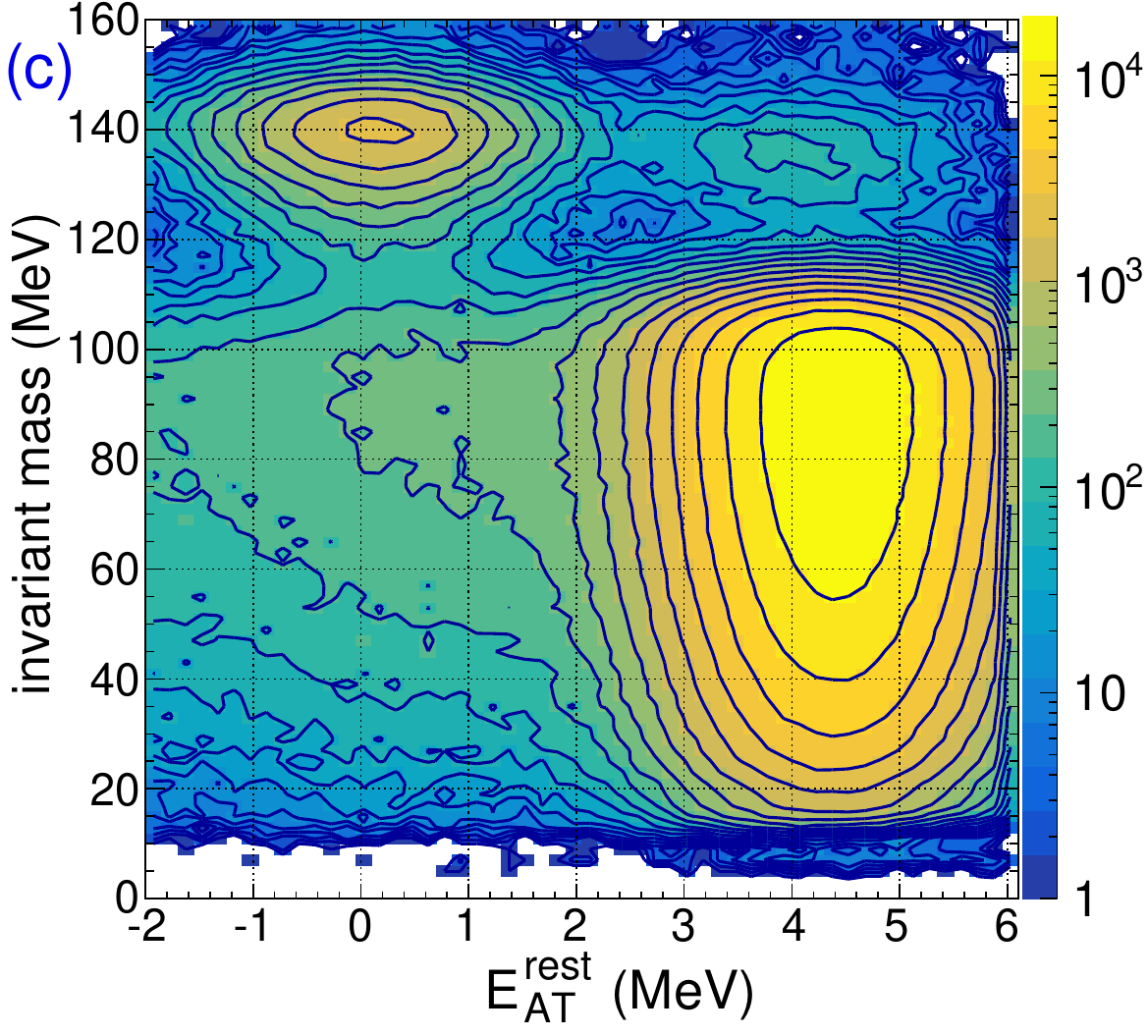}
  \includegraphics[height=0.25\textheight]{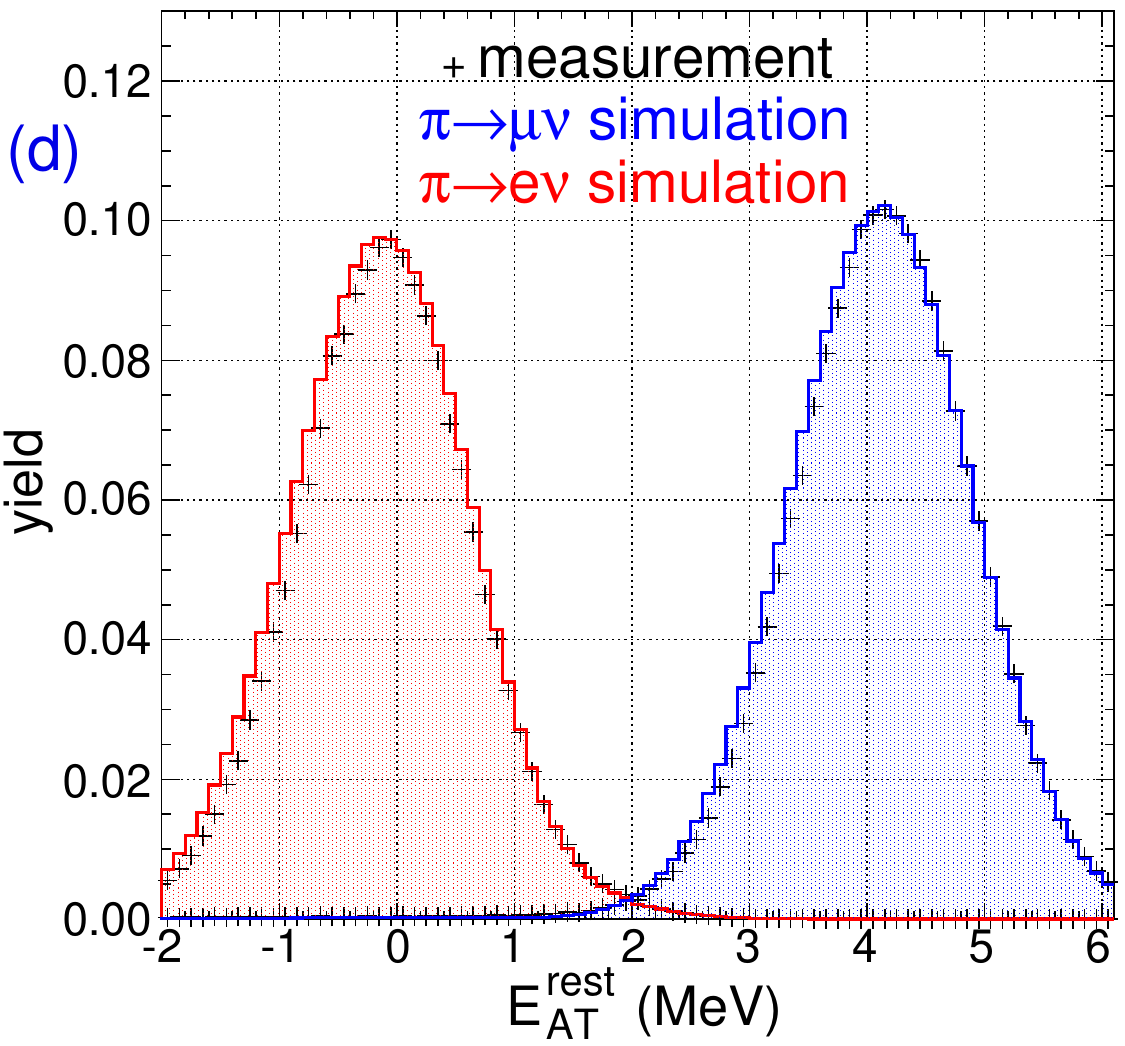}\\
  \includegraphics[height=0.25\textheight]{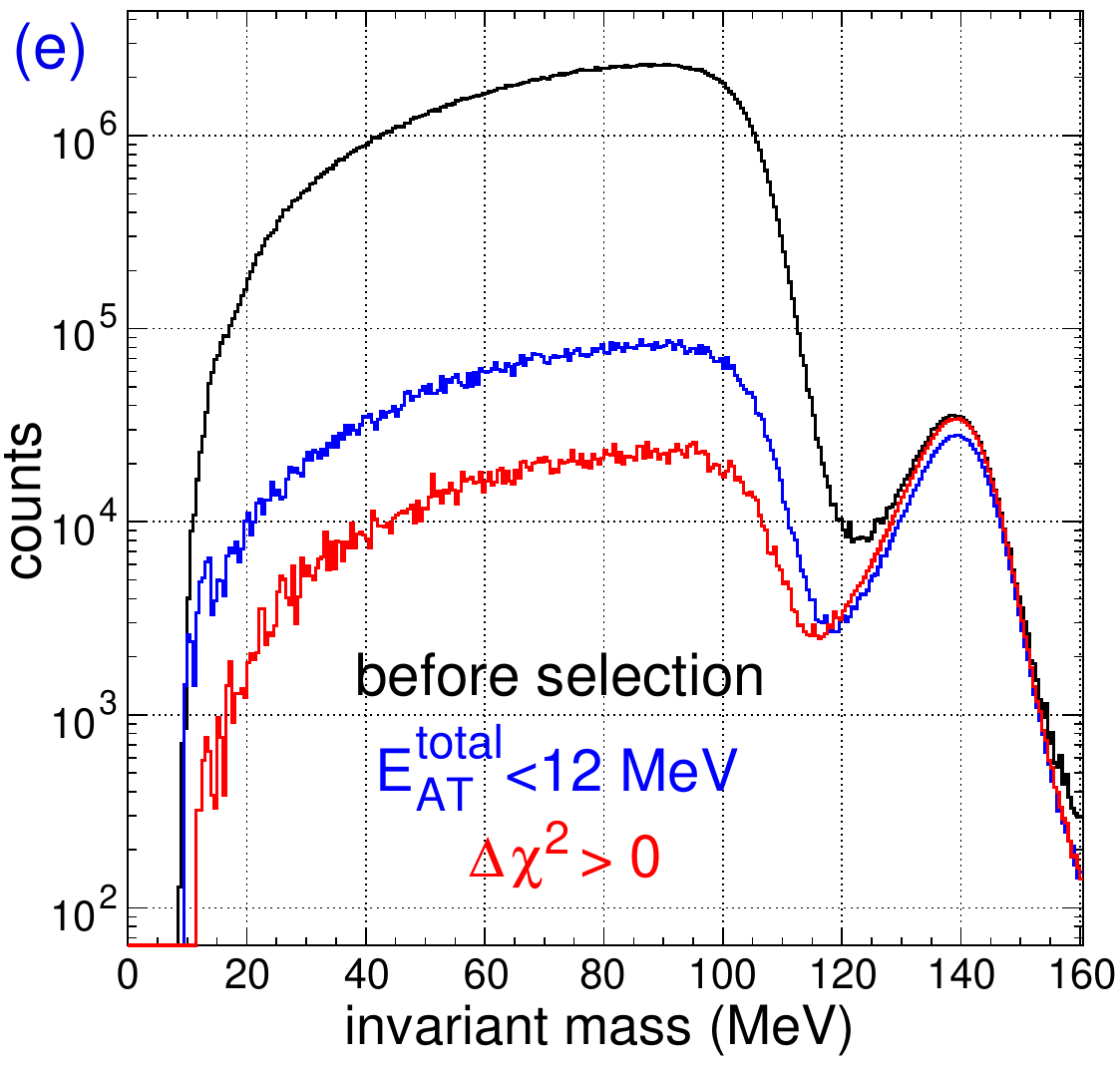}
  \caption{(a) Reconstructed invariant mass $m_0$ (see text) plotted
    against the target total energy, $E_{\text{AT}}^{\text{total}}$. (b)
    1-dimensional projection on $E_{\text{AT}}^{\text{total}}$ for the
    two main pion decay channels with comparison to simulation.  (c) and
    (d): same as (a) and (b), respectively, but using rest energy,
    $E_{\text{AT}}^{\text{rest}}$ instead of
    $E_{\text{AT}}^{\text{total}}$.  The separation of \pieii\ and \pimii\
    decays is much better than in (a) and (b).  PEN branching ratio
    analysis uses $\Delta\chi^2$, a more discriminating variable that
    builds on $E_{\text{AT}}^{\text{rest}}$.  (e) shows the effect of
    $E_{\text{AT}}^{\text{total}}$ vs.\ $\Delta\chi^2$ on the
    suppression of \pimii\ decays in the final sample of decays for
    analysis.  The ``peak'' $\pi_{e2}$ events are further defined by
    an invariant mass threshold typically set above 110\,MeV.
    \label{fig:tgt_restwave} }
\end{figure}

The extraction of the positron energy loss in AT independent of the
detailed AT waveform information is useful in separating the two main
channels of the pion decay, $\pi\to \text{e}\nu(\gamma)$ and $\pi \to
\mu\nu(\gamma)\to \text{e} \nu \bar{\nu} (\gamma)$ through the use of
higher order observables.  We recall $E_{\text{AT}}^{\text{rest}}$, the
target rest energy, obtained by subtracting the predicted energies
deposited in the AT by the pion and the positron from
$E_{\text{AT}}^{\text{total}}$ for each event.  For a $\pi \to \text{e}\nu
(\gamma)$ event, this should result in $E_{\text{AT}}^{\text{rest}} =
0$.  However, for the $\pi \to \mu \nu(\gamma) \to \text{e}\nu
\bar{\nu}(\gamma)$ decay chain, $E_{\text{AT}}^{\text{rest}} \simeq
4$\,MeV, the muon kinetic energy, as seen in
Fig.~\ref{fig:tgt_restwave}.
The invariant mass variable shown in the figure is evaluated as $m_0 = E
+|\vec{p}|c$, where $E$ is the sum of detected energy for the event in
the apparatus, while $\vec{p}$ is the vector sum of momenta associated
with all showers ($\vec{p}$ is the best available measure of the energy
of the unobserved neutrino in \pieiisg\ decay).  Invariant mass defined
in this way seamlessly accommodates radiative \pieiig\ decays alongside
the dominant variety with undetected/unseparated soft photon.

Several comments are appropriate here.
First, without the combined mTPC and MWPC tracking, PEN analysis would
not be able to evaluate the target rest energy for every event that
passes other requirements valid for the signal and normalization decay
channels (there are a relative few events that fail to produce valid
tracks).  If the analysis had to rely on the target waveforms alone, a
waveform fit based separation would not be reliable for a large subset
of closely spaced decay pulses, consequently greatly diminishing the
number of usable events.  A way around that is to use the
$E_{\text{AT}}^{\text{total}}$, the total target energy instead of
waveform fitting.  As Fig.~\ref{fig:tgt_restwave} amply demonstrates,
that does not work nearly as well as using
$E_{\text{AT}}^{\text{rest}}$.

Second, while $E_{\text{AT}}^{\text{rest}}$ produces remarkably better
$e/\mu$ decay separation compared to that achieved by
$E_{\text{AT}}^{\text{total}}$, it is not the most discriminating
variable between the \pieii\ and \pimii\ channels.  That distinction
belongs to $\Delta\chi^2$, a more complicated variable that builds on
$E_{\text{AT}}^{\text{rest}}$; its detailed discussion is beyond the
scope of the present paper.  The variable $\Delta\chi^2$ is introduced
in \cite{Poc14}, and will be further discussed in a forthcoming
publication~\cite{Gla21}.

The effects on the final selection of events used for branching ratio
analysis are seen in the final panel, Fig.~\ref{fig:tgt_restwave}(e).
\begin{figure}[t!]
  \renewcommand{\fwfr}{0.49}
  \centering
  \includegraphics[width=\fwfr\linewidth]{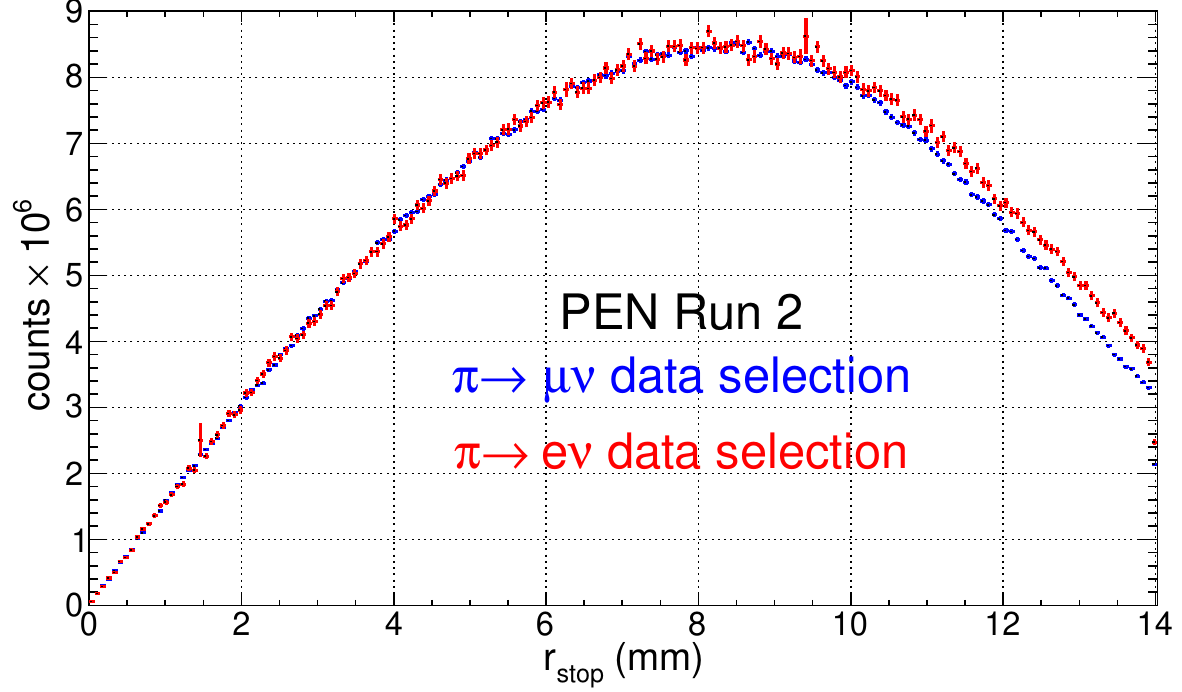}
  \includegraphics[width=\fwfr\linewidth]{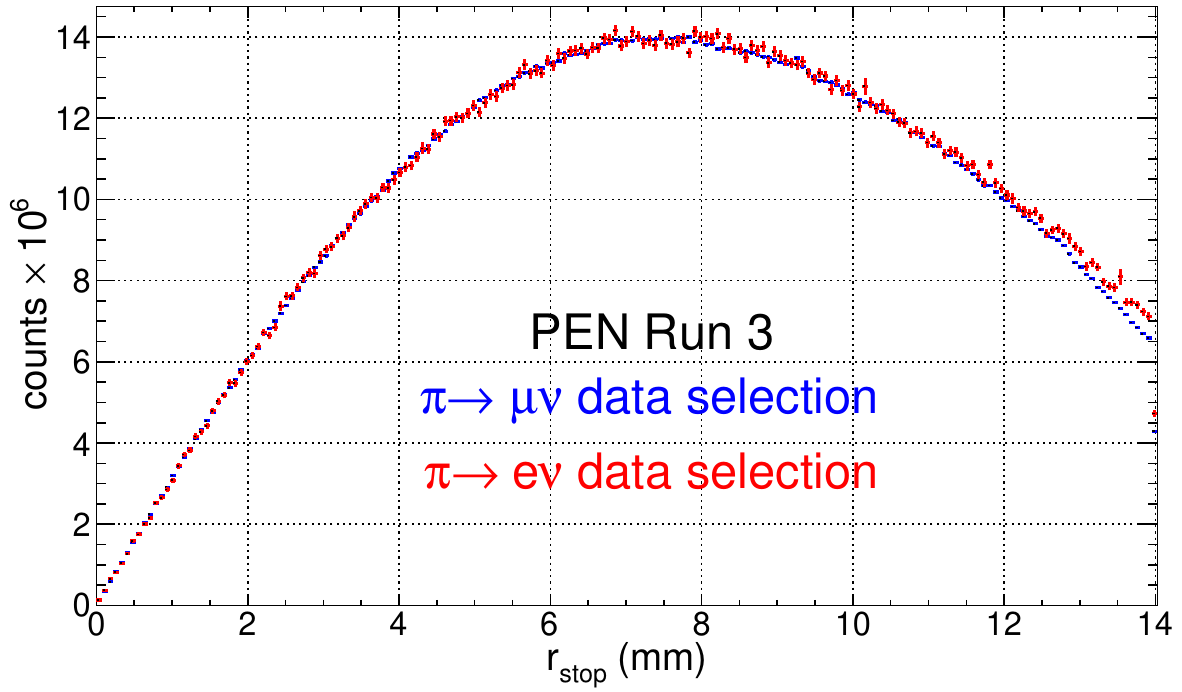} \\ 
  \vspace{3pt}
  \includegraphics[width=\fwfr\linewidth]{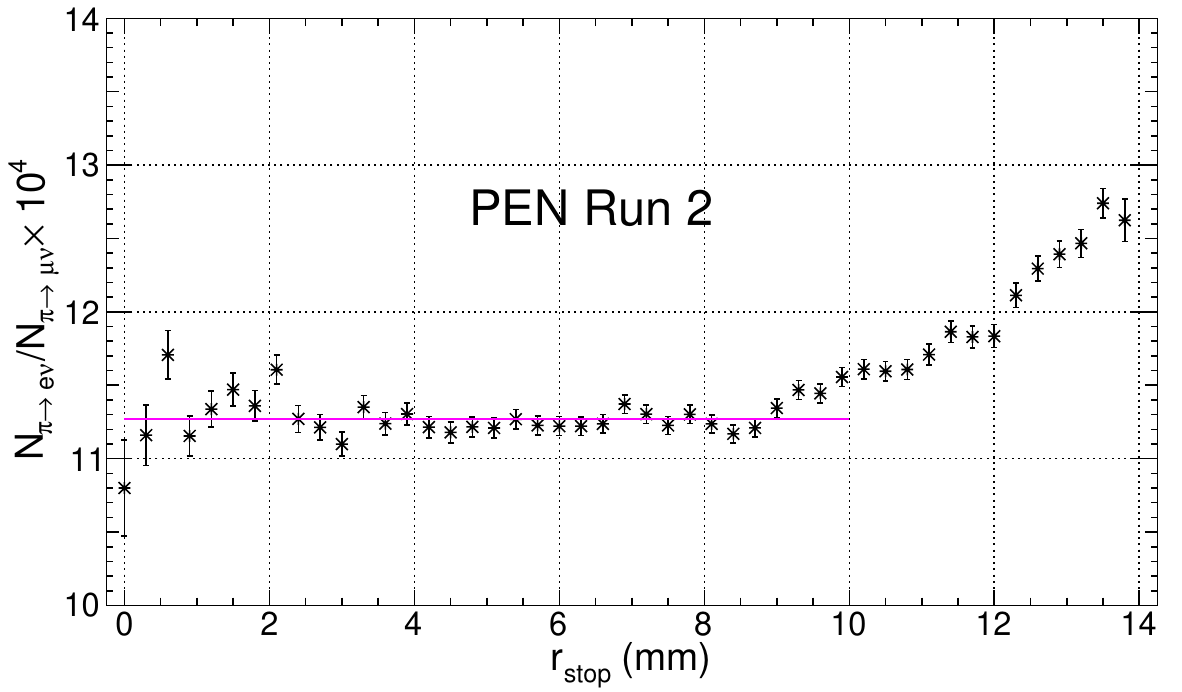}
  \includegraphics[width=\fwfr\linewidth]{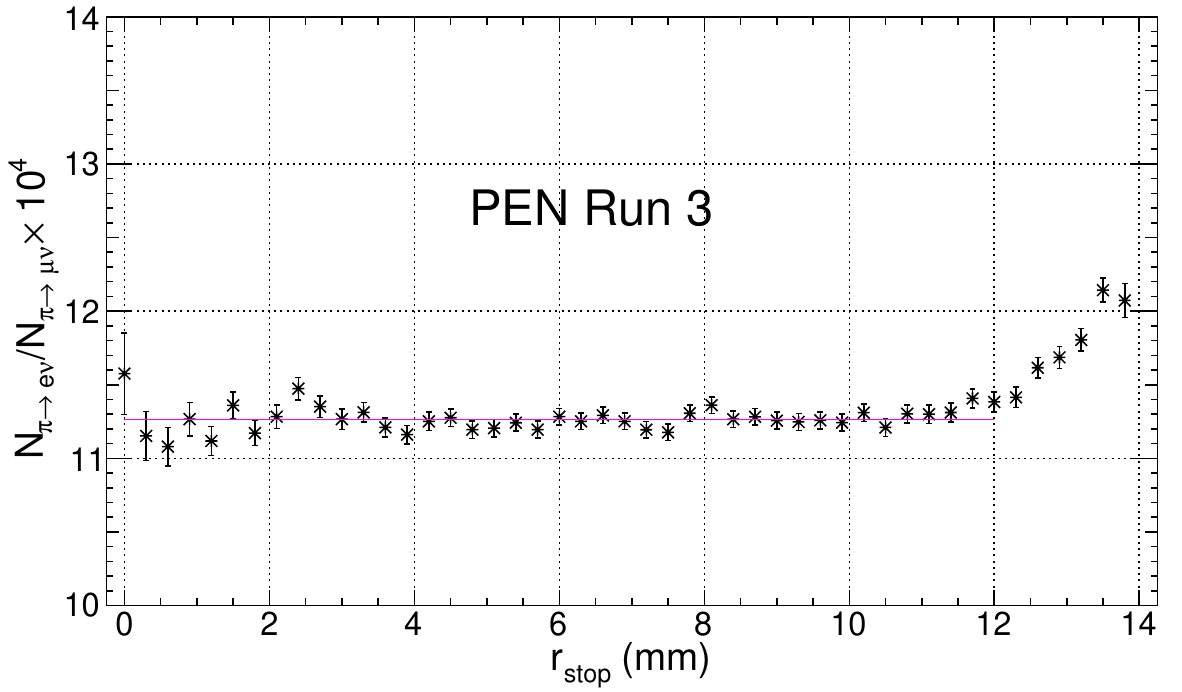} 
  \caption{Top plots: comparison of the radial distribution in the target
    for the $\pi\to\mu\,(\to e)$ decay selection (blue) and the pion
    electronic decay selection (red) for branching ratio extraction with
    the signal decay normalized and errors propagated accordingly for
    Runs\,2 (left) and 3 (right).  Bottom plots: The effect of the
    stopping distribution on the key ratio $R_0 = N[\pi\to
      \text{e}\nu(\gamma)]/N[\pi\to \mu \nu(\gamma)]$ in Runs\,2 (left)
    and 3 (right).  We note that $R_0$ is a raw number, far from a final
    branching ratio, as multiple factors are not applied to the
    expression.  The horizontal line represents the weighted average of
    the first 9.5\,mm and 12\,mm in the stopping radius within the
    target, \rstop, for Runs\,2 (left) and 3 (right), respectively.
    \label{fig:B_vs_rstop}}
\end{figure}
Clearly, compared to $E_{\text{AT}}^{\text{total}}$, the mTPC-based
$\Delta\chi^2$ cut suppresses the \pimii\ background in the ``tail'' by
an additional factor of $\sim$\,5, also lowering the invariant mass
threshold separating the ``peak'' and ``tail'' $\pi\to e\nu(\gamma)$
regions.  Finally, unlike the $E_{\text{AT}}^{\text{total}}$ cut, the
$\Delta\chi^2$ cut does not significantly reduce the peak \pieii\ yield
(the same applies to a $E_{\text{AT}}^{\text{rest}}$ cut).  Combining
all of these factors would restrict a non-mTPC analysis to the range
above the $10^{-3}$ in $\Delta\Rpiemu/\Rpiemu$.

A way to avoid target waveform fitting altogether is to perform a
branching ratio analysis based on event decay time which relies on the
characteristic time signatures of the $\pi \to e \nu(\gamma)$ decay and
the $\pi\to\mu\to e$ chain to separate the two processes.  This approach
is limited by corrections due to particle decays in flight.  Muon decay
in flight, $\mu_{\text{DIF}}$, is particularly insidious, as it occurs
at $\mathcal{O}(10^{-5})$ level, and has the same decay time signature
as the main signal, similarly restricting the achievable precision to
above $10^{-3}$.  Again, in PEN the $\mu_{\text{DIF}}$ correction
becomes tractable with the aid of mTPC-based cuts.

In addition to constructing highly discriminating observables, the mTPC
is used to ensure that the pion stops well within the target for all
events included in the analysis (top plots in Fig.~\ref{fig:B_vs_rstop}).
If a pion comes to a stop too close to the radial boundary of the
target, $r=15$\,mm, the muon produced in its decay may escape from the
target volume, obscuring the rest energy determination.  The exit of the
decay muon from the target effectively removes its decay from detection,
and leads to a deficit of counts of the $\pi \to \mu \to e$ decay chain
positrons, used for normalization in the branching ratio determination.
This deficit of recorded muon decay events systematically produces a
higher branching ratio, as demonstrated in the bottom plots of
Fig.~\ref{fig:B_vs_rstop}, which show that the raw ratio $R_0 = N[\pi\to
  \text{e}\nu(\gamma)]/N[\pi\to \mu \nu(\gamma)]$, the main input in the
branching ratio calculation, is stable for stopping radii within the
first 9.5\,mm (Run\,2), or 12.5\,mm (Run\,3), a result that is not
unexpected given the lower mass, more compact central detector setup
in Run\,3.  Pion stopping position information in the target, deduced
primarily from the mTPC beam tracking, is critical for the accurate
evaluation of the simulated acceptance corrections for events outside
the $\rstop \simeq 12.5$\,mm region for Run\,3 (or $\rstop \simeq
9.5$\,mm for Run\,2), necessary for their safe inclusion in the
analysis.  Including the large-\rstop\ events lowers the statistical
uncertainty of the branching fraction evaluation, to the overall
precision goal of $\Delta \Rpiemu / \Rpiemu \simeq 5 \times 10^{-4}$.
This is yet another essential contribution of the mTPC to the PEN
analysis.

\section{Conclusions}
The mTPC is an efficient detector, used for beam particle tracking in
the PEN experiment.  Beam particle trajectories based on mTPC data are
essential for constructing the pion stopping location in the active
target.  Precise knowledge of the stopping distribution plays a central
role in the acceptance systematics for the primary (\pieii) and
normalization ($\pi\to\mu\to e$) decay processes at the experiment's
intended precision.

In addition, the mTPC beam trajectory information enables the
construction of a number of observables critical for discriminating the
primary and normalization processes.  Reliable discrimination of the
two, based on information other than the CsI calorimeter energy, is the
central challenge of the experiment in terms of achieving its design
precision.

\goodbreak
The importance of the mTPC in PEN further extends to the basic
evaluation of the \pieii\ branching ratio, \Rpiemu, before corrections
are applied.  Event reconstruction and selection algorithms in PEN favor
\pieii\ decays over the \pimii\ decay chain at the high tail of the beam
pions' radial stopping distribution, due to imbalances in the deposited
energy in the target, or because of distortions of the decay time
distributions.  Restricting the accepted events only to low
\rstop\ values notably reduces the overall event count.  Thanks to the
ultra-realistic simulation of the mTPC detector response, the higher
\rstop\ events can be included in the branching ratio analysis without
introducing undue bias.  This significantly increases the accepted event
statistics for the $\pi\to e\nu (\gamma)$ decay, keeping it consistent
with the PEN overall precision goal.

\section*{Acknowledegments}
The PEN collaborators gratefully acknowledge support from the US
National Science Foundation, the Paul Scherrer Institute, and the
Russian Foundation for Basic Research.  The authors express their
sincere gratitude to S.M.~Korenchenko for his continued interest in this
work, and thank S.N.~Shkarovsky for the chamber field calculations.
Finally, the authors dedicate this work to the memory of our friend and
colleague Andrey Korenchenko, who developed the second version of this
unique TPC, and took part in the testing and optimization of both
chambers.



\label{}





\bibliographystyle{elsarticle-num}
\bibliography{<your-bib-database>}



\goodbreak



\end{document}